\documentclass[11pt,a4paper]{article}
\usepackage{bbm}
\pdfoutput=1
\usepackage{jheppub}
\usepackage{amsmath}
\usepackage{epsfig}
\usepackage{amssymb}
\usepackage{graphics}
\usepackage[active]{srcltx}
\usepackage{epstopdf}
\usepackage{subfigure}
\usepackage{shuffle}

\newcommand{\be}{\begin{equation}}
\newcommand{\ee}{\end{equation}}
\newcommand{\bea}{\begin{eqnarray}}
\newcommand{\eea}{\end{eqnarray}}
\newcommand{\bean}{\begin{eqnarray*}}
\newcommand{\eean}{\end{eqnarray*}}
\newcommand{\nn}{\nonumber \\}

\def\W #1{\widetilde{#1}}

\def\eref#1{(\ref{#1})}

\def\a{{\alpha}}

\def\Label#1{\label{#1}%
  \smash{\hbox to0pt{\raise1ex\hbox{\tiny[#1]}\hss}}}

\title{Transmutation operators and expansions for $1$-loop Feynman integrands}
\author[a]{Kang Zhou,}

\affiliation[a]{Center for Gravitation and Cosmology, College of Physical Science and Technology, Yangzhou University,\\
 No.180, Siwangting Road, Yangzhou, 225009, P.R. China.}

\date{\today}
\abstract{In this paper, the connections among $1$-loop Feynman integrands of a large variety of theories with massless external states are further investigated. The work includes two parts. First, we construct a new class of differential operators which transmute the $1$-loop gravitational Feynman integrands to $1$-loop Yang-Mills Feynman integrands. The new operators are commutable with the integration of loop momentum, thus the corresponding transmutational relations hold at not only the integrands level, but also the $1$-loop amplitudes level. Secondly, by using $1$-loop level transmutational relations, together with some general requirements such as gauge and Lorentz invariance, we derive the expansions of the Feynman integrands of one theory to those of other theories. The unified web for expansions is established, including a wide range of theories which are gravitational theory, Einstein-Yang-Mills theory, Einstein-Maxwell theory, Born-Infeld theory, pure Yang-Mills theory, Yang-Mills-scalar theory, special Yang-Mills-scalar theory,
Dirac-Born-Infeld theory, extended Dirac-Born-Infeld theory, special Galileon theory, non-linear sigma
model. The systematic rules for evaluating coefficients in the expansions are provided, and the duality between transmutational relations and expansions is shown.
}


\begin{document}

\maketitle \flushbottom

\section{Introduction}
\label{secintro}

The modern researches on S-matrix have exposed amazing relations and common structures within amplitudes of gauge and gravity theories,
such as the Kawai-Lewellen-Tye (KLT) relations \cite{KLT}, Bern-Carrasco-Johansson (BCJ) color-kinematics duality \cite{Bern:2008qj,Bern:2010ue,Bern:2010yg}, which
are invisible upon inspecting the traditional Feynman rules. These unexpected connections hint the existence
of some long hidden unifying relations for on-shell amplitudes. The marvelous unity was first revealed in
\cite{Cachazo:2014xea} at the level of Cachazo-He-Yuan (CHY) integrands \cite{Cachazo:2013gna,Cachazo:2013hca,Cachazo:2013iea,Cachazo:2014nsa,Cachazo:2014xea}. In the CHY framework, different
theories are defined by different CHY integrands, while they found that CHY integrands for a wide range of theories can be generated from the CHY integrand for gravity theory, through three manipulations which are compactifying, squeezing,
as well as the generalized dimensional reduction procedures \cite{Cachazo:2014xea}.
More recently, similar unifying relations for
on-shell tree amplitudes of a variety of theories, based on constructing some Lorentz and gauge invariant differential operators, were proposed by Cheung, Shen and Wen \cite{Cheung:2017ems}.
By acting these transmutation operators, one can generates the physical tree amplitudes of various theories from the gravitational tree amplitudes.
The similarity between two unified webs implies the underlying connection between two approaches. This connection has been spelled out in \cite{Zhou:2018wvn,Bollmann:2018edb,Zhou:2020umm},
by applying transmutaton operators to CHY integrals for different theories.

At the tree level, another significant reflection of these connections among amplitudes is that tree-level amplitudes of one theory can be expanded to those of other theories, which have been studied in various literatures recently, especially for the expansions of tree Einstein-Yang-Mills amplitudes to tree Yang-Mills ones \cite{Stieberger:2016lng,Schlotterer:2016cxa,Chiodaroli:2017ngp,DelDuca:1999rs,Nandan:2016pya,delaCruz:2016gnm,Fu:2017uzt,Teng:2017tbo,Du:2017kpo,
Du:2017gnh}. To obtain the coefficients in expansions, several efforts have been devoted in above literatures. Among these methods, the recent proposed approach, which based on differential operators \cite{Feng:2019tvb}, indicates that transmuting amplitudes via differential operators in \cite{Cheung:2017ems,Zhou:2018wvn,Bollmann:2018edb} and expanding amplitudes relate to each other. As discussed in \cite{Feng:2019tvb}, expansions of single-trace Einstein-Yang-Mills and gravitational amplitudes to color ordered Yang-Mills amplitudes can be reached by solving differential equations indicated by transmutation operators, together with considering the gauge invariance of gravitons. Then, by applying appropriate transmutation operators on above two expansions, expansions for amplitudes of other theories can be derived straightforwardly \cite{Feng:2019tvb,Hu:2019qdq,Zhou:2019mbe}. The unified web for expansions serves as the dual version of the web for transmutational relations, as pointed out in \cite{Zhou:2019mbe}.

It is natural to study the generalizing of these unifying relations to the loop level.
This interesting question was first considered in \cite{Bollmann:2018edb}, which exposed the strong evidence for the existence of transmutation operators
at $1$-loop level. Recently, this issue has been further studied in \cite{Zhou:2021kzv}. In \cite{Zhou:2021kzv}, the $1$-loop level differential operators which link
$1$-loop Feynman integrands of different theories together were found based on the tree level operators as well as the forward limit operation, and the complete $1$-loop level unified web was established.
On the other hand, the expansions of $1$-loop Einstein-Yang-Mills and gravitational Feynman integrands to Yang-Mills ones, and the related $1$-loop BCJ numerators, were studied in \cite{Geyer:2017ela,He:2017spx}.

In this paper, we further investigate the unifying relations among $1$-loop Feynman integrands. We first construct a new kind of transmutation operators
which also link the gravitational integrands and Einstein-Yang-Mills integrands together, in a manner different from operators found in \cite{Zhou:2021kzv}. The basic idea is the same as that in \cite{Zhou:2021kzv}, which can be summarized as follows.  Suppose the tree amplitudes of theories $A$
and $B$ are connected by the operator ${\cal O}$ as ${\cal A}_B={\cal O}\,{\cal A}_A$, we seek the $1$-loop level operator
${\cal O}_\circ$ satisfying ${\cal O}_\circ\,{\cal F}\,{\cal A}_A={\cal F}\,{\cal O}\,{\cal A}_A$, where the operator ${\cal F}$ denotes taking the forward limit.
Since the $1$-loop Feynman integrands can be obtained via
the forward limit, one can conclude that the operator ${\cal O}_\circ$ transmutes the Feynman integrand in the desired manner ${\bf I}_B={\cal O}_\circ\,{\bf I}_A$. In this paper, we make new choices of operators which transmute the tree gravitational amplitudes to the tree Einstein-Yang-Mills amplitudes, which are different from the choices in \cite{Zhou:2021kzv}. The new choices lead to a class of new $1$-loop level operators. The obtained new operators are commutable with the integration of loop momentum, thus the transmutational relation holds at not only the integrands level, but also the $1$-loop amplitudes level. These new operators play the crucial role when solving the expansions for $1$-loop gravitational Feynman integrands.

The existence of expansions for $1$-loop Feynman integrands can be observed from the $1$-loop CHY formulas, as long as the $1$-loop CHY integrands of various theories can be expanded to the $1$-loop CHY integrands of the bi-adjoint scalar theory. However, along this line it is not easy to work out the coefficients in the expansions. Clearly, to get the rules for computing coefficient, one can generate the expansions of $1$-loop Feynman integrands from the expansions of tree amplitudes, by taking the forward limit. This method is along the line tree-operator $\to$ tree-expansion $\to$ $1$-loop-expansion. Since the $1$-loop level differential operators
in \cite{Zhou:2021kzv} and this paper are constructed from tree level operators, we can ask whether or not another line tree-operator $\to$ $1$-loop-operator $\to$ $1$-loop-expansion gives rise to the correct expansions. In other words, we study if the differential equations provided by $1$-loop level transmutational relations, together with some general requirements such as Lorentz and gauge invariance, can fully determine the expansions of $1$-loop Feynman integrands, and if the tree level connection among transmutational relations and expansions exist at the $1$-loop level. As will be shown, the answer is positive. To reach this answer, we first solve the expansions of Einstein-Yang-Mills and gravitational Feynman integrands to Yang-Mills integrands, by employing the $1$-loop level differential operators. The idea is, we regard the transmutational relation ${\bf I}_B={\cal O}\,{\bf I}_A$ as a differential equation, which allows us to solve ${\bf I}_A$ from it. At this step, the general difficulty is that the full Feynman integrands always contain terms which are annihilated by the differential operators under consideration, i.e., terms which are not detectable. We fix these un-detectable parts by using some general principles/assumptions, in particular the gauge invariance requirement. Then, we perform the $1$-loop level differential operators to the expanded form of gravitational integrand
\bea
{\bf I}_{\rm GR}=\sum_i\,C_i\,{\bf I}_{{\rm YM};i}.
\eea
At the l.h.s, the differential operators transmute the gravitational Feynman integrand to the integrand of other theories. At the r.h.s, the differential operators modified the coefficients $C_i$, or transmute the Yang-Mills Feynman integrands to the integrands of other theories. Thus,
one can generate various new expansions via this manipulation.
Through the process described above, we find that the Feynman integrands of a large variety of theories can be double-expanded to bi-adjoint scalar (BAS) Feynman integrands,
and provide the systematic rules for constructing coefficients in the expansions. The established unified web includes a wide range of theories with massless external states, which are gravitational (GR) theory, Einstein-Yang-Mills (EYM) theory, Einstein-Maxwell (EM) theory, Born-Infeld (BI) theory, pure Yang-Mills (YM) theory, Yang-Mills-scalar (YMS) theory, special Yang-Mills-scalar (SYMS) theory,
Dirac-Born-Infeld (DBI) theory, extended Dirac-Born-Infeld (EDBI) theory, special Galileon (SG) theory, non-linear sigma
model (NLSM). The whole process only depend on the knowledge of transmutational relations, as well as some general principles/assumptions, without knowing any detail of Feynman integrands under consideration, and without recourse to any tool for evaluating Feynman integrands such as Feynman rules, CHY formulas, and so on. We also show the tree level duality between transmutational relations and the expansions can be generalized to the $1$-loop level, and give the mappings between classes of differential operators and classes of coefficients in expansions.

The remainder of this paper is organized as follows. In section \ref{secreview}, we give a brief introduction to the forward limit approach, as well as the tree and $1$-loop levels differential operators constructed in \cite{Cheung:2017ems} and \cite{Zhou:2021kzv}, which are crucial for subsequent discussions. In section \ref{NOP-GR-YM}, we construct the new operators which transmute GR Feynan integrands to EYM ones, and verify our construction via the $1$-loop CHY formulas.
Then, in section \ref{solve-coefficient}, we solve the expansions of EYM and GR Feynman integrands to YM ones, by employing the $1$-loop level operators together with some general principles/assumptions. The section \ref{sec-uni} devotes to providing the full unified web for expansions, and show the duality between transmutational relations and expansions. Finally, we end with a summary and discussions in section \ref{secconclu}.

\section{Background}
\label{secreview}

For reader's convenience, in this section we rapidly review the necessary background. In subsection. \ref{subsecforward},
we give a brief introduction to the forward limit method which generates the $1$-loop Feynman integrands from the tree amplitudes.
In subsection. \ref{subsecoperator}, we review the tree level differential operators which link the tree amplitudes of a wide range of theories together, as well as the $1$-loop level generalization of these tree level operators and unifying relations. Most of notations and conventions
which will be used in next sections are also introduced in this section.

\subsection{Forward limit and $1$-loop Feynman integrand}
\label{subsecforward}

As well known, the $1$-loop Feynman integrands can be generated from the corresponding tree amplitudes, via the so called forward limit procedure.
For instance, the $1$-loop CHY formulas can be obtained by applying this operation, as studied in \cite{He:2015yua,Cachazo:2015aol,Feng:2016nrf,Feng:2019xiq}. In this subsection we introduce the general idea and features of the forward limit.

The forward limit is reached as follows:
\begin{itemize}
\item Consider a
$(n+2)$-point tree amplitude ${\cal A}_{n+2}(k_+,k_-)$ including $n$ massless legs with momenta in $\{k_1,\cdots,k_n\}$ and two off-shell legs with $k_+^2=k_-^2\neq0$.
\item Take the limit $k_{\pm}\to \pm \ell$, and glue the two corresponding legs together. we denote this manipulation as ${\cal L}$. Performing ${\cal L}$ on the tree amplitude leads to
a special tree amplitude with $k_+=-k_-=\ell$, rather than a $1$-loop level object.
\item Sum over all allowed internal states of the internal particle with loop momentum $\ell$, such as polarization
vectors or tensors, colors, flavors, and so on\footnote{For theories include gauge or flavor groups, we only discuss the color ordered partial amplitudes in this paper, thus the summations over colors or flavors are hidden. }, we denote this manipulation as ${\cal E}$.
\end{itemize}
Roughly speaking, the obtained object, times the factor $1/\ell^2$ as
\bea
{1\over \ell^2}\,{\cal F}\,{\cal A}_{n+2}(k_+^{h_+},k_-^{h_-})={1\over \ell^2}\,\sum_{h}\,{\cal A}_{n+2}(\ell^{h},-\ell^{\bar{h}})\,,~~~~\label{single-dia}
\eea
contributes to the $n$-point $1$-loop Feynman integrand ${\bf I}_n$. Here we introduced the forward limit operator
\bea
{\cal F}\equiv{\cal E}\,{\cal L}\,,
\eea
to denote the operation of taking forward limit. In this paper, we denote the $1$-loop Feynman integrands by ${\bf I}$. From now on, we will
neglect the subscript $n$ of ${\bf I}$, since we will use other manners to denote the number of external legs.

\begin{figure}
  \centering
  \includegraphics[width=14cm]{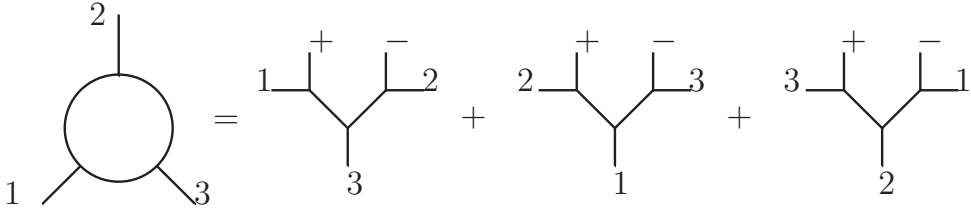} \\
  \caption{Decomposition of $1$-loop Feynman integrand.}\label{deco}
\end{figure}

For the individual Feynman diagram, the manipulation in \eref{single-dia} obviously turns the tree diagram to the $1$-loop one. However, the full
$1$-loop Feynman integrand is obtained by summing over all appropriate diagrams. Thus, let us consider what requirement should be satisfied if the resulting object of the manipulation in \eref{single-dia} can be interpreted as the correct $1$-loop Feynman integrand. It is easy to observe that after summing over all allowed tree level diagrams, each $1$-loop diagram receives contributions from tree diagrams correspond to cutting each propagator in the loop once (cutting is understood as the inverse operation of gluing legs $+$ and $-$ together, where $+$, $-$ denote external legs carry $k_+$ and $k_-$ respectively), as can be seen in Fig. \ref{deco}. Thus the statement that the operation in \eref{single-dia} generates the correct Feynman integrand holds if and only if the term for an individual $1$-loop diagram can be decomposed to terms for related tree diagrams, as shown in Fig. \ref{deco}. Such decomposition can be realized via the so called partial fraction identity \cite{He:2015yua,Baadsgaard:2015twa}:
\bea
{1\over D_1\cdots D_m}=\sum_{i=1}^m\,{1\over D_i}\Big[\prod_{j\neq i}\,{1\over D_j-D_i}\Big]\,,
\eea
which implies
\bea
{1\over \ell^2(\ell+K_1)^2(\ell+K_1+K_2)^2\cdots (\ell+K_1+\cdots+K_{m-1})^2}\simeq{1\over \ell^2}\sum_{i=1}^m\Big[\prod_{j= i}^{i+m-2}\,{1\over (\ell+K_i+\cdots+K_j)^2-\ell^2}\Big]\,.~~~~\label{loop-pro}
\eea
For each individual term at the r.h.s of the above relation, the loop momentum is shifted while result of Feynman integral is not altered.
Here $\simeq$ means the l.h.s and r.h.s are not equivalent to each other at the integrand level, but are equivalent at the integration level.
At the r.h.s of \eref{loop-pro}, we have seen the propagators with the denominates $(\ell+K_i+\cdots+K_j)^2-\ell^2$, which are different from the standard ones $(\ell+K_i+\cdots+K_j)^2$. This feature of propagators is the condition which should be satisfied if the manipulation in \eref{single-dia} provides the correct $1$-loop Feynman integrand. In CHY formulas, this requirement is satisfied via the $1$-loop level scattering equations. From the Feynman diagrams point of view, one can assume each propagator in the loop is massive with $m^2=\ell^2$. In this paper, we want to study the general relations among Feynman integrands of different theories, without referring to tools such as Feynman rules, CHY formulas, and so on. Thus we adopt a more general point of view proposed in \cite{Cachazo:2015aol}. We think the full loop momentum as a $d+1$ dimensional null vector $l^A=(\ell^\mu,e)$,
where $A\in\{0,\cdots,d-1,d\}$, and $e$ is the component in the extra $d^{\rm th}$ dimension. From the $d$ dimensional point of vies, we have
\bea
(l+K_i+\cdots+K_j)^2=(\ell+K_i+\cdots+K_j)^2-\ell^2\,,
\eea
which leads to the desired propagators. With the treatment discussed above, we can safely assume that the $1$-loop Feynman integrand ${\bf I}_\circ$ is obtained from the manipulation in \eref{single-dia}. To distinguish the full Feynman integrands and partial Feynman integrands obtained by decomposing the full ones via the partial fraction identity, from now on, we use ${\bf I}_\circ$ to denote the former ones, and
${\bf I}$ to denote the latter ones.

Clearly, the above discussion is for the full Feynman integrands without any color ordering, and now we turn to the color ordered Feynman integrands. Since we have made sure that the full $1$-loop Feynman integrand can be generated from the full tree amplitude via the forward limit operation, let us start with the full tree amplitude. Consider a theory that external particles are in the adjoint representation of the $U(N)$ group, the full tree amplitude can be expanded
using the standard color decomposition as a sum over $(n+1)!$ terms
\bea
{\cal A}_{n+2}=\sum_{\sigma_1\in S_{n+2}/\mathbb{Z}_{n+2}}\,{\rm Tr}(T^{a_{\sigma_+}}T^{a_{\sigma_1}}\cdots T^{a_{\sigma_n}}T^{a_{\sigma_-}})
{\cal A}(\sigma_+,\sigma_1,\cdots,\sigma_n,\sigma_-)\,.~~~~\label{color-deco}
\eea
Notice that at the r.h.s it is not necessary to add the subscript $n+2$ to ${\cal A}$, since the color ordering $\sigma_+,\sigma_1,\cdots,\sigma_n,\sigma_-$ already reflects the number of external legs.
Taking the forward limit of external legs requires summing over the $U(N)$ degrees of freedom of the two internal particles. This gives rise to two kinds of terms. The first comes from permutations such that legs $+$ and $-$ are adjacent, the corresponding color factors are given as
\bea
\sum_{a_+=a_-=1}^{N^2}\,\delta_{a_+a_-}\,{\rm Tr}(T^{a_{+}}T^{a_{\sigma_1}}\cdots T^{a_{\sigma_n}}T^{a_{-}})
=N{\rm Tr}(T^{a_{\sigma_1}}\cdots T^{a_{\sigma_n}})\,,
\eea
thus contributes to the $n$-point color ordered Feynman integrand ${\bf I}_\circ(\sigma_1,\cdots,\sigma_n)$.
The second case that $+$ and $-$ are not adjacent gives rise to double-trace terms. In this paper, we only consider the single-trace terms, since the double-trace terms are determined by the single-trace ones \cite{Bern:1996je}, as can be proved by employing the tree level
Kleiss-Kuijf relation together with the forward limit operation \cite{Kleiss:1988ne}. For the single trace case, the above discussion shows that the partial integrand
obtained form taking the forward limit for ${\cal A}(+,\sigma_1,\cdots,\sigma_n,-)$ contributes to ${\bf I}_\circ(\sigma_1,\cdots,\sigma_n)$. To find the full decomposition of ${\bf I}_\circ(\sigma_1,\cdots,\sigma_n)$, we use the clear observation that several original color orderings give rise to the same trace factor after summing over $a_+$ and $a_-$, due to the cyclic symmetry of the trace factors. Collecting theses color orderings together, one finds that after taking the forward limit, the decomposition \eref{color-deco} can be organized as
\bea
{\cal F}\,{\cal A}_{n+2}=\sum_{\sigma_1\in S_{n+2}/\mathbb{Z}_{n+2}}\,{\rm Tr}(T^{a_{\sigma_1}}\cdots T^{a_{\sigma_n}})\,\sum_{j=0}^{n-1}\,{\cal F}\,
{\cal A}(+,\sigma_{1+j},\cdots,\sigma_{n+j},-)+({\rm double-trace})\,.
\eea
Consequently, the full color ordered Feynman integrand can be expanded as the following cyclic summation
\bea
{\bf I}_\circ(\sigma_1,\cdots,\sigma_n)=\sum_{j=0}^{n-1}\,
{\bf I}(+,\sigma_{1+j},\cdots,\sigma_{n+j},-)\,,~~~~\label{sum-cyclic}
\eea
where the partial color ordered integrands ${\bf I}(+,\sigma_{1+j},\cdots,\sigma_{n+j},-)$ are obtained from the color ordered tree amplitudes via the standard forward limit procedure in \eref{single-dia}, namely,
\bea
{\bf I}(+,\sigma_{1+j},\cdots,\sigma_{n+j},-)={1\over\ell^2}\,{\cal F}\,
{\cal A}(+,\sigma_{1+j},\cdots,\sigma_{n+j},-)\,.
\eea
The cyclic summation in \eref{sum-cyclic} indicates that each propagator in the loop has been cut once, thus ${\bf I}_\circ(\sigma_1,\cdots,\sigma_n)$ and $
{\bf I}(+,\sigma_{1+j},\cdots,\sigma_{n+j},-)$ are also related via the partial fraction identity.

The forward limit is well defined for the ${\cal N}=4$ SYM theory. For other theories, a quite general feature is, the obtained Feynman integrand suffer from divergence in the forward limit. Fortunately, the singular parts is found to be physically irrelevant, at least for theories under consideration in this paper. From the Feynman diagrams point of view, the singular parts generated by the forward limit correspond to tadpole diagrams, as well as babble diagrams for external legs, which do not contribute to the $S$-matrix. From the CHY point of view, the singular parts can be bypassed by employing the following observation \cite{Cachazo:2015aol}: as long as the CHY integrand is homogeneous in $\ell^\mu$,
the singular solutions contribute to the scaleless integrals which vanish under the dimensional regularization. The homogeneity in $\ell^\mu$
are satisfied by all theories under consideration in this paper. Thus, in this paper, we just assume that the singular parts generated by the forward limit are excluded by an appropriate way.

\subsection{Transmutation operators at tree and $1$-loop levels}
\label{subsecoperator}

The tree level differential operators proposed by Cheung, Shen and Wen transmute tree amplitudes of one theory to those of other theories \cite{Cheung:2017ems,Zhou:2018wvn,Bollmann:2018edb}. Three kinds of basic operators are defined as follows:
\begin{itemize}
\item (1) Trace operator:
\bea
{\cal T}^\epsilon_{ij}\equiv \partial_{\epsilon_i\cdot\epsilon_j}\,,
\eea
where $\epsilon_i$ is the polarization vector of $i^{\rm th}$ external leg. The up index $\epsilon$ means the operators are defined through polarization vectors in $\{\epsilon_i\}$. Since the graviton carries the polarization tensor $\epsilon^{\mu\nu}=\epsilon^u\W\epsilon^\nu$,
the operators can always be defined via $\{\epsilon_i\}$ or $\{\W\epsilon_i\}$\footnote{Here the
 gravity theory has to be understood in a generalized version,
i.e., Einstein gravity theory couples to a dilaton and $2$-forms.}.
\item (2) Insertion operator:
\bea
{\cal I}^\epsilon_{ikj}\equiv \partial_{\epsilon_k\cdot k_i}-\partial_{\epsilon_k\cdot k_j}\,,~~~~\label{defin-insertion}
\eea
where $k_i$ denotes the momentum of the $i^{\rm th}$ external leg. When applying to physical amplitudes, the insertion operator ${\cal I}^\epsilon_{ik(i+1)}$ inserts the external leg $k$ between external legs $i$ and $i+1$ in the color-ordering $\cdots,i,i+1,\cdots$. For general ${\cal I}^\epsilon_{ikj}$ with $i<j$, one can use the definition \eref{defin-insertion} to decompose ${\cal I}^\epsilon_{ikj}$ as
\bea
{\cal I}^\epsilon_{ikj}={\cal I}^\epsilon_{ik(i+1)}+{\cal I}^\epsilon_{(i+1)k(i+2)}+\cdots+{\cal I}^\epsilon_{(j-1)kj}\,.
\eea
In the above expression, each ${\cal I}^\epsilon_{ak(a+1)}$ on the r.h.s can be interpreted as inserting the leg $k$ between $a$ and $(a+1)$.
Consequently, the effect of applying ${\cal I}^\epsilon_{ikj}$ can be understood as inserting $k$ between $i$ and $j$ in the color-ordering $\cdots,i,\cdots,j,\cdots$, and summing over all possible positions together.
\item (3) Longitudinal operator:
\bea
{\cal L}^\epsilon_i\equiv \sum_{j\neq i}\,(k_i\cdot k_j)\partial_{\epsilon_i\cdot k_j}\,,~~~~~~~~
{\cal L}^\epsilon_{ij}\equiv -(k_i\cdot k_j)\partial_{\epsilon_i\cdot \epsilon_j}\,.
\eea
\end{itemize}

By using products of these three kinds of basic operators, one can transmute amplitudes of one theory to those of other theories. Three combinatory operators which are products of basic operators are defined as follows:
\begin{itemize}
\item (1) For a length-$m$ ordered set $\vec{\pmb a}_m=\langle a_1,\cdots,a_m\rangle$ of external particles, the operator ${\cal T}^\epsilon_{\vec{\pmb a}_m}$
is given as\footnote{In this paper, we adopt the convention that the operator at l.h.s acts after the operator at r.h.s. From the mathematical
point of view, the order of operators is irrelevant, since all operators are commutable with each other. We choose the order of operators in the
definition to emphasize the interpretation of each one.}
\bea
{\cal T}^\epsilon_{\vec{\pmb a}_m}\equiv \Big(\prod_{i=2}^{m-1}\,{\cal I}^\epsilon_{a_1a_ia_{i+1}}\Big)\,{\cal T}^\epsilon_{a_1a_m}\,.~~~~\label{defin-T}
\eea
In this paper, we use ${\pmb a}_m=\{ a_1,\cdots,a_m\}$ to denote an un-ordered set with length $m$, and $\vec{\pmb a}_m=\langle a_1,\cdots,a_m\rangle$ for an ordered set with length $m$. For amplitudes/Feynman integrands carry color orderings, some times
we write down the orderings explicitly if necessary, and some times we use $\vec{\pmb a}_m$ to denote orderings.
The combinatory operator ${\cal T}^\epsilon_{\vec{\pmb a}_m}$ fixes $a_1$ and $a_m$ at two ends in the color-ordering via the operator ${\cal T}^\epsilon_{a_1a_m}$, and inserts other elements between them through insertion operators.
The operator ${\cal T}^\epsilon_{\vec{\pmb a}_m}$ is also called the trace operator since it generates the color-ordering $a_1,a_2,\cdots,a_m$. The interpretation of insertion operators indicates that ${\cal T}^\epsilon_{\vec{\pmb a}_m}$ has various equivalent choices, for example
\bea
& &{\cal T}^\epsilon_{\vec{\pmb a}_m}= \Big(\prod_{i=m-1}^{2}\,{\cal I}^\epsilon_{a_{i-1}a_ia_m}\Big)\,{\cal T}^\epsilon_{a_1a_m}\,,\nn
& &{\cal T}^\epsilon_{\vec{\pmb a}_m}= \Big(\prod_{i=3}^{m-3}\,{\cal I}^\epsilon_{a_2a_ia_{i+1}}\Big)\, {\cal I}^\epsilon_{a_{2}a_{m-2}a_{m-1}}\,{\cal I}^\epsilon_{a_{1}a_2a_{m-1}}\,{\cal I}^\epsilon_{a_{m-1}a_ma_1}\,{\cal T}^\epsilon_{a_1a_{m-1}}\,,
\eea
and so on. The second example provided above shows that it is not necessary to choose the first operator to be ${\cal T}^\epsilon_{a_1a_m}$. In other words, two reference legs in the color ordering can be chosen arbitrary.
\item (2) For $n$-point amplitudes, the operator ${\cal L}^\epsilon$
is defined as
\bea
{\cal L}^\epsilon\equiv\prod_i\,{\cal L}^\epsilon_i,~~~~~~~~\bar{{\cal L}}^\epsilon\equiv\sum_{\rho\in{\rm pair}}\,\prod_{i,j\in\rho}\,{\cal L}^\epsilon_{ij}\,.~~~~\label{defin-L}
\eea
Two definitions ${\cal L}^\epsilon$ and $\bar{{\cal L}}^\epsilon$ are not equivalent to each other at the algebraic level. However, when acting on proper on-shell
physical amplitudes, two combinations
${\cal L}^\epsilon\cdot{\cal T}^\epsilon_{ab}$ and $\bar{{\cal L}}^\epsilon\cdot{\cal T}^\epsilon_{ab}$, with subscripts of ${\cal L}^\epsilon_i$ and ${\cal L}^\epsilon_{ij}$
run through all nodes in $\{1,2,\cdots,n\}\setminus\{a,b\}$, give the same effect which can be interpreted physically.
\item (3) For a length-$2m$ set, the operator ${\cal T}^\epsilon_{{\cal X}_{2m}}$ is defined as
\bea
{\cal T}^\epsilon_{{\cal X}_{2m}}\equiv\sum_{\rho\in{\rm pair}}\,\prod_{i_k,j_k\in\rho}\,\delta_{I_{i_k}I_{j_k}}{\cal T}^\epsilon_{i_kj_k}\,,~~~~\label{TX1}
\eea
where $\delta_{I_{i_k}I_{j_k}}$ forbids the interaction between particles with different flavors. For the special case $2m$ particles do not carry any flavor, the operator ${\cal T}^\epsilon_{X_{2m}}$ is defined by removing $\delta_{I_{i_k}I_{j_k}}$,
\bea
{\cal T}^\epsilon_{X_{2m}}\equiv\sum_{\rho\in{\rm pair}}\,\prod_{i_k,j_k\in\rho}\,{\cal T}^\epsilon_{i_kj_k}\,.~~~\label{TX2}
\eea
\end{itemize}
The explanation for the notation $\sum_{\rho\in{\rm pair}}\,\prod_{i_k,j_k\in\rho}$ is in order. Let $\Gamma$ be the set of all partitions of the set $\{1,2,\cdots, 2m\}$ into pairs without regard to the order.
An element in $\Gamma$ can be written as
\bea
\rho=\{(i_1,j_1),(i_2,j_2),\cdots,(i_m,j_m)\}\,,
\eea
with conditions $i_1<i_2<\cdots<i_m$ and $i_t<j_t,\,\forall t$. Then, $\prod_{i_k,j_k\in\rho}$ stands for the product of ${\cal T}^\epsilon_{i_kj_k}$
for all pairs $(i_k,j_k)$ in $\rho$, and $\sum_{\rho\in{\rm pair}}$ denotes the summation over all partitions.

The combinatory operators given above link tree amplitudes of a wide range of theories together, by transmuting the GR amplitudes into amplitudes of other theories, formally expressed as
\bea
{\cal A}={\cal O}^\epsilon{\cal O}^{\W\epsilon}{\cal A}^{\epsilon,\W\epsilon}_{\rm GR}\,.~~~~\label{fund-uni-diff}
\eea
Operators ${\cal O}^\epsilon$ and ${\cal O}^{\W\epsilon}$ for different theories are listed in Table \ref{tab:unifying}.
\begin{table}[!h]
\begin{center}
\begin{tabular}{c|c|c}
Amplitude& ${\cal O}^\epsilon$  & ${\cal O}^{\W\epsilon}$ \\
\hline
${\cal A}_{{\rm GR}}^{\epsilon,\W\epsilon}(\pmb{a}^h_n)$ & $\mathbb{I}$ & $\mathbb{I}$  \\
${\cal A}_{{\rm sEYM}}^{\epsilon,\W\epsilon}(\vec{\pmb a}_m;\pmb{a}^h_{n-m})$  & ${\cal T}^{\epsilon}_{\vec{\pmb a}_m}$ & $\mathbb{I}$ \\
${\cal A}_{{\rm EMf}}^{\epsilon,\W\epsilon}(\pmb{a}^p_{2m};\pmb{a}^h_{n-2m})$  & ${\cal T}^{\epsilon}_{{\cal X}_{2m}}$  & $\mathbb{I}$\\
${\cal A}_{{\rm EM}}^{\epsilon,\W\epsilon}(\pmb{a}^p_{2m};\pmb{a}^h_{n-2m})$  & ${\cal T}^{\epsilon}_{X_{2m}}$ & $\mathbb{I}$ \\
${\cal A}_{{\rm BI}}^{\W\epsilon}(\pmb{a}^p_n)$  & ${\cal L}^{\epsilon}\,{\cal T}^{\epsilon}_{ab}$ & $\mathbb{I}$\\
${\cal A}_{{\rm YM}}^{\W\epsilon}(\vec{\pmb a}^g_n)$ &   ${\cal T}^{\epsilon}_{\vec{\pmb a}_n}$ & $\mathbb{I}$ \\
${\cal A}_{{\rm sYMS}}^{\W\epsilon}(\vec{\pmb a}^s_m;\pmb{a}^g_{n-m}\parallel\vec{\pmb a}^A_n)$ & ${\cal T}^{\epsilon}_{\vec{\pmb a}_n}$ & ${\cal T}^{\W\epsilon}_{\vec{\pmb a}_m}$ \\
${\cal A}_{{\rm SYMS}}^{\W\epsilon}({\pmb a}^s_{2m};\pmb{a}^g_{n-m}\parallel\vec{\pmb a}^A_n)$ & ${\cal T}^{\epsilon}_{\vec{\pmb a}_n}$ & ${\cal T}^{\W\epsilon}_{{\cal X}_{2m}}$ \\
${\cal A}_{{\rm NLSM}}(\vec{\pmb a}^s_n)$ & ${\cal T}^{\epsilon}_{\vec{\pmb a}_n}$ & ${\cal L}^{\W\epsilon}\,{\cal T}^{\W\epsilon}_{a'b'}$ \\
${\cal A}_{{\rm BAS}}(\vec{\pmb a}^s_n\parallel\vec{\pmb s}^s_n)$ &  ${\cal T}^{\epsilon}_{\vec{\pmb a}_n}$ & ${\cal T}^{\W\epsilon}_{\vec{\pmb s}_n}$ \\
${\cal A}_{{\rm DBI}}^{\W\epsilon}(\pmb{a}^s_{2m};\pmb{a}^p_{n-2m})$ &${\cal L}^{\epsilon}\,{\cal T}^{\epsilon}_{ab}$ & ${\cal T}^{\W\epsilon}_{{\cal X}_{2m}}$ \\
${\cal A}_{{\rm EDBI}}^{\W\epsilon}(\vec{\pmb{a}}^s_{m};\pmb{a}^p_{n-2m})$ &${\cal L}^{\epsilon}\,{\cal T}^\epsilon_{ab}$ & ${\cal T}^{\W\epsilon}_{\vec{\pmb a}_m}$ \\
${\cal A}_{{\rm SG}}(\pmb{a}^s_n)$ &  ${\cal L}^{\epsilon}\,{\cal T}^{\epsilon}_{ab}$ & ${\cal L}^{\W\epsilon}\,{\cal T}^{\W\epsilon}_{a'b'}$ \\
\end{tabular}
\end{center}
\caption{\label{tab:unifying}Unifying relations for differential operators at tree level.}
\end{table}
In this table, all amplitudes include $n$ external particles, as can be seen by adding the lengths of sets of external legs together.

Let us explain the notations in Table. \ref{tab:unifying} in turn. GR denotes the gravity, sEYM denotes the single-trace Einstein-Yang-Mills theory\footnote{In \cite{Cheung:2017ems,Zhou:2018wvn,Bollmann:2018edb}, the operators which generate the general multiple-trace tree EYM amplitudes are also considered. In Table \ref{tab:unifying} we did not exhibit these operators, since we will not use them throughout this paper.}, EM denotes the Einstein-Maxwell theory and EMf denotes the Einstein-Maxwell theory that photons carry flavors, BI denotes the Born-Infeld theory, YM denotes the pure Yang-Mills theory, sYMS denotes the single-trace Yang-Mills-scalar theory, SYMS denotes the special Yang-Mills-scalar theory, NLSM denotes the non-linear sigma model, BAS denotes the bi-adjoint scalar theory, DBI denotes the Dirac-Born-Infeld theory, EDBI denotes the extended Dirac-Born-Infeld theory, and SG denotes the special Galileon theory. The symbol $\mathbb{I}$ stands for the identical operator. Up indexes $h$, $p$, $g$ and $s$
denote gravitons, photons, gluons and scalars. For instance, ${\pmb a}^h_n$ is the un-ordered set of gravitons with length $n$,
$\vec{\pmb a}^g_m$ is the ordered set of gluons with length $m$. The total number of external legs is denoted by $n$, each set with length $m$
is a subset of external legs. We use ${\cal A}_{{\rm SYMS}}^{\W\epsilon}({\pmb a}^s_{2m};\pmb{a}^g_{n-m}\parallel\vec{\pmb a}^A_n)$ as the example to explain notations $\vec{{\pmb a}}^A_n$, $;$, and $\parallel$. For amplitudes include more than one kind of particles, such as scalars and gluons in the example, $\vec{{\pmb a}}^A_n$ stands for the color ordering among all external legs, without distinguishing the kinds of them.
The additional color ordering among all external particles is presented at the r.h.s of $\parallel$. Notation $;$ is used to separate different kinds of external particles, with the convention that particles at the l.h.s of $;$ carry lower spin. In our example, the l.h.s of $;$ is the set of scalars while the r.h.s is the set of gluons. The up index of
${\cal A}$ denotes the polarization vectors of external particles. In the cases amplitudes include external gravitons, the rule is: the previous polarization vectors are only carried by gravitons,
while the later ones are carried by all particles. For instance, in the notation ${\cal A}_{{\rm EMf}}^{\epsilon,\W\epsilon}(\pmb{a}^p_{2m};\pmb{a}^h_{n-2m})$, $\epsilon_i$ are only carried by gravitons, while $\W\epsilon_i$ are carried by both photons and gravitons. For the BAS amplitude, we have used $\vec{\pmb a}_n$ and $\vec{\pmb s}_n$ to distinguish two color orderings among external legs. In next sections, when considering more than one color orderings simultaneously, we frequently use $\vec{\pmb s}$ in addition to $\vec{\pmb a}$, to avoid the ambiguity.

In Table \ref{tab:unifying}, two sectors of operators labeled by polarization vectors $\epsilon$ and $\W\epsilon$ are exchangeable. As an example, YM amplitudes carry the polarization vectors $\epsilon$ can be generated by
\bea
{\cal A}^{\epsilon}_{\rm YM}(\vec{\pmb a}^g_n)={\cal T}^{\W\epsilon}_{\vec{\pmb a}_n}{\cal A}^{\epsilon,\W\epsilon}_{\rm GR}(\pmb{a}^h_n)\,.
\eea
All relations among amplitudes of different theories can be extracted from Table \ref{tab:unifying}. For example, from relations
\bea
& &{\cal A}^\epsilon_{\rm BI}(\pmb{a}^p_n)={\cal L}^{\W\epsilon}\,{\cal T}^{\W\epsilon}_{ab}\,{\cal A}^{\epsilon,\W\epsilon}_{\rm GR}(\pmb{a}^h_n)\,,\nn
& &{\cal A}_{\rm NLSM}(\vec{\pmb a}^s_n)={\cal T}^\epsilon_{\vec{\pmb a}_n}\Big({\cal L}^{\W\epsilon}\,{\cal T}^{\W\epsilon}_{ab}\Big){\cal A}^{\epsilon,\W\epsilon}_{\rm GR}(\pmb{a}^h_n)\,,
\eea
one can get
\bea
{\cal A}_{\rm NLSM}(\vec{\pmb a}^s_n)={\cal T}^\epsilon_{\vec{\pmb a}_n}{\cal A}^\epsilon_{\rm BI}(\pmb{a}^p_n)\,.
\eea
Thus, the full unified web for tree amplitudes of different theories is involved in Table \ref{tab:unifying}.

\begin{table}[!h]
\begin{center}
\begin{tabular}{c|c|c}
Feynman integrand& ${\cal O}^\epsilon_\circ$  & ${\cal O}^{\W\epsilon}_\circ$ \\
\hline
${\bf I}_{{\rm GR};\circ}^{\epsilon,\W\epsilon}(\pmb{a}^h_n)$ & $\mathbb{I}$ & $\mathbb{I}$  \\
${\bf I}_{{\rm ssEYM}}^{\epsilon,\W\epsilon}(+^g,\vec{\pmb a}^g_m,-^g;\pmb{a}^h_{n-m})$ & $\mathbb{I}$ & ${\cal T}^{\W\epsilon}_{+\vec{\pmb a}_m-}$  \\
${\bf I}_{{\rm EMf};\circ}^{\epsilon,\W\epsilon}(\pmb{a}^p_{2m};\pmb{a}^h_{n-2m})$ & $\mathbb{I}$ & ${\cal T}^{\W\epsilon}_{{\cal X}_{2m}}(N\,\W{\cal D}+1)$  \\
${\bf I}_{{\rm EM};\circ}^{\epsilon,\W\epsilon}(\pmb{a}^p_{2m};\pmb{a}^h_{n-2m})$ & $\mathbb{I}$ & ${\cal T}^{\W\epsilon}_{X_{2m}}(\W{\cal D}+1)$  \\
${\bf I}_{{\rm BI};\circ}^\epsilon(\pmb{a}^p_n)$ & $\mathbb{I}$ & ${\cal L}^{\W\epsilon}\,\W{\cal D}$ \\
${\bf I}_{{\rm YM}}^\epsilon(+\vec{\pmb a}^g_n-)$ & $\mathbb{I}$ & ${\cal T}^{\W\epsilon}_{+\vec{\pmb a}_n-}$  \\
${\bf I}_{{\rm ssYMS}}^{\W\epsilon}(+^s,\vec{\pmb a}^s_m,-^s;\pmb{a}^g_{n-m}\parallel+^A,\vec{\pmb a}^A_n,-^A)$ & ${\cal T}^{\epsilon}_{+\vec{\pmb a}_n-}$ & ${\cal T}^{\W\epsilon}_{+\vec{\pmb a}_m-}$ \\
${\bf I}_{{\rm SYMS}}^{\W\epsilon}({\pmb a}^s_{2m};\pmb{a}^g_{n-m}\parallel+^A,\vec{\pmb a}^A_n,-^A)$ & ${\cal T}^{\epsilon}_{+\vec{\pmb a}_n-}$ & ${\cal T}^{\W\epsilon}_{{\cal X}_{2m}}(N\W D+1)$ \\
${\bf I}_{{\rm NLSM};\circ}(\vec{\pmb a}^s_n)$ & ${\cal T}^{\epsilon}_{+\vec{\pmb a}_n-}$ & ${\cal L}^{\W\epsilon}\,\W{\cal D}$ \\
${\bf I}_{{\rm BAS}}(+^s,\vec{\pmb a}^s_n,-^s\parallel+^s,\vec{\pmb s}^s_n,-^s)$ &  ${\cal T}^{\epsilon}_{+\vec{\pmb a}_n-}$ & ${\cal T}^{\W\epsilon}_{+\vec{\pmb s}_n-}$ \\
${\bf I}_{{\rm DBI};\circ}^{\W\epsilon}(\pmb{a}^s_{2m};\pmb{a}^p_{n-2m})$ &${\cal L}^{\epsilon}\,{\cal D}$ & ${\cal T}^{\W\epsilon}_{{\cal X}_{2m}}(N\,\W{\cal D}+1)$ \\
${\bf I}_{{\rm ssEDBI}}^{\W\epsilon}(+^s,\vec{\pmb{a}}^s_{m},-^s;\pmb{a}^p_{n-2m})$ &${\cal L}^{\epsilon}\,{\cal D}$ & ${\cal T}^{\W\epsilon}_{+\vec{\pmb a}_m-}$ \\
${\bf I}_{{\rm SG};\circ}(\pmb{a}^s_n)$ &  ${\cal L}^{\epsilon}\,{\cal D}$ & ${\cal L}^{\W\epsilon}\,\W{\cal D}$ \\
\end{tabular}
\end{center}
\caption{${\cal O}^\epsilon_\circ$  and ${\cal O}^{\W\epsilon}_\circ$ for various theories.\label{tab:unifying-loop}}
\end{table}
\begin{figure}
  \centering
  \includegraphics[width=10cm]{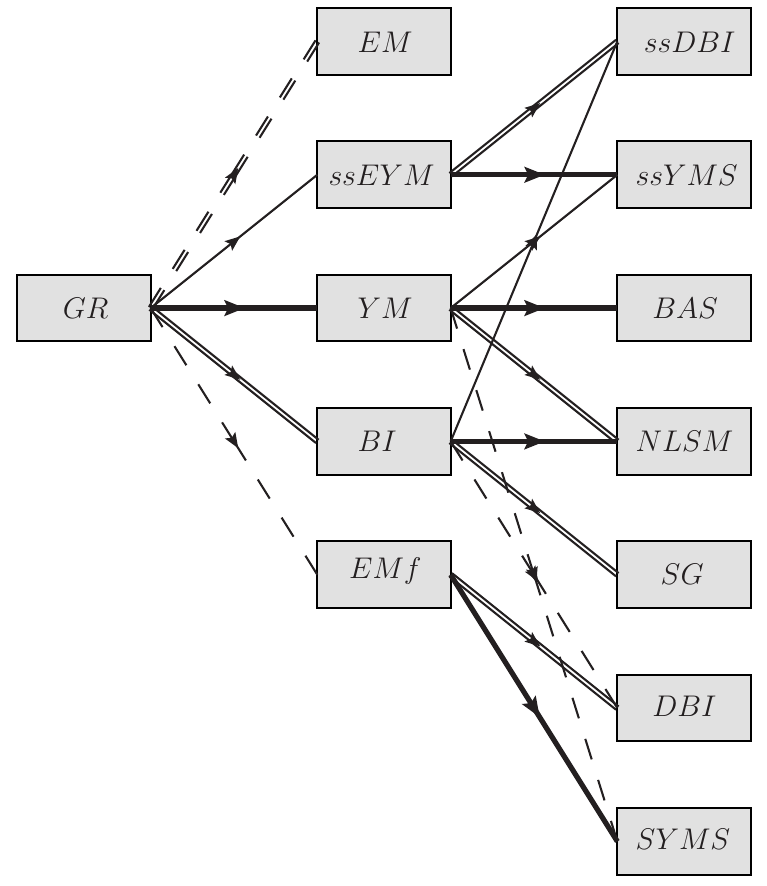} \\
  \caption{Unified web for $1$-loop Feynman integrands. The bold straight line represents the operator ${\cal T}_{\vec{\pmb a}_n;\circ}$, the straight line represents the operator ${\cal T}_{\vec{\pmb a}_m;\circ}$ with $0\leq m<n$, the double straight line represents the operator ${\cal L}\,{\cal D}$, the dashed line represents the operator ${\cal T}_{{\cal X}_{2m}}(N\,{\cal D}+1)$, the double dashed line represents the operator ${\cal T}_{X_{2m}}({\cal D}+1)$. Here we omitted the up index $\epsilon$ or $\W\epsilon$, it means these operators can be defined via polarization vectors in both $\{\epsilon_i\}$ or $\{\W\epsilon_i\}$.}\label{op}
\end{figure}

The above tree level unifying relation can be generalized to $1$-loop Feynman integrands, as studied in \cite{Zhou:2021kzv}. The basic idea is as follows. Suppose the tree amplitudes of theories $A$
and $B$ are connected by the operator ${\cal O}$ as ${\cal A}_B={\cal O}\,{\cal A}_A$, we seek the $1$-loop level operator
${\cal O}_\circ$ satisfying ${\cal O}_\circ\,{\cal F}\,{\cal A}_A={\cal F}\,{\cal O}\,{\cal A}_A$.
Since the $1$-loop Feynman integrands are obtained via
the forward limit as ${\bf I}_A=(1/\ell^2){\cal F}\,{\cal A}_A$ and ${\bf I}_B=(1/\ell^2){\cal F}\,{\cal A}_B$, one can conclude that the operator ${\cal O}_\circ$ transmutes the Feynman integrand as ${\bf I}_B={\cal O}_\circ\,{\bf I}_A$.

Using the above idea, the $1$-loop analog of the tree level unifying relation \eref{fund-uni-diff} is found to be
\bea
{\bf I}={\cal O}^\epsilon_{\circ}\,{\cal O}^{\W\epsilon}_{\circ}\,{\bf I}^{\epsilon,\W\epsilon}_{\rm GR}\,.~~~~\label{uni-diff-loop}
\eea
The $1$-loop level operators ${\cal O}^\epsilon_{\circ}$ and ${\cal O}^{\W\epsilon}_{\circ}$ for different theories are listed in
Table. \ref{tab:unifying-loop}. The full web of transmutational relations is summarized in Fig. \ref{op}. Similar as the tree level case, all information
in the web Fig. \ref{op} is included in Table. \ref{tab:unifying-loop}.

In Table. \ref{tab:unifying-loop}, ssEYM denotes the special part of sEYM integrand that only a virtual gluon propagating in the loop.
Similarly, ssYMS denotes the special sYMS integrand with a virtual scalar running in the loop, and ssEDBI is the special EDBI integrand with a virtual scalar in the loop. Integrands ${\bf I}_{\circ}$ with the subscript $\circ$ are full $1$-loop Feynman integrands, while ${\bf I}$ without $\circ$ are partial Feynman integrands, as introduced in the previous subsection. We used $+,\vec{\pmb a}_m,-$ and $+,\vec{\pmb a}_n,-$ to denote the color orderings of partial integrands, where $+$, $-$ are external legs carry $k_+$ and $k_-$ respectively before taking the forward limit. After doing the cyclic summation over the $1$-loop level equivalent color orderings, we use $\vec{\pmb a}_m$ or $\vec{\pmb a}_n$
instead of $+,\vec{\pmb a}_m,-$ or $+,\vec{\pmb a}_n,-$. For example,
\bea
{\bf I}_{{\rm BAS}}(\vec{\pmb a}^s_n\parallel+^s,\vec{\pmb s}^s_n,-^s)&=&\sum_{\pi_c}\,{\bf I}_{{\rm BAS}}(+^s,\pi_c(\vec{\pmb a}^s_n),-^s\parallel+^s,\vec{\pmb s}^s_n,-^s)\,,\nn
{\bf I}_{{\rm BAS};\circ}(\vec{\pmb a}^s_n\parallel\vec{\pmb s}^s_n)&=&\sum_{\pi_c}\,{\bf I}_{{\rm BAS}}(\pi_c(\vec{\pmb a}^s_n)\parallel+^s,\pi_c(\vec{\pmb s}^s_n),-^s)\,,~~~\label{sum-cyc}
\eea
where $\pi_c$ denotes the cyclic permutation throughout this paper. Notice that ${\bf I}_{{\rm BAS};\circ}(\vec{\pmb a}^s_n\parallel\vec{\pmb s}^s_n)$
in the second line of \eref{sum-cyc} is the full BAS $1$-loop Feynman integrand, rather than a partial one.
The operator ${\cal T}^\epsilon_{+\vec{\pmb a}_m-}$ is given as
\bea
{\cal T}^\epsilon_{+\vec{\pmb a}_m-}\equiv\Big(\prod_{i=1}^{m-1}\,{\cal I}^\epsilon_{+a_ia_{i+1}}\Big)\,{\cal I}^\epsilon_{+a_m-}\,{\cal D}\,.~~~~\label{defin-T-loop}
\eea
The operator ${\cal D}$ is defined in the following way. We think the Lorentz vectors before taking the forward limit as follows, the momenta in $\{k_1,\cdots,k_n,\ell\}$ and polarization vectors in $\{\epsilon_1,\cdots,\epsilon_n\}$ lie in the $d$ dimensional space where $d$ is regarded as a constant, while the polarization vectors $\epsilon_+$ and $\epsilon_-$ are in the $D$ dimensional space where $D$
is regarded as a variable. We can set $D=d$ finally to obtain a physically acceptable object. Then we define
\bea
{\cal D}\equiv\partial_D\,.
\eea
For gravitons, we regard $D=\sum_r\epsilon^r_+\cdot\epsilon^r_-+2$ and $\W D=\sum_r\W\epsilon^r_+\cdot\W\epsilon^r_-+2$
as two independent variables, namely, $\partial_D\W D=0$, $\partial_{\W D}D=0$. The insertion operators are
\bea
{\cal I}^\epsilon_{+a_i-}\equiv\partial_{\epsilon_i\cdot\ell}\,,~~~~~~~~{\cal I}^\epsilon_{+a_ia_{i+1}}\equiv\partial_{\epsilon_{a_i}\cdot\ell}
-\partial_{\epsilon_{a_i}\cdot k_{a_{i+1}}}\,.
\eea
Notice that here $\ell$ is understood as $k_+$, since one can always think $k_-=-\ell$ as being removed via the momentum conservation law.
The operators ${\cal T}^\epsilon_{{\cal X}_{2m}}$, ${\cal T}^\epsilon_{X_{2m}}$ and ${\cal L}^\epsilon$ are the same as the tree level ones.
When applying ${\cal L}^\epsilon$ at the $1$-loop level, the operator ${\cal L}^\epsilon_i$ should include $\partial_{\epsilon_i\cdot k_+}=\partial_{\epsilon_i\cdot\ell}$.

\section{New operators for GR to YM and YM to BAS}
\label{NOP-GR-YM}

In \cite{Zhou:2021kzv}, the differential operators transmute the $1$-loop GR Feynman integrand to the YM partial Feynman integrands are constructed as follows.
At tree level, one can use the trace operator ${\cal T}^\epsilon_{+-}$ to turn the external gravitons $+^h$ and $-^h$ to gluons, and fix legs $+$ and $-$ at two ends in the color
ordering, i.e., choose $+$ and $-$ as two reference legs. Then, use the insertion operators, one can turn other gravitons to gluons, and insert them between
$+$ and $-$ to generate the full color ordering. The obtained color ordered tree amplitude is ${\cal A}^{\W\epsilon}_{\rm YM}(+^g,\vec{\pmb a}^g_n,-^g)$.
Based on the above tree level manipulation, by seeking the operator ${\cal O}^\epsilon_\circ$ satisfying ${\cal O}^\epsilon_\circ{\cal F}{\cal A}^{\epsilon,\W\epsilon}_{\rm GR}={\cal F}{\cal O}^\epsilon{\cal A}^{\epsilon,\W\epsilon}_{\rm GR}$, one can construct the
corresponding $1$-loop level operator ${\cal O}^\epsilon_\circ$ which transmutes the GR integrand ${\bf I}^{\epsilon,\W\epsilon}_{{\rm GR};\circ}({\pmb a}^h_n)$
to the partial YM one ${\bf I}^{\W\epsilon}_{{\rm YM}}(+^g,\vec{\pmb a}^g_n,-^g)$.
However, to generates the tree color ordered YM amplitude ${\cal A}^{\W\epsilon}_{\rm YM}(+^g,\vec{\pmb a}^g_n,-^g)$, the above tree level
manipulation is not the only way. Actually, one can turn arbitrary two
gravitons to gluons at the first step, and insert other legs between them to arrive at the final result, via the
insertion operators. Thus it is natural to ask, if we make the choices of reference legs different from $+$, $-$ at the tree level, what $1$-loop
level operators can be constructed? what effects will these new operators have when acting on ${\bf I}^{\epsilon,\W\epsilon}_{\rm GR;\circ}({\pmb a}^h_n)$?

In this section, we show that generating two reference legs from one graviton in $\{+^h,-^h\}$ and another one in ${\pmb a}^h_n$ leads to a kind of new
$1$-loop level operators. When acting on ${\bf I}^{\epsilon,\W\epsilon}_{\rm GR;\circ}({\pmb a}^h_n)$, these new operators beautifully
provide $1$-loop ssEYM integrands. These operators also link the YM integrands and the ssYMS integrands together, as indicated by the tree level transmutational relations. In subsection \ref{construction of the operator}, we discuss how to construct such new operators. In subsection \ref{verify}, we verify our construction by using CHY formulas.

\subsection{Construction of new operators}
\label{construction of the operator}

In this subsection, a new kind of operators which transmute the GR Feynman integrands to ssEYM Feynman integrands will be constructed.
The general idea is similar as in \cite{Zhou:2021kzv}:  we seek the $1$-loop level operator ${\cal O}^\epsilon_\circ$ satisfying ${\cal O}^\epsilon_\circ{\cal F}{\cal A}^{\epsilon,\W\epsilon}_{\rm GR}={\cal F}{\cal O}^\epsilon{\cal A}^{\epsilon,\W\epsilon}_{\rm GR}$.
We start by considering the single-trace tree sEYM amplitude ${\cal A}^{\epsilon,\W\epsilon}_{\rm sEYM}(+^g,a^g,-^g;{\pmb a}^h_n\setminus a^h)$, with three external gluons $+^g$, $-^g$ and $a^g$. It can be rewritten as ${\cal A}^{\epsilon,\W\epsilon}_{\rm sEYM}(a^g,-^g,+^g;{\pmb a}^h_n\setminus a^h)$, due to the cyclic symmetry of the
color ordering. The rewritten form indicates that such amplitude can be generated from the tree GR amplitude via differential operators
as
\bea
{\cal A}^{\epsilon,\W\epsilon}_{\rm sEYM}(a^g,-^g,+^g;{\pmb a}^h_n\setminus a^h)&=&{\cal I}^\epsilon_{a-+}\,{\cal T}^\epsilon_{a+}\,{\cal A}^{\epsilon,\W\epsilon}_{\rm GR}({\pmb a}^h_n\cup \{+^h,-^h\})\nn
&=&(\partial_{\epsilon_-\cdot k_a}-\partial_{\epsilon_-\cdot l})\,\partial_{\epsilon_a\cdot\epsilon_+}\,{\cal A}^{\epsilon,\W\epsilon}_{\rm GR}({\pmb a}^h_n\cup \{+^h,-^h\})\,.~~~~\label{tree-3p}
\eea
The color ordering is generated by choosing $a$ and $+$ as two reference legs first, then inserting the leg $-$ between them.
Here the operator $\partial_{\epsilon_-\cdot l}$ is understood as $\partial_{\epsilon_-\cdot k_+}$.
Unfortunately, there is no $1$-loop level operator ${\cal O}^\epsilon_\circ$ satisfies
\bea
{\cal O}^\epsilon_\circ\,{\cal F}\,{\cal A}^{\epsilon,\W\epsilon}_{\rm GR}({\pmb a}^h_n\cup \{+^h,-^h\})={\cal F}\,{\cal I}^\epsilon_{a-+}\,{\cal T}^\epsilon_{a+}\,{\cal A}^{\epsilon,\W\epsilon}_{\rm GR}({\pmb a}^h_n\cup \{+^h,-^h\})\,.
\eea
To see this, we first consider the piece $\partial_{\epsilon_-\cdot k_a}\partial_{\epsilon_a\cdot\epsilon_+}$ in the operator ${\cal I}^\epsilon_{a-+}\,{\cal T}^\epsilon_{a+}$. At the tree level, this piece of operator turns $(\epsilon_-\cdot k_a)(\epsilon_a\cdot\epsilon_+)$
to $1$, and annihilates all terms do not contain the Lorentz invariant $(\epsilon_-\cdot k_a)(\epsilon_a\cdot\epsilon_+)$, due to the observation that the amplitude is linear in each polarization vector. Under the the action of ${\cal E}$, tree level object $(\epsilon_-\cdot k_a)(\epsilon_a\cdot\epsilon_+)$ behaves as
\bea
\sum_r\,(\epsilon_a\cdot\epsilon^r_+)(\epsilon^r_-\cdot k_a)=\epsilon_a\cdot k_a\,,
\eea
thus the on-shell condition $\epsilon_a\cdot k_a=0$ indicates that the first piece of the operator does not make sense at the $1$-loop level. Notice that in general the summation over $\epsilon^r_+\epsilon^r_-$ should be
\bea
\sum_r\,(\epsilon^r_+)^\mu(\epsilon^r_-)^\nu=\eta^{\mu\nu}-{\ell^\mu q^\nu+\ell^\nu q^\mu\over\ell\cdot q}\,,
\eea
where the null $q$ satisfies $\epsilon^r_+\cdot q=\epsilon^r_-\cdot q=0$. Here we are allowed to drop the $q$-dependent
terms, since their contributions vanish on the solution to the scattering equations, see in \cite{Roehrig:2017gbt}. Now we turn to another piece
$\partial_{\epsilon_-\cdot l}\partial_{\epsilon_a\cdot\epsilon_+}$. At tree level, this piece turns $(\epsilon_-\cdot l)(\epsilon_a\cdot\epsilon_+)$
to $1$, and annihilates all terms do not contain $(\epsilon_-\cdot l)(\epsilon_a\cdot\epsilon_+)$. However, $\epsilon_-\cdot l$ vanishes under the action of ${\cal L}$,
since $\epsilon_-\cdot k_-=0$ and $k_-=-k_+=-\ell$. Thus, the second piece of the operator also makes no sense at the $1$-loop level.

\begin{figure}
  \centering
  \includegraphics[width=12cm]{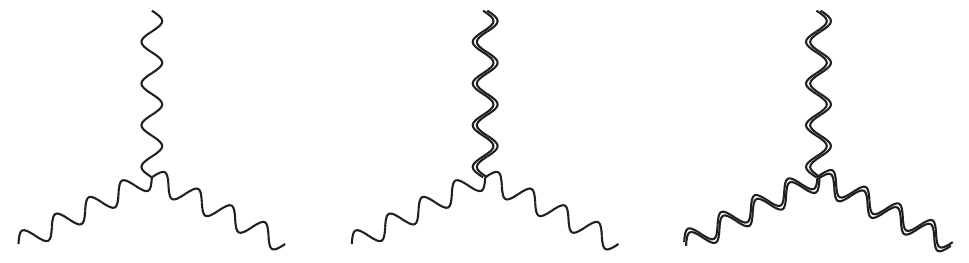} \\
  \caption{Three vertices of EYM theory, the single wavy lines denote gluons while the double wavy lines denote gravitons. }\label{vertex}
\end{figure}

Although the tree level relation \eref{tree-3p} does not give rise to any physically meaningful operator at $1$-loop level, the above method leads to well defined and meaningful operators when considering tree sEYM amplitudes with at least $4$ external gluons. Consider the $4$-gluon amplitude
${\cal A}^{\epsilon,\W\epsilon}_{\rm sEYM}(+^g,a^g,b^g,-^g;{\pmb a}^h_n\setminus \{a^h,b^h\})$, using the cyclic symmetry of color ordering as well as the tree level differential operators, we have
\bea
{\cal A}^{\epsilon,\W\epsilon}_{\rm sEYM}(a^g,b^g,-^g,+^g;{\pmb a}^h_n\setminus \{a^h,b^h\})&=&{\cal I}^\epsilon_{b-+}\,{\cal I}^\epsilon_{ab+}\,{\cal T}^\epsilon_{a+}\,{\cal A}^{\epsilon,\W\epsilon}_{\rm GR}({\pmb a}^h_n\cup \{+^h,-^h\})\nn
&=&(\partial_{\epsilon_-\cdot k_b}-\partial_{\epsilon_-\cdot\ell})\,(\partial_{\epsilon_b\cdot k_a}-\partial_{\epsilon_b\cdot\ell})\,\partial_{\epsilon_a\cdot\epsilon_+}\,{\cal A}^{\epsilon,\W\epsilon}_{\rm GR}({\pmb a}^h_n\cup \{+^h,-^h\})\,.
\eea
The above manipulation is understood as turn gravitons $a^h$ and $+^h$ to gluons and regard them as the reference legs, then turn the graviton
$b^h$ to a gluon and insert it between $a$ and $+$, and turn the graviton $-^h$ to a gluon and insert it between $b$ and $+$ finally. The object $\epsilon_-\cdot\ell$ vanishes under the action of ${\cal L}$, as in the previous case, thus we focus on the $\partial_{\epsilon_-\cdot k_b}$
part in ${\cal I}^\epsilon_{b-+}$. At the tree level, the combinatory operator $\partial_{\epsilon_-\cdot k_b}\partial_{\epsilon_a\cdot\epsilon_+}$
turns $(\epsilon_-\cdot k_b)(\epsilon_a\cdot\epsilon_+)$ to $1$, and annihilates all terms do not contain the above object. Under the action of ${\cal E}$, this object behaves as
\bea
\sum_r\,(\epsilon_a\cdot\epsilon^r_+)(\epsilon^r_-\cdot k_b)=\epsilon_a\cdot k_b\,,
\eea
thus the effect of the operator $\partial_{\epsilon_-\cdot k_b}\partial_{\epsilon_a\cdot\epsilon_+}$ at the tree level is equivalent to $\partial_{\epsilon_a\triangleright k_b}$ at the $1$-loop level. In other words, we have found
\bea
\partial_{\epsilon_a\triangleright k_b}\,{\cal F}\,{\cal A}^{\epsilon,\W\epsilon}_{\rm GR}({\pmb a}^h_n\cup \{+^h,-^h\})
={\cal F}\,{\cal I}^\epsilon_{b-+}\,{\cal T}^\epsilon_{a+}\,{\cal A}^{\epsilon,\W\epsilon}_{\rm GR}({\pmb a}^h_n\cup \{+^h,-^h\})\,.~~~~\label{tree-4p1}
\eea
From now on, we use $A\triangleright B$ to denote $A\cdot B$ arised
from $\sum_r (A\cdot \epsilon^r_+)(\epsilon^r_-\cdot B)$, and $A\triangleleft B$ to denote $A\cdot B$
from $\sum_r (A\cdot \epsilon^r_-)(\epsilon^r_+\cdot B)$. The observation that $\epsilon_-\cdot\ell$ vanishes under the action of ${\cal L}$
is used in \eref{tree-4p1}. When the contribution from summing over $\epsilon^r_+\epsilon^r_-$ is included in $\epsilon_a\triangleright k_b$,
the operator ${\cal I}^\epsilon_{ab+}$ can not act on $\epsilon_b\triangleright k_a$, $\epsilon_b\triangleright k_l$, or $\epsilon_b\triangleleft k_a$, $\epsilon_b\triangleleft k_l$, therefore obviously commutable with ${\cal F}$. Then we arrive at the relation
\bea
{\cal I}^\epsilon_{ab+}\,\partial_{\epsilon_a\triangleright k_b}\,{\cal F}\,{\cal A}^{\epsilon,\W\epsilon}_{\rm GR}({\pmb a}^h_n\cup \{+^h,-^h\})
={\cal F}\,{\cal I}^\epsilon_{b-+}\,{\cal I}^\epsilon_{ab+}\,{\cal T}^\epsilon_{a+}\,{\cal A}^{\epsilon,\W\epsilon}_{\rm GR}({\pmb a}^h_n\cup \{+^h,-^h\})\,,~~~~\label{tree-4p2}
\eea
which suggests
\bea
{\cal I}^\epsilon_{ab+}\,\partial_{\epsilon_a\triangleright k_b}\,{\bf I}^{\epsilon,\W\epsilon}_{\rm GR;\circ}({\pmb a}^h_n)
={\bf I}^{\epsilon,\W\epsilon}_{\rm ssEYM}(+^g,a^g,b^g,-^g;{\pmb a}^h_n\setminus\{a^h,b^h\})\,.~~~~\label{tree-4p3}
\eea
Here ${\bf I}^{\epsilon,\W\epsilon}_{\rm ssEYM}$ denotes the special single-trace EYM partial Feynman integrand with a virtual gluon running in the loop, as introduced in subsection \ref{subsecoperator}. The reason for interpreting the object at the r.h.s of \eref{tree-4p3} as such special sEYM one can be explained as follows.
The EYM theory includes three interaction vertices in Figure \ref{vertex}, these vertices indicate that for the tree sEYM amplitude including external gluons $+^g$ and $-^g$, one can always start from one of them, go along the gluon lines, and arrive at another one. It means, after taking the forward limit for external gluons $+^g$ and $-^g$,
a closed loop contains only gluon lines is obtained. This observation fixes the virtual particle in the loop to be a gluon. This configuration ensures the statement that the object at the r.h.s of \eref{tree-4p3} is nothing but the
$1$-loop ssEYM partial Feynman
integrand with a virtual gluon in the loop.
Notice that ${\bf I}^{\epsilon,\W\epsilon}_{\rm ssEYM}(+^g,a^g,b^g,-^g;{\pmb a}^h_n\setminus\{a^h,b^h\})$ is a partial Feynman integrand contributes to the full
Feynman integrand, and the full one is obtained via the cyclic summation as
\bea
{\bf I}^{\epsilon,\W\epsilon}_{\rm ssEYM;\circ}(a^g,b^g;{\pmb a}^h_n\setminus\{a^h,b^h\})&=&{\bf I}^{\epsilon,\W\epsilon}_{\rm ssEYM}(+^g,a^g,b^g,-^g;{\pmb a}^h_n\setminus\{a^h,b^h\})\nn
& &+{\bf I}^{\epsilon,\W\epsilon}_{\rm ssEYM}(+^g,b^g,a^g,-^g;{\pmb a}^h_n\setminus\{a^h,b^h\})\,.
\eea

At this step, a problem occurs. In the $1$-loop integrand ${\bf I}^{\epsilon,\W\epsilon}_{\rm GR;\circ}({\pmb a}^h_n)$, the tree level information
associated to $\epsilon^r_+$ and $\epsilon^r_-$ is lost, thus $\epsilon_a\triangleright k_b$ or $\epsilon_a\triangleleft k_b$ can not be distinguished
from $\epsilon_a\cdot k_b$. It means the operator $\partial_{\epsilon_a\triangleright k_b}$ is not well defined at the $1$-loop level, and we should replace it by $\partial_{\epsilon_a\cdot k_b}$. Such replacement will be made later. Before doing this, let us continue the discussion for the operator ${\cal I}^\epsilon_{ab+}\,\partial_{\epsilon_a\triangleright k_b}$ and its generalization.

The operator ${\cal I}^\epsilon_{ab+}\,\partial_{\epsilon_a\triangleright k_b}$ can be simplified to $\partial_{\epsilon_b\cdot k_a}\,\partial_{\epsilon_a\triangleright k_b}$, since $\partial_{\epsilon_b\cdot \ell}\,\partial_{\epsilon_a\triangleright k_b}$
gives no contribution at the $1$-loop level. To observe this, it is convenient to use the CHY formula. A brief introduction to CHY formulas can be seen in Appendix. \ref{sec-CHY}. The operator ${\cal I}^\epsilon_{ab+}\,\partial_{\epsilon_a\triangleright k_b}$ does not act on the measure of CHY contour integration, thus can be applied to the CHY integrands directly. As can be seen in Appendix. \ref{sec-CHY}, in this paper the convention for the reduced Pfaffian of $\Psi$ is ${\bf Pf}'\Psi={(-)\over z_{+-}}{\bf Pf}\Psi'$ where $\Psi'=\Psi^{+-}_{+-}$.
We use the notation $S^{ij\cdots}_{lk\cdots}$ to denote the matrix which is generated from the matrix $S$ by removing $i^{\rm th},\,j^{\rm th}\,\cdots$ rows, and $l^{\rm th},\,k^{\rm th}\,\cdots$ columns.
Because of the choice $\Psi'=\Psi^{+-}_{+-}$, when acting $\partial_{\epsilon_b\cdot \ell}$ on ${\cal F}{\bf Pf}'\Psi$, the non-vanishing
contribution only from acting on the diagonal element $C_{bb}$ in the block $C$ of the matrix $\Psi$. However, the operator $\partial_{\epsilon_a\triangleright k_b}$ turns
$\epsilon_a\triangleright k_b$ to $1$ and annihilates all terms do not contain $\epsilon_a\triangleright k_b$, thus eliminates $b^{\rm th}$ rows and columns in $\Psi'$. Thus $C_{bb}$ does not exist in $\partial_{\epsilon_a\triangleright k_b}{\cal F}{\bf Pf}'\Psi$,
therefore the operator $\partial_{\epsilon_b\cdot \ell}$ annihilates $\partial_{\epsilon_a\triangleright k_b}{\bf I}^{\epsilon,\W\epsilon}_{\rm GR;\circ}({\pmb a}^h_n)$.
The above observation allows us to simplify the relation \eref{tree-4p3} to a compact form
\bea
\partial_{\epsilon_b\cdot k_a}\,\partial_{\epsilon_a\triangleright k_b}\,{\bf I}^{\epsilon,\W\epsilon}_{\rm GR;\circ}({\pmb a}^h_n)
={\bf I}^{\epsilon,\W\epsilon}_{\rm ssEYM}(+^g,a^g,b^g,-^g;{\pmb a}^h_n\setminus\{a^h,b^h\})\,.~~~~\label{tree-4p4}
\eea

The discussion about the tree sEYM amplitude ${\cal A}^{\epsilon,\W\epsilon}_{\rm sEYM}(+^g,a^g,b^g,-^g;{\pmb a}^h_n\setminus \{a^h,b^h\})$ can be generalized
to the tree sEYM amplitudes with more external gluons straightforwardly. Along the similar line, one arrive at the relation
\bea
\Big(\prod_{i=1}^{m-1}\,\partial_{\epsilon_{a_{i+1}}\cdot k_{a_i}}\Big)\,\partial_{\epsilon_{a_1}\triangleright k_{a_m}}\,{\bf I}^{\epsilon,\W\epsilon}_{\rm GR;\circ}({\pmb a}^h_n)
={\bf I}^{\epsilon,\W\epsilon}_{\rm ssEYM}(+^g,a^g_1,\cdots,a^g_m,-^g;{\pmb a}^h_n\setminus\{a^h_1,\cdots,a^h_m\})\,,~~~~\label{tree-mp1}
\eea
with $m\geq2$.

Now we come back to the problem that at the $1$-loop level the well defined operator is $\partial_{\epsilon_{a_1}\cdot k_{a_m}}$
rather than $\partial_{\epsilon_{a_1}\triangleright k_{a_m}}$. Motivated by the relation \eref{tree-mp1}, it is natural to act the operator ${\cal C}^\epsilon_{\vec{\pmb a}_m}$ on ${\bf I}^{\epsilon,\W\epsilon}_{\rm GR;\circ}({\pmb a}^h_n)$, where the cyclical operator ${\cal C}^\epsilon_{\vec{\pmb a}_m}$ for the ordered set $\vec{\pmb a}_m=\langle a_1,\cdots,a_m\rangle$ with $m\geq2$ is defined as
\bea
{\cal C}^\epsilon_{\vec{\pmb a}_m}\equiv\prod_{i=1}^m\,\partial_{k_{a_i}\cdot\epsilon_{a_{i+1}}}=\partial_{C^\epsilon_{\vec{\pmb a}_m}}\,,~~~~\label{defin-C}
\eea
where
\bea
C^\epsilon_{\vec{\pmb a}_m}=\prod_{i=1}^m\,k_{a_i}\cdot\epsilon_{a_{i+1}}\,.~~~\label{defin-C2}
\eea
The second equality in \eref{defin-C} holds as long as the Feynman integrand is linear in each polarization vector.
When saying this is a cyclical operator, we mean ${\cal C}^\epsilon_{\vec{\pmb a}_m}$ is invariance under the arbitrary cyclic permutation of the ordered set $\vec{{\pmb a}}_m$.
Suppose one perform $\partial_{\epsilon_{a_1}\cdot k_{a_m}}$ on ${\bf I}^{\epsilon,\W\epsilon}_{\rm GR;\circ}({\pmb a}^h_n)$, this operator acts on both $\epsilon_{a_1}\triangleright k_{a_m}$
and $\epsilon_{a_1}\triangleleft k_{a_m}$, as well as ordinary $\epsilon_{a_1}\cdot k_{a_m}$ in ${\cal A}^{\epsilon,\W\epsilon}_{\rm GR}({\pmb a}^h_n\cup\{+^h,-^h\})$. To collect all contributions together, we will workout the effect of acting $\Big(\prod_{i=1}^{m-1}\,\partial_{\epsilon_{a_{i+1}}\cdot k_{a_i}}\Big)\partial_{\epsilon_{a_1}\triangleleft k_{a_m}}$ on ${\bf I}^{\epsilon,\W\epsilon}_{\rm GR;\circ}({\pmb a}^h_n)$, and argue that
the operator ${\cal C}^\epsilon_{\vec{\pmb a}_m}$
annihilates ${\cal A}^{\epsilon,\W\epsilon}_{\rm GR}({\pmb a}^h_n\cup\{+^h,-^h\})$.

To study the effect of the operator $\Big(\prod_{i=1}^{m-1}\,\partial_{\epsilon_{a_{i+1}}\cdot k_{a_i}}\Big)\partial_{\epsilon_{a_1}\triangleleft k_{a_m}}$, let us go back to the $4$-gluon example, and use the cyclic symmetry of color ordering, as well as the tree level transmutation operators, to understand the $4$-gluon tree amplitude as
\bea
{\cal A}^{\epsilon,\W\epsilon}_{\rm sEYM}(-^g,+^g,a^g,b^g;{\pmb a}^h_n\setminus\{a^h,b^h\})&=&{\cal I}^\epsilon_{-+a}\,{\cal I}^\epsilon_{-ab}\,{\cal T}^\epsilon_{-b}\,{\cal A}^{\epsilon,\W\epsilon}_{\rm GR}({\pmb a}^h_n\cup\{+^h,-^h\})\nn
&=&(-\partial_{\epsilon_+\cdot\ell}-\partial_{\epsilon_+\cdot k_a})\,
(-\partial_{\epsilon_a\cdot\ell}-\partial_{\epsilon_a\cdot k_b})\,\partial_{\epsilon_-\cdot\epsilon_b}\,{\cal A}^{\epsilon,\W\epsilon}_{\rm GR}({\pmb a}^h_n\cup\{+^h,-^h\})\,.~~~~\label{tree-4p5}
\eea
Here the color ordering is generated by choosing reference legs $-$ and $b$, inserting the leg $a$ between $-$ and $b$, and inserting $+$
between $-$ and $a$. The operators $-\partial_{\epsilon_i\cdot\ell}$ for $i=+,a$ are understood as $\partial_{\epsilon_i\cdot k_-}$.
In the expression \eref{tree-4p5}, the operator $\partial_{\epsilon_+\cdot k_a}\partial_{\epsilon_-\cdot\epsilon_b}$ acts on $(\epsilon_+\cdot k_a)(\epsilon_-\cdot\epsilon_b)$, thus yields the $1$-loop level operator which acts on $\epsilon_b\triangleleft k_a\equiv\sum_r(\epsilon_b\cdot\epsilon^r_-)(\epsilon^r_+\cdot k_a)$.
Follow the
discussion for the ${\cal A}^{\epsilon,\W\epsilon}_{\rm sEYM}(a^g,b^g,-^g,+^g;{\pmb a}^h_n\setminus\{a^h,b^h\})$ case, and use the completely similar method, one obtain
\bea
(-\partial_{\epsilon_a\cdot k_b})\,(-\partial_{\epsilon_b\triangleleft k_a})\,{\bf I}^{\epsilon,\W\epsilon}_{\rm GR;\circ}({\pmb a}^h_n)
={\bf I}^{\epsilon,\W\epsilon}_{\rm ssEYM}(+^g,a^g,b^g,-^g;{\pmb a}^h_n\setminus\{a^h,b^h\})\,,
\eea
and equivalently,
\bea
\partial_{\epsilon_b\cdot k_a}\,\partial_{\epsilon_a\triangleleft k_b}\,{\bf I}^{\epsilon,\W\epsilon}_{\rm GR;\circ}({\pmb a}^h_n)
=(-)^2\,{\bf I}^{\epsilon,\W\epsilon}_{\rm ssEYM}(+^g,b^g,a^g,-^g;{\pmb a}^h_n\setminus\{a^h,b^h\})\,.~~~~\label{tree-4p6}
\eea
It is straightforward to generalize the relation \eref{tree-4p6} to
\bea
\Big(\prod_{i=1}^{m-1}\,\partial_{\epsilon_{a_{i+1}}\cdot k_{a_i}}\Big)\,\partial_{\epsilon_{a_1}\triangleleft k_{a_m}}\,{\bf I}^{\epsilon,\W\epsilon}_{\rm GR;\circ}({\pmb a}^h_n)
=(-)^m\,{\bf I}^{\epsilon,\W\epsilon}_{\rm ssEYM}(+^g,a^g_m,\cdots,a^g_1,-^g;{\pmb a}^h_n\setminus{\pmb a}^h_m)\,,~~~~\label{tree-mp2}
\eea
with $m\geq2$.

Then, we prove that the cyclical operator ${\cal C}^\epsilon_{\vec{\pmb a}_m}$ annihilates ${\cal A}^{\epsilon,\W\epsilon}_{\rm GR}({\pmb a}^h_n\cup\{+^h,-^h\})$. To reach the goal, we again use the CHY formula. Here we only mention the conclusion,
\bea
{\cal C}^\epsilon_{\vec{\pmb a}_m}\,{\cal A}^{\epsilon,\W\epsilon}_{\rm GR}({\pmb a}^h_n\cup\{+^h,-^h\})=0\,.~~~\label{c-ani-A}
\eea
The details are postponed to the next subsection.

With the conclusions \eref{tree-mp1}, \eref{tree-mp2} and \eref{c-ani-A}, now we are ready to determine the consequence of acting ${\cal C}^\epsilon_{\vec{\pmb a}_m}$ on ${\bf I}^{\epsilon,\W\epsilon}_{\rm GR;\circ}({\pmb a}^h_n)$. Since there are only one $\epsilon_+$ and one $\epsilon_-$ at the tree level, among $m$
operators $\partial_{\epsilon_{a_i}\cdot k_{a_{i-1}}}$, at most one of them can act on $\epsilon_{a_i}\triangleright k_{a_{i-1}}$ or
$\epsilon_{a_i}\triangleleft k_{a_{i-1}}$, the remaining operators must act on original $\epsilon_{a_i}\cdot k_{a_{i-1}}$ in ${\cal A}^{\epsilon,\W\epsilon}_{\rm GR}({\pmb a}^h_n\cup\{+^h,-^h\})$. On the other hand, if non of them acts on $\epsilon_{a_i}\triangleright k_{a_{i-1}}$ or
$\epsilon_{a_i}\triangleleft k_{a_{i-1}}$, the equality \eref{c-ani-A} indicates the vanishing of the result. Thus, the non-vanishing contributions
come from $\Big(\prod_{j=i}^{i-2}\,\partial_{\epsilon_{a_{j+1}}\cdot k_{a_j}}\Big)\partial_{\epsilon_{a_i}\triangleright k_{a_{i-1}}}$, as well as $\Big(\prod_{j=i}^{i-2}\,\partial_{\epsilon_{a_{j+1}}\cdot k_{a_j}}\Big)\partial_{\epsilon_{a_i}\triangleleft k_{a_{i-1}}}$, for all $i\in\{1,\cdots,m\}$. Applying \eref{tree-mp1} and \eref{tree-mp2},
we finally find
\bea
{\cal C}^\epsilon_{\vec{\pmb a}_m}\,{\bf I}^{\epsilon,\W\epsilon}_{\rm GR;\circ}({\pmb a}^h_n)={\bf I}^{\epsilon,\W\epsilon}_{{\rm ssEYM};\circ}(\vec{{\pmb a}}^g_m;{\pmb a}^h_n\setminus{\pmb a}^h_m)+(-)^{m}{\bf I}^{\epsilon,\W\epsilon}_{{\rm ssEYM};\circ}(\overleftarrow{\pmb a}^g_m;{\pmb a}^h_n\setminus{\pmb a}^h_m)\,,~~~~{\rm with}~m\geq2\,,~~~~\label{loop-mp-main}
\eea
where $\overleftarrow{\pmb a}_m$ is the reversed set of $\vec{{\pmb a}}_m$, i.e., $\overleftarrow{\pmb a}_m=\langle a_m,\cdots,a_1\rangle$, and
\bea
{\bf I}^{\epsilon,\W\epsilon}_{{\rm ssEYM};\circ}(\vec{{\pmb a}}^g_m;{\pmb a}^h_n\setminus{\pmb a}^h_m)&=&\sum_{\pi_c}\,{\bf I}^{\epsilon,\W\epsilon}_{{\rm ssEYM}}(+,\pi_c(\vec{{\pmb a}}^g_m),-;{\pmb a}^h_n\setminus{\pmb a}^h_m)\,,\nn
{\bf I}^{\epsilon,\W\epsilon}_{{\rm ssEYM};\circ}(\overleftarrow{\pmb a}^g_m;{\pmb a}^h_n\setminus{\pmb a}^h_m)&=&\sum_{\pi_c}\,{\bf I}^{\epsilon,\W\epsilon}_{{\rm ssEYM}}(+,\pi_c(\overleftarrow{\pmb a}^g_m),-;{\pmb a}^h_n\setminus{\pmb a}^h_m)\,,
\eea
where $\pi_c$ are the cyclic permutations. Now we have found the $1$-loop level operators ${\cal C}^\epsilon_{\vec{\pmb a}_m}$, which transmute the GR Feynman integrands to the ssEYM ones, as can be seen in \eref{loop-mp-main}. The r.h.s of \eref{loop-mp-main} is invariant under the cyclic permutation of $\vec{\pmb a}_m$, as indicated by the cyclic symmetry of the operator ${\cal C}^\epsilon_{\vec{\pmb a}_m}$. We emphasize that ${\bf I}^{\epsilon,\W\epsilon}_{{\rm ssEYM};\circ}(\vec{{\pmb a}}^g_m;{\pmb a}^h_n\setminus{\pmb a}^h_m)$ and ${\bf I}^{\epsilon,\W\epsilon}_{{\rm ssEYM};\circ}(\overleftarrow{\pmb a}^g_m;{\pmb a}^h_n\setminus{\pmb a}^h_m)$ appear in \eref{loop-mp-main} are full $1$-loop color ordered ssEYM Feynman integrands, rather than partial ones without the cyclic summation.

Some discussions are in order. First, the operators ${\cal C}^\epsilon_{\vec{{\pmb a}}_m}$ do not act on any Lorentz invariants those include
the loop momentum $\ell$, thus are commutable with the integration of $\ell$. This observation implies that the relation \eref{loop-mp-main}
holds at not only the integrands level, but also the $1$-loop amplitudes level.

Secondly, the individual operator $\partial_{\epsilon_{a_{i+1}}\cdot k_{a_i}}$ breaks the gauge invariance of the
leg $a_i$, but the full operator ${\cal C}^\epsilon_{\vec{{\pmb a}}_m}$ will not, since $\epsilon_{a_i}$ in ${\bf I}^{\epsilon,\W\epsilon}_{\rm GR;\circ}({\pmb a}^h_n)$ will be removed by $\partial_{\epsilon_{a_i}\cdot k_{a_{i-1}}}$. More explicitly, let us consider the Ward's identity
operator for the leg $a_i$,
\bea
{\cal W}^\epsilon_{a_i}\equiv\sum_V\,k_{a_i}\cdot V\,\partial_{\epsilon_{a_i}\cdot V}\,,~~~~\label{ward-iden}
\eea
where $V$ denotes either momenta or polarization vectors contract with $\epsilon_{a_i}$.
The gauge invariance indicates ${\cal W}^\epsilon_{a_i}{\bf I}^{\epsilon,\W\epsilon}_{\rm GR;\circ}({\pmb a}^h_n)=0$,
therefore $\partial_{\epsilon_{a_{i+1}}\cdot k_{a_i}}{\cal W}^\epsilon_{a_i}{\bf I}^{\epsilon,\W\epsilon}_{\rm GR;\circ}({\pmb a}^h_n)=0$,
which is equivalent to
\bea
(\partial_{\epsilon_{a_{i+1}}\cdot k_{a_i}}\,{\cal W}^\epsilon_{a_i})\,{\bf I}^{\epsilon,\W\epsilon}_{\rm GR;\circ}({\pmb a}^h_n)+
{\cal W}^\epsilon_{a_i}\,\partial_{\epsilon_{a_{i+1}}\cdot k_{a_i}}\,{\bf I}^{\epsilon,\W\epsilon}_{\rm GR;\circ}({\pmb a}^h_n)=0\,.
\eea
The first term at the l.h.s gives the non-vanishing contribution $\partial_{\epsilon_{a_{i+1}}\cdot \epsilon_{a_i}}{\bf I}^{\epsilon,\W\epsilon}_{\rm GR;\circ}({\pmb a}^h_n)$, thus the second term does not vanish. It means $\partial_{\epsilon_{a_{i+1}}\cdot k_{a_i}}\,{\bf I}^{\epsilon,\W\epsilon}_{\rm GR;\circ}({\pmb a}^h_n)$ is not gauge invariant for the leg $a_i$. However, replacing $\partial_{\epsilon_{a_{i+1}}\cdot k_{a_i}}$ by ${\cal C}^\epsilon_{\vec{{\pmb a}}_m}$ yields
\bea
({\cal C}^\epsilon_{\vec{{\pmb a}}_m}\,{\cal W}^\epsilon_{a_i})\,{\bf I}^{\epsilon,\W\epsilon}_{\rm GR;\circ}({\pmb a}^h_n)+
{\cal W}^\epsilon_{a_i}\,{\cal C}^\epsilon_{\vec{{\pmb a}}_m}\,{\bf I}^{\epsilon,\W\epsilon}_{\rm GR;\circ}({\pmb a}^h_n)=0\,.
\eea
Both two term at the l.h.s vanish manifestly, due to the observation that the Feynman integrand is linear in each polarization vector.
Thus we claim that the operators ${\cal C}^\epsilon_{\vec{{\pmb a}}_m}$ preserve the gauge invariance.

Thirdly, at the tree level, the operators which transmute the tree GR amplitudes to the tree sEYM amplitudes, also transmute the color ordered YM ones to
the sYMS ones, as can be seen in Table. \ref{tab:unifying}. Thus, along the same line which yields \eref{loop-mp-main}, one finds that the operators ${\cal C}^\epsilon_{\vec{{\pmb a}}_m}$ also transmute the $1$-loop YM partial Feynman integrands to the ssYMS partial integrands as follows,
\bea
{\cal C}^\epsilon_{\vec{\pmb a}_m}\,{\bf I}^{\epsilon}_{\rm YM}(+^g,\vec{{\pmb a}}^g_n,-^g)={\bf I}^{\epsilon}_{{\rm ssYMS}}(\vec{{\pmb a}}^s_m;{\pmb a}^g_n\setminus{\pmb a}^g_m\parallel+^A,\vec{\pmb a}^A_n,-^A)+(-)^m{\bf I}^{\epsilon}_{{\rm ssYMS}}(\overleftarrow{\pmb a}^s_m;{\pmb a}^g_n\setminus{\pmb a}^g_m\parallel+^A,\vec{\pmb a}^A_n,-^A)\,.~~~~\label{loop-mp-YM}
\eea
Doing the cyclic summation for the color orderings $+,\vec{\pmb a}_n,-$, one obtain
\bea
{\cal C}^\epsilon_{\vec{\pmb a}_m}\,{\bf I}^{\epsilon}_{\rm YM;\circ}(\vec{{\pmb a}}^g_n)={\bf I}^{\epsilon}_{{\rm ssYMS};\circ}(\vec{{\pmb a}}^s_m;{\pmb a}^g_n\setminus{\pmb a}^g_m\parallel\vec{\pmb a}^A_n)+(-)^m{\bf I}^{\epsilon}_{{\rm ssYMS};\circ}(\overleftarrow{\pmb a}^s_m;{\pmb a}^g_n\setminus{\pmb a}^g_m\parallel\vec{\pmb a}^A_n)\,,~~~~\label{loop-mp-YM-f}
\eea
which links the full color ordered YM Feynman integrands and full double-color ordered ssYMS integrands together.

Finally, we explain the reason for generating one reference leg from $\{+^h,-^h\}$ and another one from ${\pmb a}^h_n$. The choice that both two reference legs
are from $\{+^h,-^h\}$ was already studied in \cite{Zhou:2021kzv}. If one choose both two reference legs from ${\pmb a}^h_n$, the result is complicated and hard to interpret. Choosing both two reference legs from ${\pmb a}^h_n$ means, both $+$ and $-$ should be inserted into the color ordering at the tree level, via the insertion operators. Suppose
we have the color ordering $\cdots,a,b,\cdots$, we can insert $+$ and $-$ between them to create the new color ordering
$(\cdots,a,+,-,b,\cdots)$ (recall that in this paper we only consider the single-trace color ordered Feynman integrands, with $+$ and $-$ adjancent, as mentioned in subsection. \ref{subsecforward} of section. \ref{secreview}), the corresponding insertion operators are
\bea
{\cal I}^\epsilon_{+-b}{\cal I}^\epsilon_{a+b}=(\partial_{\epsilon_-\cdot\ell}-\partial_{\epsilon_-\cdot k_b})(\partial_{\epsilon_+\cdot k_a}-\partial_{\epsilon_+\cdot k_b})\,.
\eea
Under the action of ${\cal L}$, one can eliminate $\partial_{\epsilon_-\cdot\ell}$ in ${\cal I}^\epsilon_{+-b}$. Under the action of
${\cal E}$, we have $\sum_r(k_b\cdot\epsilon^r_-)(\epsilon^r_+\cdot k_a)=k_b\cdot k_a$, $\sum_r(k_b\cdot\epsilon^r_-)(\epsilon^r_+\cdot k_a)=k_b\cdot k_b=0$. Thus the tree level ${\cal I}^\epsilon_{+-b}{\cal I}^\epsilon_{a+b}$ equivalent to $\partial_{k_b\triangleleft k_a}$ at the $1$-loop level. Since $k_b\triangleleft k_a$, $k_b\triangleright k_a$ and $k_b\cdot k_a$ can not be distinguished in the Feynman integrand, we should use $\partial_{k_a\cdot k_b}$ instead of $\partial_{k_b\triangleleft k_a}$. But the operator $\partial_{k_a\cdot k_b}$ is not a desired one, due to the double-copy structure. When performing $\partial_{\epsilon_a\cdot V}$
to ${\bf I}^{\epsilon,\W\epsilon}_{\rm GR;\circ}({\pmb a}^h_n)$, the operator acts only on the one copy, since another one depends on $\{\W\epsilon_i\}$ rather than $\{\epsilon_i\}$. This property allows us to reduce the GR Feynman integrand to the YM ones, by applying the operators in the form $\partial_{\epsilon_a\cdot V}$. However, the operator $\partial_{k_a\cdot k_b}$ will act on both two copies, thus the resulting object is complicated, without the clear physical interpretation. Of course, one can still
try to workout the result, and try to interpret it. Then another difficulty occurs, it is hard to find a tool to verify the result. For operators
in the form $\partial_{\epsilon_a\cdot V}$, it is convenient to verify their properties by employing the CHY formula, as we will do in the next subsection. The reason is, the operators $\partial_{\epsilon_a\cdot V}$ will not alter the CHY measure, thus can be applied to the CHY integrands directly. However, the operator $\partial_{k_a\cdot k_b}$ can act on the measure, thus is not commutable with CHY contour integration. For this case, the CHY formula is not an effective tool.

\subsection{Verification via CHY formulas}
\label{verify}

This subsection devotes to verify the results in the previous subsection by using CHY formulas. The reader can skim this part if
only care about the results.

In the previous subsection, we have used the CHY formula twice. The first time, we employed
the CHY formula to eliminate $\partial_{\epsilon_b\cdot\ell}$ in ${\cal I}^\epsilon_{ab+}$; the second time, we used the CHY formula to prove
that ${\cal C}^\epsilon_{\vec{{\pmb a}}_m}$ annihilates ${\cal A}^{\epsilon,\W\epsilon}_{\rm GR}({\pmb a}^h_n\cup\{+,-\})$, the details were postponed to the current subsection. The reason for choosing CHY formula is as follows. The differential operators only act on Lorentz invariants contain external momenta and polarization vectors,
thus it is convenient to choose a description of amplitudes which is independent of the internal off-shell virtual particles, to calculate and prove the related properties of operators. From this point of view, the CHY formula is a nice candidate. Furthermore, the operators under consideration
are in the form $\partial_{\epsilon_a\cdot V}$, thus are commutable with the CHY contour integration. Thus we can act the operators on the CHY integrands directly without altering the measure. Based on the same reason, in this subsection,
we use CHY formulas to verify the main result \eref{loop-mp-main} in the previous subsection. The introduction to CHY formulas is given in Appendix. \ref{sec-CHY}.

Without lose of generality, let us assume $\vec{\pmb a}_m=\langle1,\cdots,m\rangle$. Since the operator ${\cal C}^\epsilon_{\vec{{\pmb a}}_m}$ only depends on polarization vectors in $\{\epsilon_i\}$, it only
acts on ${\cal F}\,{\bf Pf}'\Psi(\epsilon_i,k_i,z_i)$, with out altering the ${\cal F}\,{\bf Pf}'\Psi(\W\epsilon_i,k_i,z_i)$ part. As pointed out in the previous subsection, the non-vanishing contributions arise from acting
${\cal C}^\epsilon_{\vec{{\pmb a}}_m}$ on terms contain $\epsilon_i\triangleright k_{i-1}$, or $\epsilon_i\triangleleft k_{i-1}$ with $i\in\{1,\cdots,m\}$, as indicated by the equality \eref{c-ani-A}. We will prove the equality \eref{c-ani-A} at the end of this subsection. For now, let us just accept it and continue. Based on the discussion above, now we consider the effect of acting ${\cal C}^\epsilon_{\vec{{\pmb a}}_m}$ on terms contain $\epsilon_1\triangleright k_m$, the treatments for other terms are analogous.

When performing to $\epsilon_1\triangleright k_m$, the operator $\partial_{\epsilon_1\cdot k_m}$ included in ${\cal C}^\epsilon_{\vec{{\pmb a}}_m}$ acts on $\epsilon_1\cdot\epsilon_+$ and $\epsilon_-\cdot k_m$ in ${\bf Pf}'\Psi(\epsilon_i,k_i,z_i)$,
thus transmutes ${\cal F}{\bf Pf}\Psi'$ as
\bea
{\cal F}\,{\bf Pf}\,\Psi'\,\to\,{1\over z_{1+}z_{m-}}\,{\cal F}\,{\bf Pf}\,(\Psi')^{1m(2n+1)(2n+2)}_{1m(2n+1)(2n+2)}\,,~~~~\label{trans-psi}
\eea
where the later matrix is obtained from $\Psi'$ by deleting $1^{\rm th}$, $m^{\rm th}$, $(2n+1)^{\rm th}$, $(2n+2)^{\rm th}$ rows and columns.
More explicitly, using the definition of Pfaffian, one can expand ${\bf Pf}\Psi'$ as
\bea
{\bf Pf}\Psi'=\sum_{\a\in\Pi}\,{\bf sgn}(\sigma_{\a})\,(\Psi')_{a_1b_1}(\Psi')_{a_2b_2}\cdots(\Psi')_{a_nb_n}\,.~~~~\label{pfa-insertion}
\eea
We dived terms in the summation at the r.h.s of \eref{pfa-insertion} into two classes, terms in the first class are those do not contain both
$(\Psi')_{1(2n+1)}$ and $(\Psi')_{m(2n+2)}$, while those in the second class contain both
$(\Psi')_{1(2n+1)}$ and $(\Psi')_{m(2n+2)}$.
The operator $\partial_{\epsilon_1\cdot k_m}$ annihilates all terms in the first class. Notice that since the matrix $\Psi'$ is generated from the original $\Psi$ by removing $(n+1)^{\rm th}$ and $(n+2)^{\rm th}$ rows and columns, $\partial_{\epsilon_1\cdot\epsilon_+}$ and $\partial_{\epsilon_-\cdot k_m}$ can not act on diagonal elements $C_{(n+1)(n+1)}$
and $C_{(n+2)(n+2)}$ of the block $C$. While acting on terms in the second class,
the operator $\partial_{\epsilon_1\cdot k_m}$ transmutes ${\cal F}[(\Psi')_{1(2n+1)}(\Psi')_{m(2n+2)}]$ as
\bea
\partial_{\epsilon_1\cdot k_m}\,{\cal F}\Big[(\Psi')_{1(2n+1)}(\Psi')_{m(2n+2)}\Big]=\partial_{\epsilon_1\cdot k_m}\,{\epsilon_1\triangleright k_m\over z_{1+}z_{m-}}={1\over z_{1+}z_{m-}}\,.
\eea
Now we collect terms in the second class together, and remove $(\Psi')_{1(2n+1)}(\Psi')_{m(2n+2)}$ in them.
By the definition of Pfaffian, the obtained terms after performing the above manipulation can be regrouped as ${\bf Pf}(\Psi')^{1m(2n+1)(2n+2)}_{1m(2n+1)(2n+2)}$, up to an irrelevant overall sign arises from the change of ${\bf sgn}(\sigma_{\a})$.
Thus we get the result \eref{trans-psi}.

Then, we perform the the operator $\partial_{k_1\cdot \epsilon_2}$ to ${\cal F}{\bf Pf}(\Psi')^{1m(2n+1)(2n+2)}_{1m(2n+1)(2n+2)}$.
Since the matrix $(\Psi')^{1m(2n+1)(2n+2)}_{1m(2n+1)(2n+2)}$ does not include $1^{\rm th}$ rows and columns, $\partial_{k_1\cdot \epsilon_2}$
only acts on the diagonal element $C_{22}$, thus transmutes ${\cal F}{\bf Pf}(\Psi')^{1m(2n+1)(2n+2)}_{1m(2n+1)(2n+2)}$ as
\bea
{\cal F}\,{\bf Pf}\,(\Psi')^{1m(2n+1)(2n+2)}_{1m(2n+1)(2n+2)}\,\to\,{1\over z_{21}}\,{\cal F}\,{\bf Pf}\,(\Psi')^{12m(2n+1)(2n+2)}_{12m(2n+1)(2n+2)}\,,
\eea
up to an overall sign.
Now the recursive pattern occurs, the operator $\partial_{k_2\cdot \epsilon_3}$ only acts on $C_{33}$, due to the observation
$(\Psi')^{12m(2n+1)(2n+2)}_{12m(2n+1)(2n+2)}$ does not include $2^{\rm th}$ rows and columns. Thus, by iterating the above manipulation,
one finally arrive at
\bea
{\cal F}\,{\bf Pf}\,\Psi'\,\to\, {1\over z_{1+}z_{m-}}\,\prod_{i=1}^{m-1}\,{1\over z_{(i+1)i}}\,,
\eea
therefore
\bea
{\cal F}\,{\bf Pf}'\Psi(\epsilon_i,k_i,z_i)\,\to\, \,PT(+,1,\cdots,m,-) \,,
\eea
up to an irrelevant overall sign.

The resulting object $PT(+,1,\cdots,m,-)$, together with ${\cal F}\,{\bf Pf}'\Psi(\W\epsilon_i,k_i,z_i)$, give rise to the ssEYM
$1$-loop CHY integrand, which leads to the partial Feynman integrand ${\bf I}^{\epsilon,\W\epsilon}_{{\rm ssEYM}}(+^g,1^g,\cdots,m^g,-^g;{\pmb a}^h_n\setminus{\pmb a}^h_m)$. Similar manipulation shows that acting ${\cal C}^\epsilon_{\vec{{\pmb a}}_m}$ on terms contain $\epsilon_1\triangleleft k_m$ yields
$(-)^m{\bf I}^{\epsilon,\W\epsilon}_{{\rm ssEYM}}(+^g,m^g,\cdots,1^g,-^g;{\pmb a}^h_n\setminus{\pmb a}^h_m)$. The relative sign $(-)^m$ arises as follows.
The terms contain $\epsilon_1\triangleright k_m$ are those contain
\bea
{\cal F}\Big[(\Psi')_{1(2n+1)}(\Psi')_{m(2n+2)}\Big]={\epsilon_1\triangleright k_m\over z_{1+}z_{m-}}\,,
\eea
while terms contain $\epsilon_1\triangleleft k_m$ are those contain
\bea
{\cal F}\Big[(\Psi')_{1(2n+2)}(\Psi')_{m(2n+1)}\Big]={\epsilon_1\triangleleft k_m\over z_{1-}z_{m+}}\,,
\eea
the difference between these two objects determines the relative sign.
First
\bea
& &{1\over z_{1+}z_{m-}}\,\prod_{i=1}^{m-1}\,{1\over z_{(i+1)i}}=(-)^m\,PT(+,1,\cdots,m,-)\,,\nn
& &{1\over z_{m+}z_{1-}}\,\prod_{i=1}^{m-1}\,{1\over z_{(i+1)i}}=(-)\,PT(+,m,\cdots,1,-)\,,
\eea
comparing them gives a relative sign $(-)^{m-1}$.
Secondly, comparing elements $(\Psi')_{1(2n+1)}(\Psi')_{m(2n+2)}$ with elements $(\Psi')_{1(2n+2)}(\Psi')_{m(2n+1)}$, the permutation from
the ordering $1(2n+1)m(2n+2)$ to $1(2n+2)m(2n+1)$ contributes a relative $(-)$, due to the definition of Pfaffian.

The above results can be generalized to
$\epsilon_i\triangleright k_{i-1}$ or $\epsilon_i\triangleleft k_{i-1}$ for arbitrary $i\in\{1,\cdots,m\}$ via the replacement $1\to i$, due to the cyclic symmetry. Collecting all pieces together,
we get the desired conclusion \eref{loop-mp-main}. Since our method only transmutes ${\cal F}\,{\bf Pf}'\Psi(\epsilon_i,k_i,z_i)$, one can claim that the verification for the relation \eref{loop-mp-YM} which links YM and ssYMS Feynman integrands together has also been verified.

Now we turn to the equality \eref{c-ani-A}. The reason why we postpone it to the end of this subsection is that the technique is similar as that
mentioned above. For our purpose, it is sufficient to show that ${\cal C}^\epsilon_{\vec{\pmb a}_m}\,{\bf Pf}'\Psi(\epsilon_i,k_i,z_i)=0$. Let us consider the $4$-gluon example,
i.e., act $\partial_{\epsilon_b\cdot k_a}\partial_{\epsilon_a\cdot k_b}$ on ${\bf Pf}\Psi'$.
The operator $\partial_{\epsilon_a\cdot k_b}$ acts on both $C_{aa}$ and $C_{ba}$ in the block $C$. For the first case, $\partial_{\epsilon_a\cdot k_b}$ transmutes ${\bf Pf}\Psi'$ as
\bea
{\bf Pf}\,\Psi'\,\to\,{-1\over z_{ba}}\,{\bf Pf}\,(\Psi')^{a(n+a)}_{a(n+a)}\,,
\eea
up to a sign, while for the second case we have
\bea
{\bf Pf}\,\Psi'\,\to\,{1\over z_{ba}}\,{\bf Pf}\,(\Psi')^{b(n+a)}_{b(n+a)}\,,
\eea
again up to a sign. The relative sign will be considered latter.
For $(\Psi')^{a(n+a)}_{a(n+a)}$, the operator $\partial_{\epsilon_b\cdot k_a}$ only acts on $C_{bb}$, thus provides
\bea
{-1\over z_{ba}}{\bf Pf}\,(\Psi')^{a(n+a)}_{a(n+a)}\,\to\,{1\over z_{ab}z_{ba}}\,{\bf Pf}\,(\Psi')^{a(n+a)b(n+b)}_{a(n+a)b(n+b)}\,,~~~~\label{psi-1}
\eea
up to a sign.
For $(\Psi')^{b(n+a)}_{b(n+a)}$, the operator $\partial_{\epsilon_b\cdot k_a}$ only acts on $C_{ab}$, thus gives
\bea
{1\over z_{ba}}{\bf Pf}\,(\Psi')^{b(n+a)}_{b(n+a)}\,\to\,{1\over z_{ab}z_{ba}}\,{\bf Pf}\,(\Psi')^{b(n+a)a(n+b)}_{b(n+a)a(n+b)}={1\over z_{ab}z_{ba}}\,{\bf Pf}\,(\Psi')^{a(n+a)b(n+b)}_{a(n+a)b(n+b)}\,,~~~\label{psi-2}
\eea
up to a sign.
It seems that after performing $\partial_{\epsilon_b\cdot k_a}\partial_{\epsilon_a\cdot k_b}$, we arrive at the Pfaffians of two equivalent matrices, with the same coefficient. However, we have not consider the relative sign until now. Notice that \eref{psi-1} is obtained by turning elements $(\Psi')_{a(n+a)}$ and $(\Psi')_{b(n+b)}$ to $1$, and eliminating all terms in ${\bf Pf}\Psi'$ those do not contain both two elements, while \eref{psi-2} is obtained by turning $(\Psi')_{b(n+a)}$ and $(\Psi')_{a(n+b)}$ to $1$, and eliminating all terms do not contain both of them. Comparing $(\Psi')_{a(n+a)}(\Psi')_{b(n+b)}$
with $(\Psi')_{b(n+a)}(\Psi')_{a(n+b)}$, the permutation from the ordering $a(n+a)b(n+b)$ to $b(n+a)a(n+b)$ is odd. Thus, when considering $\partial_{\epsilon_b\cdot k_a}\partial_{\epsilon_a\cdot k_b}{\bf Pf}\Psi'$, contributions from
\eref{psi-1} and \eref{psi-2} cancel each other, due to the definition of Pfaffian. Consequently, ${\cal C}^\epsilon_{\vec{\pmb a}_2}\,{\bf Pf}'\Psi=0$
for $\vec{{\pmb a}}_2=\langle a,b\rangle$. The above argument can be generalized to the general ordered set $\vec{{\pmb a}}_m=\langle a_1,\cdots,a_m\rangle$ directly, thus we get the conclusion \eref{c-ani-A}.

\section{Solving coefficients in expansions}
\label{solve-coefficient}

At the tree level, the unifying relations among different theories can also be reflected by expansions, i.e, the amplitude of one theory can be expanded to amplitudes of another theory. Especially, all theories in Table. \ref{tab:unifying} can be expanded to BAS amplitudes, with double copied coefficients.
The unified web for expansions serves as the dual version of the web for transmutational relations, based on three reasons. The first one, one can construct the
full web of expansions only through transmutation operators, as well as few very natural and general requirements such as Lorentz invariance, gauge invariance, and so on, without knowing any detail about tree amplitudes themselves.
The second reason, two unified webs include the same list of theories. Finally, there are one-to-one mappings between classes of differential operators and classes of coefficients in expansions.

At the $1$-loop level, one can take the forward limit of the expanded formulas of tree amplitudes to get the expansions of $1$-loop Feynman integrands. Then, one can also verify the transmutational relations, by applying operators to the expanded
formulas of $1$-loop Feynman integrands. However, inspired by the tree level story, a more interesting question is, can the $1$-loop level web
of expansions be constructed from the $1$-loop transmutational relations, together with other appropriate general principles and assumptions? The answer is positive, as will be shown in this and next sections. The plain is as follows. In this section, we will use the differential operators together with few principles/assumptions to solve the expansions of $1$-loop ssEYM partial Feynman integrands and GR Feynman integrands to YM partial Feynman integrands. In the next section,
we will act differential operators to the expanded formulas of GR Feynman integrands, to generate expansions of Feynman integrands of other theories, and construct the complete unified web of expansions.

It is worth to list the principles and assumptions beside differential operators, which will be used in this section to solve expansions:
\begin{itemize}
\item (1) {\bf Lorentz invariance}
\item (2) {\bf Gauge invariance}
\item (3) {\bf Property of GR Feynman integrands:} We assume that each external graviton $i^h$ carries the polarization tensor $\epsilon^{\mu\nu}_i=\epsilon^\mu_i\W\epsilon^\nu_i$, and the gravitational Feynman integrands carry no color ordering.
\item (4) {\bf Double-copy structure:} We assume that each polarization vector in the set $\{\epsilon_i\}$ can not contract with another polarization vector in the set $\{\W\epsilon_i\}$.
\item (5) {\bf On-shell condition:} We assume that $\epsilon_i\cdot k_i=0$ for each external leg $i$.
\item (6) {\bf Linearity in $\epsilon_i$:} We assume that the Feynman integrand is linear in each polarization vector $\epsilon_i$.
\item (7) {\bf Forward limit:} We assume that the $1$-loop integrands are generated from the tree amplitudes via the forward limit.
\end{itemize}
Notice that the first six principles/assumptions are also used for deriving the expansions of tree amplitudes, while the last one only makes
sense at the $1$-loop level. Let us give a brief discussion about the third assumption. It seems that one should make similar assumptions for other theories, but indeed it is not necessary, since the assumption for the GR Feynman integrands, together with differential operators, uniquely fix the information about polarization vectors and color orderings for the Feynman integrands of other theories. For example, the relation
\bea
{\cal T}^\epsilon_{+\vec{\pmb a}^g_n-}\,{\bf I}^{\epsilon,\W\epsilon}_{\rm GR}({\pmb a}^h_n)={\bf I}^{\W\epsilon}_{\rm YM}(+^g,\vec{\pmb a}^g_n,-^g)
\eea
indicates each external gluon $i^g$ carries the polarization vector $\W\epsilon_i$, and the YM
partial integrand carries the color ordering $+,\vec{\pmb a}_n,-$, if each graviton carries $\epsilon_i\W\epsilon_i$ and the gravitational Feynman integrand carries no color ordering. Notice that the above relation also implies
\bea
{\rm dim}\Big({\bf I}^{\epsilon,\W\epsilon}_{\rm GR}({\pmb a}^h_n)\Big)={\rm dim}\Big({\bf I}^{\W\epsilon}_{\rm YM}(+^g,\vec{\pmb a}^g_n,-^g)\Big) +n\,,
\eea
where ${\rm dim}(A)$ denotes the number of mass dimensions of $A$.
The information carried by transmutation operators, together with the seven general principles/assumptions mentioned above, fully determine the expansions of ssEYM partial Feynman integrands and GR Feynman integrands to YM partial integrands, with polynomial coefficients.

Now we briefly discuss the basis for expansions.
At the tree level, with $n+2$ external legs, one can take the basis as $n!$ color ordered YM amplitudes with two legs fixed at two ends in the color ordering, and expand the EYM and GR amplitudes to these YM amplitudes with polynomial coefficients. Such basis is called the KK basis, since its completeness is ensured by the
Kleiss-Kuijf relation \cite{Kleiss:1988ne}. Suppose we fix legs $+$ and $-$ at two ends in the color ordering to generate the tree level KK basis, taking the forward limit naturally leads to the $1$-loop
KK basis including YM partial Feynman integrands ${\bf I}^{\W\epsilon}_{\rm YM}(+^g,\sigma(\vec{\pmb a}^g_n),-^g)$, where $\sigma$ denotes the permutations. This is the choice of basis in this section.
In the next section, we will generalize the $1$-loop KK basis to color ordered partial Feynman integrands of other theories.

The main idea in this section can be summarized as follows. Suppose a $1$-loop level operator ${\cal O}_\circ$ transmutes the Feynman integrand of theory $A$
to that of theory $B$ as follows
\bea
{\cal O}_\circ\,{\bf I}_A={\bf I}_B\,,~~~~\label{equ-diff}
\eea
where ${\bf I}_A$ can be either a full integrand or a partial one, and ${\bf I}_B$ is analogous. We regard \eref{equ-diff}
as a differential equation, rather than a transmutational relation. Then, one can solve ${\bf I}_A$
by solving the differential equation. The general feature of the solution is, it contains an un-fixed term which is annihilated by the operator
${\cal O}_\circ$. Terms vanish under the action of ${\cal O}_\circ$ are called un-detectable terms for the operator ${\cal O}_\circ$,
while terms survive under the action of ${\cal O}_\circ$ are called detectable terms. The un-detectable terms should be determined via other methods, such as imposing the gauge invariance, and so on. By applying the method described above, in subsection.\ref{exp-EYM} we solve the expansions of ssEYM partial Feynman integrands to other ssEYM ones with less gravitons. The details are similar as those in \cite{Feng:2019tvb} at tree level. In subsection.\ref{exp-GR}, we solve the expansions of GR
Feynman integrands to ssEYM partial integrands. The details are totally new, without any tree level analog. In subsection. \ref{subsec-coe}, we use the results in subsection.\ref{exp-EYM} and subsection.\ref{exp-GR} to
get the expansions of GR
and ssEYM Feynman integrands to YM ones, and give the rules for constructing the coefficients.

\subsection{Expansion of EYM}
\label{exp-EYM}

By employing the differential operators which link tree amplitudes of different theories together,
one can find the recursive expansions of sEYM tree amplitudes, as can be seen in \cite{Feng:2019tvb}.
In this subsection, we generalize such recursive expansions to the $1$-loop level.

Considering the ssEYM partial Feynman integrand ${\bf I}^{\epsilon,\W\epsilon}_{\rm ssEYM}(+^g,\vec{\pmb a}^g_{n-m},-^g;{\pmb a^h}_{m})$ with $n-m$ gluons and $m$ gravitons. For latter convenience, we denote the ordered set of gluons $\vec{\pmb a}^g_{n-m}$ as $\vec{\pmb a}_{n-m}=\langle1,\cdots,n-m\rangle$, and label the gravitons in ${\pmb a}^h_{m}$ as ${\pmb a}_{m}=\{h_1,\cdots,h_m\}$. The Lorentz invariance, together with the assumption that polarization vectors in $\{\epsilon_i\}$ and $\{\W\epsilon_i\}$ can not contract with each other, indicate the polarization vector $\epsilon_{h_m}$
can only appear in the following
combinations, which are $\epsilon_{h_m}\cdot k_b$, $\epsilon_{h_m}\cdot k_{h_g}$,
and $\epsilon_{h_m}\cdot\epsilon_{h_g}$, with $g\in\{1,\cdots,m-1\}$.
Since ${\bf I}^{\epsilon,\W\epsilon}_{\rm ssEYM}(+^g,\vec{\pmb a}^g_{n-m},-^g;{\pmb a}^h_{m})$ is assumed to be linear in each polarization vector, the partial integrand ${\bf I}^{\epsilon,\W\epsilon}_{\rm ssEYM}(+^g,\vec{\pmb a}^g_{n-m},-^g;{\pmb a}^h_{m})$ can be expanded as
\bea
{\bf I}^{\epsilon,\W\epsilon}_{\rm ssEYM}(+^g,\vec{\pmb a}^g_{n-m},-^g;{\pmb a}^h_{m})&=&(\epsilon_{h_m}\cdot \ell)\,B_++\sum_{b=1}^{n-m}\,(\epsilon_{h_m}\cdot k_b)\,B_b\nn
& &
+\sum_{g=1}^{m-1}\,(\epsilon_{h_m}\cdot k_{h_g})\,(\epsilon_{h_g}\cdot C_g)+\sum_{g=1}^{m-1}\,(\epsilon_{h_m}\cdot \epsilon_{h_g})\,D_g\,.~~~~\label{exp-EYM-KK-mh2} \eea
At the tree level, one can regard one of $k_b$ or $k_{h_g}$ as the un-independent one due to the momentum conservation, and remove the corresponding $\epsilon_{h_m}\cdot k_b$ or $\epsilon_{h_m}\cdot k_{h_g}$, as can be seen in \cite{Feng:2019tvb}. In the current $1$-loop case, we do not do this procedure at the r.h.s of \eref{exp-EYM-KK-mh2}, based on the following reason. At the tree level, if one replace any $\epsilon_i\cdot k_j$ by $-\epsilon_i\cdot(\sum_{l\neq j} k_l)$, the resulting objects after performing the insertion operators will not be modified, as can be observed from the definition of operators.
At the $1$-loop level, similar statement does not hold for our $1$-loop level insertion operators ${\cal I}^\epsilon_{+a_ia_{i+1}}$, as well as the new operators ${\cal C}^\epsilon_{\vec{\pmb a}_m}$. Thus, in this and next subsections, we do not use the momentum conservation to change the representations for external momenta. It means, we solve the special expanded formulas with special representations of external momenta.
On the other hand, we think the momentum $k_-=-\ell$ as being removed, via $k_-=-k_+=-\ell$, and $\epsilon_{h_m}\cdot\ell$ in \eref{exp-EYM-KK-mh2} should be understood as $\epsilon_{h_m}\cdot k_+$, since the effects of acting ${\cal I}^\epsilon_{+a_i-}$, ${\cal I}^\epsilon_{+a_ia_{i+1}}$ and ${\cal C}^\epsilon_{\vec{\pmb a}_m}$ will not be altered if replacing $k_-$ in the Feynman integrands by $-k_+$.

Our aim is to use the differential equations provided by transmutational relations, together with the gauge invariance requirement, to solve coefficients
$B_+$, $B_b$, $C_g^\mu$, $D_g$ in \eref{exp-EYM-KK-mh2}. The desired solutions are formulas of these coefficients consisted by Lorentz invariants such as $\epsilon_{h_m}\cdot k_b$, $\epsilon_{h_m}\cdot k_{h_g}$,
$\epsilon_{h_m}\cdot\epsilon_{h_g}$, and physically meaningful objects such as ssEYM Feynman integrands. We do not care about the exact expressions of  these physical objects appear in solutions. As will be seen, such solutions of $B_+$, $B_b$, $C_g^\mu$, $D_g$ naturally give the recursive expansion of ${\bf I}^{\epsilon,\W\epsilon}_{\rm ssEYM}(+^g,\vec{\pmb a}^g_{n-m},-^g;{\pmb a}^h_{m})$ to partial ssEYM integrands with less gravitons.

The first line at the r.h.s of \eref{exp-EYM-KK-mh2} can be detected by the insertion operators.
The insertion operators act on Lorentz invariants $\epsilon_{h_m}\cdot \ell$ and $\epsilon_{h_m}\cdot k_b$,
thus extract coefficients $B_{+}$ and $B_{b}$. Since the insertion operators transmute the l.h.s of \eref{exp-EYM-KK-mh2}
to ssEYM partial integrands with less gravitons, one can identify $B_{+}$ and $B_{b}$
as the combinations of these partial integrands. Now let us follow the above idea to solve $B_{+}$ and $B_{b}$.
For convenience, we denote nodes $+$ and $-$ as $0$ and $n-m+1$, respectively.
Acting ${\cal I}^\epsilon_{ih_m(i+1)}$
with $i\in\{0,\cdots ,n-m\}$ on the
l.h.s of \eref{exp-EYM-KK-mh2} gives
\bea
{\cal I}^\epsilon_{ih_m(i+1)}\,{\bf I}^{\epsilon,\W\epsilon}_{\rm ssEYM}(+^g,\vec{\pmb a}^g_{n-m},-^g;{\pmb a}^h_{m})
&=&{\bf I}^{\epsilon,\W\epsilon}_{\rm ssEYM}(0^g,\cdots,i^g,h^g_m,(i+1)^g,\cdots,(n-m+1)^g;{\pmb a}^h_m\setminus h^h_m)\,.
~~~~\label{exp-EYM-KK-1hLHS-3}
\eea
While acting on the r.h.s,  these operators annihilate the second line, and transmute the first line as follows,
\bea {\cal I}^\epsilon_{ih_m(i+1)}\,\Big(\sum_{b=1}^{n-m}\,(\epsilon_{h_m}\cdot k_b)\,B_b\Big) =\left\{ \begin{array}{ll} B_i- B_{i+1}\,, & ~~~~{\rm if}~i\leq n-m-1\,, \\ B_i\,,& ~~~~{\rm
if}~i=n-m\,. \end{array} \right.~~~~\label{exp-EYM-KK-1hRHS2-3} \eea
When applying ${\cal I}^\epsilon_{(n-m)h_m(n-m+1)}$, we have used the assumption that $k_-=-\ell$ has been removed by momentum conservation thus the effective part of the operator  is $\partial_{\epsilon_{h_m}\cdot k_{n-m}}$.
Comparing the l.h.s result \eref{exp-EYM-KK-1hLHS-3} with the r.h.s result
\eref{exp-EYM-KK-1hRHS2-3} provides
\bea
B_{n-m}={\bf I}^{\epsilon,\W\epsilon}_{\rm ssEYM}(0^g,\cdots,(n-m)^g,h^g_m,(n-m+1)^g;{\pmb a}^h_m\setminus h^h_m)\,, \eea
and
\bea
B_i=B_{i+1}+{\bf I}^{\epsilon,\W\epsilon}_{\rm ssEYM}(0^g,\cdots,i^g,h^g_m,(i+1)^g,\cdots,(n-m+1)^g;{\pmb a}^h_m\setminus h^h_m)\,,~~~~{\rm for}~i\in\{0,\cdots,n-m-1\}\,. \eea
Thus $B_b$ with $b\in\{0,\cdots,n-m\}$ can be calculated recursively as
\bea
B_b&=&\sum_{i=b}^{n-m}\,{\bf I}^{\epsilon,\W\epsilon}_{\rm ssEYM}(+^g,1^g,\cdots,i^g,h^g_m,(i+1)^g,\cdots,(n-m)^g,-^g;{\pmb a}^h_m\setminus h^h_m)\,.\eea
Substituting the above solution into \eref{exp-EYM-KK-mh2}, the first line at the r.h.s is obtained as
\bea \sum_{b=0}^{n-m}\,(\epsilon_{h_m}\cdot k_b)\,B_b
=\sum_{\shuffle}\,(\epsilon_{h_m}\cdot Y_{h_m})\, {\bf I}^{\epsilon,\W\epsilon}_{\rm ssEYM}(+^g,h^g_m\shuffle\vec{\pmb a}^g_{n-m},-^g;{\pmb a}^h_m\setminus h^h_m)\,.~~~~\label{exp-EYM-KK-1h4} \eea
The combinatory momentum $Y_i$ is defined as the summation of momenta of gluons at the l.h.s of the leg $i^g$ in the color ordering \cite{Fu:2017uzt}.
The summation over all
possible shuffles $\shuffle$ of two ordered sets $\vec{\pmb a}$ and $\vec{\pmb s}$ is the summation over all permutations of $\vec{\pmb a}\cup\vec{\pmb s}$ those preserving the orderings
of $\vec{\pmb a}$ and $\vec{\pmb s}$. For example, $\langle1,2\rangle\shuffle\langle3,4\rangle$ includes the following ordered sets:
$\langle2,3,4,5\rangle$, $\langle2,4,3,5\rangle$, $\langle2,4,5,3\rangle$, $\langle4,2,3,5\rangle$, $\langle4,2,5,3\rangle$, $\langle4,5,2,3\rangle$.

Now we arrive at
\bea
{\bf I}^{\epsilon,\W\epsilon}_{\rm ssEYM}(+^g,\vec{\pmb a}^g_{n-m},-^g;{\pmb a}^h_{m})
&=&\sum_{\shuffle}\,(\epsilon_{h_m}\cdot Y_{h_m})\, {\bf I}^{\epsilon,\W\epsilon}_{\rm ssEYM}(+,h^g_m\shuffle\vec{\pmb a}^g_{n-m},-^g;{\pmb a}^h_m\setminus h^h_m)\nn
& &+\sum_{g=1}^{m-1}\,(\epsilon_{h_m}\cdot k_{h_g})\,(\epsilon_{h_g}\cdot C_g)+\sum_{g=1}^{m-1}\,(\epsilon_{h_m}\cdot \epsilon_{h_g})\, D_g\,.~~~~\label{exp-EYM-KK-mh3}
\eea
To continue, we study the relation between $C_g^\mu$ and $D_g$, by imposing the gauge invariance of the graviton $h_g$. Notice that direct replacing $\epsilon_{h_g}\to k_{h_g}$
makes the treatment complicated, since each term includes $\epsilon_{h_g}$. To single out the $C_g^\mu$ and $D_g$ terms, a convenient way is to use the
ward's identity operator defined in \eref{ward-iden}. The key point is, the gauge invariance condition
\bea
{\cal W}^\epsilon_{h_g}\,{\bf I}^{\epsilon,\W\epsilon}_{\rm ssEYM}(+^g,\vec{\pmb a}^g_{n-m},-^g;{\pmb a}^h_{m})=0
\eea
indicates
\bea
{\cal I}^\epsilon_{h_gh_m-}\,{\cal W}^\epsilon_{h_g}\,{\bf I}^{\epsilon,\W\epsilon}_{\rm ssEYM}(+^g,\vec{\pmb a}^g_{n-m},-^g;{\pmb a}^h_{m})=0\,,
\eea
and the later one is equivalent to
\bea
{\cal W}^\epsilon_{h_g}\,{\cal I}^\epsilon_{h_gh_m-}\,{\bf I}^{\epsilon,\W\epsilon}_{\rm ssEYM}(+^g,\vec{\pmb a}^g_{n-m},-^g;{\pmb a}^h_{m})+\Big({\cal I}^\epsilon_{h_gh_m-}\,{\cal W}^\epsilon_{h_g}\Big)\,{\bf I}^{\epsilon,\W\epsilon}_{\rm ssEYM}(+^g,\vec{\pmb a}^g_{n-m},-^g;{\pmb a}^h_{m})=0\,,~~~~\label{commu-2}
\eea
then we get the equation
\bea
{\cal W}^\epsilon_{h_g}\,{\cal I}_{h_gh_m-}\, {\bf I}^{\epsilon,\W\epsilon}_{\rm ssEYM}(+^g,\vec{\pmb a}^g_{n-m},-^g;{\pmb a}^h_{m})
+\partial_{\epsilon_{h_m}\cdot\epsilon_{h_g}}\, {\bf I}^{\epsilon,\W\epsilon}_{\rm ssEYM}(+^g,\vec{\pmb a}^g_{n-m},-^g;{\pmb a}^h_{m})=0\,.~~~~\label{commu-3}
\eea
Notice that the assumption $k_-=-\ell$ is removed via the momentum conservation has been used again.
Substituting the r.h.s of \eref{exp-EYM-KK-mh3} into the above equation \eref{commu-3}, we obtain
$D_g=-(k_{h_g}\cdot C_g)$, thus
\bea
(\epsilon_{h_m}\cdot k_{h_g})\,(\epsilon_{h_g}\cdot C_g)+(\epsilon_{h_m}\cdot \epsilon_{h_g})\,D_g= \epsilon_{h_m}\cdot f_{h_g}\cdot C_g\,.~~~~\label{BD-Bmh}
\eea
and
\bea
{\bf I}^{\epsilon,\W\epsilon}_{\rm ssEYM}(+^g,\vec{\pmb a}^g_{n-m},-^g;{\pmb a}^h_{m})&=&\sum_{\shuffle}\,(\epsilon_{h_m}\cdot Y_{h_m})\, {\bf I}^{\epsilon,\W\epsilon}_{\rm ssEYM}(+^g,h^g_m\shuffle\vec{\pmb a}^g_{n-m},-^g;{\pmb a}^h_m\setminus h^h_m)\nn
& &+\sum_{g=1}^{m-1}\,(\epsilon_{h_m}\cdot f_{h_g}\cdot C_g)\,,~~~~\label{exp-EYM-KK-mh4} \eea
where the anti-symmetric strength tensors are defined as $f_i^{\mu\nu}\equiv k^\mu_i\epsilon^\nu_i-\epsilon^\mu_i k^\nu_i$, $\W f_i^{\mu\nu}\equiv k^\mu_i\W\epsilon^\nu_i-\W\epsilon^\mu_i k^\nu_i$.

Until now, there is one remaining class of coefficients $C_g^\mu$ has not been fixed.
To solve it, we first need to find the equations satisfied by $C_g^\mu$. This goal can be reached by applying two insertion operators,
one acts on $\epsilon_{h_m}\cdot k_{h_q}$ at the r.h.s of  \eref{exp-EYM-KK-mh4}, therefore selects the term $\epsilon_{h_q}\cdot C_q$, and another one acts on $\epsilon_{h_q}\cdot C_q$
to provide the equation for solving $C_q^\mu$. Two operators must transmute the l.h.s of  \eref{exp-EYM-KK-mh4} in an appropriate way so that
the obtained object is physically meaningful, since otherwise we can not get the expected solution.
Based on the above discussion, the combinatory operators ${\cal I}^{\epsilon}_{jh_q(j+1)}{\cal I}^{\epsilon}_{jh_mh_q}$ with $j\in\{0,\cdots, n-m\}$
are nice candidates. Again, we have denoted nodes $+$ and $-$ as $0$ and $n-m+1$, respectively.

When applying ${\cal I}^{\epsilon}_{jh_q(j+1)}{\cal I}^{\epsilon}_{jh_mh_q}$ to the l.h.s of \eref{exp-EYM-KK-mh4}, we use the observation
$[{\cal I}^{\epsilon}_{jh_q(j+1)},{\cal I}^{\epsilon}_{jh_mh_q}]=0$ to get the physically meaningful result
\bea
& &{\cal I}^\epsilon_{jh_q(j+1)}\,{\cal I}^\epsilon_{jh_mh_q} \,{\bf I}^{\epsilon,\W\epsilon}_{\rm ssEYM}(+^g,\vec{\pmb a}^g_{n-m},-^g;{\pmb a}^g_{m})\nn
&=&{\cal I}^\epsilon_{jh_mh_q}\,{\cal I}^\epsilon_{jh_q(j+1)}\,{\bf I}^{\epsilon,\W\epsilon}_{\rm ssEYM}(0^g,\vec{\pmb a}^g_{n-m},(n-m+1)^g;{\pmb a}^h_{m})\nn
&=& {\bf I}^{\epsilon,\W\epsilon}_{\rm EYM}(0^g,\cdots,j^g,h^g_m,h^g_q,(j+1)^g,\cdots,(n-m+1)^g;{\pmb a}^h_m\setminus\{h^h_m,h^h_q\})\,.
\eea
While acting on the r.h.s, the operator ${\cal I}^{\epsilon}_{jh_q(j+1)}{\cal I}^{\epsilon}_{jh_mh_q}$ leads to
\bea
& &{\cal I}^\epsilon_{jh_q(j+1)}\Big(\sum_{i=j}^{n-m}\,
{\bf I}^{\epsilon,\W\epsilon}_{\rm ssEYM}(0^g,\cdots,i^g,h^g_m,(i+1)^g,\cdots,(n-m+1)^g;{\pmb a}^h_m\setminus h^h_m)-(\epsilon_{h_q}\cdot C_q)\Big)\nn
&=&\sum_{\shuffle}\, {\bf I}^{\epsilon,\W\epsilon}_{\rm ssEYM}(0^g,\cdots,j^g,h^g_q,h^g_m\shuffle\langle (j+1)^g,\cdots,(n-m)^g\rangle,(n-m+1)^g;{\pmb a}^h_m\setminus \{h^h_m,h^h_q\}) \nn
& &+ {\bf I}^{\epsilon,\W\epsilon}_{\rm ssEYM}(0^g,\cdots,j^g,h^g_m,h^g_q,(j+1)^g,\cdots,(n-m+1)^a;{\pmb a}^h_m\setminus \{h^h_m,h^h_q\})
 -{\cal I}^\epsilon_{jh_q(j+1)}(\epsilon_{h_q}\cdot C_q)\,,
\eea
where the splitting ${\cal I}^\epsilon_{jh_q(j+1)}={\cal I}^\epsilon_{jh_qh_m}+{\cal I}^\epsilon_{h_mh_q(j+1)}$ due to the definition
has been used when acting on ${\bf I}^{\epsilon,\W\epsilon}_{\rm ssEYM}(0^g,\cdots,j^g,h^g_m,(j+1)^g,\cdots,(n-m+1)^g;{\pmb a}^h_m\setminus h_m^h)$.
Comparing two sides gives the desired equation
\bea
{\cal I}^\epsilon_{jh_q(j+1)}(\epsilon_{h_q}\cdot C_q)=\sum_{\shuffle}\, {\bf I}^{\epsilon,\W\epsilon}_{\rm ssEYM}(0^g,\cdots,j^g,h^g_q,h^g_m\shuffle\{(j+1)^g,\cdots,(n-1)^g\},(n-m+1)^g;{\pmb a}^h_m\setminus \{h^h_m,h^h_q\})\,,
~~~~\label{Bqh}
\eea
which holds for arbitrary $j\in\{0,\cdots,n-m\}$.
The equation \eref{Bqh} bears strong similarity with the insertion relation for $(n-1)$-point partial integrands
\bea
& &{\cal I}^\epsilon_{jh_q(j+1)} {\bf I}^{\epsilon,\W\epsilon}_{\rm ssEYM}(0^g,\cdots,(n-m+1)^g;{\pmb a}^h_{m-1})\nn
&=&{\bf I}^{\epsilon,\W\epsilon}_{\rm ssEYM}(0^g,\cdots,j^g,h^g_q,(j+1)^g,\cdots,(n-m+1)^g;{\pmb a}^h_{m-1}\setminus h^h_q)\,,
\eea
therefore it is natural to expect the technic used in the previous part of this subsection can be applied to the current case.
The Lorentz invariant $(\epsilon_{h_q}\cdot C_q)$ contains polarization vectors $\epsilon_{h_p}$ with $p\neq q$,
thus can be divided as
\bea
\epsilon_{h_q}\cdot C_q&=&\sum_{b=0}^{n-m}\,(\epsilon_{h_q}\cdot k_b)\,B'_b
+\sum_{h_p\in{\pmb a}^h_m\setminus \{h_m,h_q\}}\,\Big((\epsilon_{h_q}\cdot k_{h_p})\,(\epsilon_{h_p}\cdot C'_p)
+(\epsilon_{h_q}\cdot \epsilon_{h_p})\,D'_p\Big)\,.~~~~\label{exp-ec}
\eea
It is worth to emphasize that the above expansion does not include the $\epsilon_{h_q}\cdot k_{h_m}$ term, due to the following reason.
Combining this term with the coefficient of $\epsilon_{h_q}\cdot C_q$ in \eref{exp-EYM-KK-mh4} gives the combination
$(\epsilon_{h_m}\cdot k_{h_q})(\epsilon_{h_q}\cdot k_{h_m})$, which is forbidden by the observation \eref{c-ani-A},
since the partial integrand ${\bf I}^{\epsilon,\W\epsilon}_{\rm ssEYM}(+^g,\vec{\pmb a}^g_{n-m},-^g;{\pmb a}^h_m)$ does not include
any $\epsilon_i\triangleright k_j$ or $\epsilon_i\triangleleft k_j$. The strict proof can be seen in
\eref{proof-c-anni} and the related discussions in the next subsection, for the more general case.
The coefficients $B'_b$ can be solved by using ${\cal I}^\epsilon_{jh_q(j+1)}$, which is found to be
\bea
\sum_{b=0}^{n-1}\,(\epsilon_{h_q}\cdot k_b)B'_b
=\sum_{\shuffle}\,(\epsilon_{h_q}\cdot Y_{h_q})\, {\bf I}^{\epsilon,\W\epsilon}_{\rm ssEYM}(+^g,\{h^g_q,h^g_m\}\shuffle\vec{\pmb a}^g_{n-m},-^g;{\pmb a}^h_m\setminus\{h^h_m,h^h_q\})\,.~~~~\label{exp-EYM-KK-1h5}
\eea
This serves as the analog of \eref{exp-EYM-KK-1h4}.
The gauge invariance condition ${\cal W}_{h_p} {\bf I}^{\epsilon,\W\epsilon}_{\rm ssEYM}(+^g,\vec{\pmb a}^g_{n-m},-^g;{\pmb a}^h_m)=0$ requires $(\epsilon_{h_q}\cdot C_q)$ to be gauge invariant for the leg $h_p$,
thus one can impose the gauge invariance to relate coefficients of $(\epsilon_{h_q}\cdot k_{h_p})$
and $(\epsilon_{h_q}\cdot \epsilon_{h_p})$ together as
\bea
(\epsilon_{h_q}\cdot k_{h_p})\,(\epsilon_{h_p}\cdot C'_p)
+(\epsilon_{h_q}\cdot \epsilon_{h_p})\,D'_p=\epsilon_{h_q}\cdot f_{h_p}\cdot C'_p\,.~~~~\label{BD-Bmh-2}
\eea
This equality serves as the analog of \eref{BD-Bmh}.

Substituting \eref{exp-EYM-KK-1h5} and \eref{BD-Bmh-2} into \eref{exp-EYM-KK-mh4} yields
\bea
{\bf I}^{\epsilon,\W\epsilon}_{\rm ssEYM}(+^g,\vec{\pmb a}^g_{n-m},-^g;{\pmb a}^h_{m})&=&\sum_{\shuffle}\,(\epsilon_{h_m}\cdot Y_{h_m})\, {\bf I}^{\epsilon,\W\epsilon}_{\rm ssEYM}(+^g,h^g_m\shuffle\vec{\pmb a}^g_{n-m},-^g;{\pmb a}^h_m\setminus h^h_m)\nn
& &+\sum_{g=1}^{m-1}\,\sum_{\shuffle}\,(\epsilon_{h_m}\cdot f_{h_g}\cdot Y_{h_g})\, {\bf I}^{\epsilon,\W\epsilon}_{\rm ssEYM}(+^g,\langle h^g_g,h^g_m\rangle\shuffle\vec{\pmb a}^g_{n-m},-^;{\pmb a}^h_m\setminus\{h^h_m,h^h_g\})\nn
& &+\sum_{g=1}^{m-1}\,\sum_{h_p\in{\pmb a}^h_m\setminus \{h_m,h_g\}}\,\epsilon_{h_m}\cdot f_{h_g}\cdot f_{h_p}\cdot C'_p\,.~~~~\label{exp-EYM-KK-mh5} \eea
Solving the coefficients $(C'_p)^\mu$ is completely analogous as solving $C_g^\mu$. Now the recursive pattern is manifested.
By iterating the above manipulation, the full expansion of ${\bf I}^{\epsilon,\W\epsilon}_{\rm ssEYM}(+^g,\vec{\pmb a}^g_{n-m},-^g;{\pmb a}^h_{m})$ can finally be obtained as
\bea
{\bf I}^{\epsilon,\W\epsilon}_{\rm ssEYM}(+^g,\vec{\pmb a}^g_{n-m},-^g;{\pmb a}^h_{m})&=&\sum_{\vec{\pmb s}:{\pmb s}\subseteq{\pmb a}^h_m\setminus h_m}\,\sum_{\shuffle}\,
K^\epsilon_{\vec{\pmb s}}\, {\bf I}^{\epsilon,\W\epsilon}_{\rm ssEYM}(+^g,\langle \vec{\pmb s}^g,h^g_m\rangle\shuffle\vec{\pmb a}^g_{n-m},-;{\pmb a}^h_m\setminus\{h^h_m,{\pmb s}^h\})\,,~~~~\label{exp-EYM-final} \eea
where the summation is over all ordered sets $\vec{\pmb s}$ with ${\pmb s}\subseteq{\pmb a}^h_m\setminus h_m$,
and $K^\epsilon_{\vec{\pmb s}}$ is defined as
\bea
K^\epsilon_{\vec{\pmb s}}=\epsilon_{h_m}\cdot f_{s_{|{\pmb s}|}}\cdots f_{s_1}\cdot Y_{s_1}\,,
\eea
for any $\vec{\pmb s}=\langle s_1,\cdots,s_{|{\pmb s}|} \rangle$. The set ${\pmb s}$ is allowed to be empty.

Using the expansion \eref{exp-EYM-final} recursively, one can expand any ssEYM partial integrand to YM partial integrands without any external graviton, the coefficients of these YM partial integrands will be discussed in subsection. \ref{subsec-coe}.

To derive the main result \eref{exp-EYM-final}, all seven principles/assumptions listed at the beginning of this section are used. The using
of first six are manifest, while the last one, is necessary when excluding the $\epsilon_{h_q}\cdot k_{h_m}$ term in \eref{exp-ec}, via the tree
level equality \eref{c-ani-A}.
In the expansion \eref{exp-EYM-final}, the gauge invariance for each graviton in the set ${\pmb a}^h_m\setminus h_m$ is manifest,
since the tensor $f_{h_i}^{\mu\nu}$ vanishes under the replacement $\epsilon_{h_i}\to k_{h_i}$. However, the gauge invariance for the graviton $h_m$ is hidden. Similar phenomenons happen at the tree level when solving the expansions of sEYM and GR amplitudes. This is a quite general feature if we start with the expansions in the form of \eref{exp-EYM-KK-mh2}. In the next subsection, we will use a different starting point to
derive the expansions of $1$-loop GR Feynman integrands in which the gauge invariance for each external graviton is manifest.

\subsection{Expansion of GR}
\label{exp-GR}

To solve the expansions of the GR Feynman integrands, the form \eref{exp-EYM-KK-mh2} is not the best choice for the starting point. First, this form is not convenient for applying our
operators ${\cal D}$ and ${\cal C}^\epsilon_{\vec{\pmb a}_m}$. Secondly, a general feature of using \eref{exp-EYM-KK-mh2} is that the gauge invariance for the external leg $h_m$ is not manifest. However, for the GR Feynman integrands without any color ordering, it is natural to expect a form without any special external leg. To realize the goal, we use the conclusion in \cite{Dong:2021qai} that if one impose the gauge invariance for all legs in ${\pmb a}^h_n$, then each term of the tree GR amplitude ${\cal A}^{\epsilon,\W\epsilon}_{\rm GR}({\pmb a}^h_n\cup\{+,-\})$ always contain a monomial
in the form
\bea
\omega^\epsilon(+^h,\vec{\pmb a}^h,-^h|{\rm signs})\equiv \epsilon_+\cdot v_{a_1}\,\Big(\prod_{i=2}^{|{\pmb a}|}\,\bar{v}_{a_{i-1}}\cdot v_{a_i}\Big)\,\bar{v}_{a_{|{\pmb a}|}}\cdot\epsilon_-\,.~~~~\label{monomial-chain}
\eea
Here ${\pmb a}$ is a subset of ${\pmb a}^h_n$ which is allowed to be empty, $v_{i}$ denotes either $k_i$ or $\epsilon_i$, with $\bar{v}_i$ the other one, i.e., $(v_i,\bar{v}_i)=(k_i,\epsilon_i)$ or $(\epsilon_i, k_i)$,
and the first/second choice is denoted by a $+$ or $-$ sign.
The proof of this formula only based on general considerations for tree amplitudes, which are first six of seven principles/assumptions
mentioned at the beginning of this section, as well as the counting of the number of mass dimensions.
We emphasize that the power counting can be made with out recourse to Feynman rules, CHY formulas, or other tools, since the relation
\bea
{\cal T}^\epsilon_{\vec{\pmb a}_n}\,{\cal A}^{\epsilon,\W\epsilon}_{{\rm GR}}({\pmb a}^h_n)&=&{\cal A}^{\W\epsilon}_{\rm YM}(\vec{\pmb a}^g_n)
\eea
is sufficient to determine
\bea
{\rm dim}\Big({\cal A}^{\epsilon,\W\epsilon}_{{\rm GR}}({\pmb a}^h_n)\Big)={\rm dim}\Big({\cal A}^{\W\epsilon}_{\rm YM}(\vec{\pmb a}^g_n)\Big)+(n-2)\,,
\eea
which indicates each BCJ numerator contains $n-2$ powers of momenta.
When taking the forward limit, the monomials $\omega^\epsilon(+^h,\vec{\pmb a}^h,-^h)$ behave as
\bea
\sum_r\,\epsilon^r_+\cdot\epsilon^r_-=d-2\,,~~~~~~~~{\rm if}~{\pmb a}=\emptyset\,,
\eea
and
\bea
\sum_r\,\epsilon^r_+\cdot v_{a_1}\,\Big(\prod_{i=2}^{|{\pmb a}|}\,\bar{v}_{a_{i-1}}\cdot v_{a_i}\Big)\,\bar{v}_{a_{|{\pmb a}|}}\cdot\epsilon^r_-
=\prod_{i=1}^{|{\pmb a}|}\,\bar{v}_{a_{i-1}}\cdot v_{a_i}\,.~~~~{\rm if}~{\pmb a}\neq\emptyset\,.
\eea
Thus, the $1$-loop GR Feynman integrand can be expressed as
\bea
{\bf I}^{\epsilon,\W\epsilon}_{\rm GR;\circ}({\pmb a}^h_n)&=&(d-2)\,{\bf I}^{\epsilon,\W\epsilon}_{\emptyset}+\sum_{\vec{\pmb a}/\pi_c}\,\Big(\prod_{i=1}^{|{\pmb a}|}\,\bar{v}_{a_{i-1}}\cdot v_{a_i}\Big)\,{\bf I}^{\epsilon,\W\epsilon}_{\vec{\pmb a}}({\rm sign})\nn
&=&(d-2)\,{\bf I}^{\epsilon,\W\epsilon}_{\emptyset}+\sum_{\vec{\pmb a}/\pi_c}\,C^\epsilon_{\vec{\pmb a}}\,{\bf I}^{\epsilon,\W\epsilon}_{\vec{\pmb a}}(-)+R\,.~~~~\label{exp-GR-1}
\eea
In the above formula, the summation is over all subsets ${\pmb a}\subseteq{\pmb a}^h_n$ satisfying ${\pmb a}\neq\emptyset$, and all un-cyclic permutations for each $\vec{{\pmb a}}$.
In the second line, we have collected together terms contain $C^\epsilon_{\vec{\pmb a}}$, and denoted the remaining terms by $R$. Notice that
the monomial $C^\epsilon_{\vec{\pmb a}}$ defined in \eref{defin-C2} stisfies the form $\prod_{i=1}^{|{\pmb a}|}\,\bar{v}_{a_{i-1}}\cdot v_{a_i}$, with the choice $(v_{i},\bar{v}_i)=(\epsilon_i,k_i)$ for each $i$, i.e., the sign is $-$ for each $i$. This is the reason why we use the sign $-$ in ${\bf I}^{\epsilon,\W\epsilon}_{\vec{\pmb a}}(-)$. The reason for organizing ${\bf I}^{\epsilon,\W\epsilon}_{\rm GR;\circ}({\pmb a}^h_n)$ as in the second line is
that the $(d-2)$ term is detectable for the operator ${\cal D}$, while $C^\epsilon_{\vec{\pmb a}}$ terms are detectable for the operators
${\cal C}^\epsilon_{\vec{\pmb a}_m}$. In the remainder of this subsection, we will show solving coefficients ${\bf I}^{\epsilon,\W\epsilon}_{\emptyset}$ and ${\bf I}^{\epsilon,\W\epsilon}_{\vec{\pmb a}}({\rm sign})$ naturally gives the expansion of $1$-loop GR Feynman integrand to ssEYM integrands.

The coefficients ${\bf I}^{\epsilon,\W\epsilon}_{\emptyset}$ and ${\bf I}^{\epsilon,\W\epsilon}_{\vec{\pmb a}}(-)$
can be determined via the transmutational relations \eref{loop-mp-main}, as well as
\bea
{\cal D}\,{\bf I}^{\epsilon,\W\epsilon}_{\rm GR;\circ}({\pmb a}^h_n)={\bf I}^{\epsilon,\W\epsilon}_{\rm ssEYM;\circ}({\pmb a}^h_n)\,,~~~~\label{D-GR}
\eea
which can be seen by setting ${\pmb a}_m=\emptyset$ in the second line of Table. \ref{tab:unifying-loop}. Here we have used the
observation ${\bf I}^{\epsilon,\W\epsilon}_{\rm ssEYM;\circ}({\pmb a}^h_n)={\bf I}^{\epsilon,\W\epsilon}_{\rm ssEYM}(+^g,-^g;{\pmb a}^h_n)$,
since no summation over cyclic permutations is required when ${\pmb a}_m=\emptyset$.
Applying \eref{D-GR} fixes ${\bf I}^{\epsilon,\W\epsilon}_{\emptyset}$
to be\footnote{Here we used the observation that the parameter $d$ only arises from $\sum_r\epsilon^r_+\cdot \epsilon^r_-$, since there is no other Lorentz invariant depends on the number of dimensions of space-time explicitly.}
\bea
{\bf I}^{\epsilon,\W\epsilon}_{\emptyset}={\bf I}^{\epsilon,\W\epsilon}_{\rm ssEYM;\circ}({\pmb a}^h_n)\,.
\eea
The transmutational relation \eref{loop-mp-main} indicates that
\bea
C^\epsilon_{\vec{\pmb a}}\,\Big[{\bf I}^{\epsilon,\W\epsilon}_{{\rm ssEYM};\circ}(\vec{{\pmb a}}^g;{\pmb a}^h_n\setminus{\pmb a}^h)+(-)^{|{\pmb a}|}{\bf I}^{\epsilon,\W\epsilon}_{{\rm ssEYM};\circ}(\overleftarrow{\pmb a}^g;{\pmb a}^h_n\setminus{\pmb a}^h)\Big]\,\in\,{\bf I}^{\epsilon,\W\epsilon}_{\rm GR;\circ}({\pmb a}^h_n)\,,
\eea
this motivate us to naively identify ${\bf I}^{\epsilon,\W\epsilon}_{\vec{\pmb a}}(-)$ as
\bea
{\bf I}^{\epsilon,\W\epsilon}_{\vec{\pmb a}}(-)={\bf I}^{\epsilon,\W\epsilon}_{{\rm ssEYM};\circ}(\vec{{\pmb a}}^g;{\pmb a}^h_n\setminus{\pmb a}^h)+(-)^{|{\pmb a}|}{\bf I}^{\epsilon,\W\epsilon}_{{\rm ssEYM};\circ}(\overleftarrow{\pmb a}^g;{\pmb a}^h_n\setminus{\pmb a}^h)\,.
\eea
and write down
\bea
{\bf I}^{\epsilon,\W\epsilon}_{\rm GR;\circ}({\pmb a}^h_n)=(d-2){\bf I}^{\epsilon,\W\epsilon}_{\rm ssEYM;\circ}({\pmb a}^h_n)+\sum_{\vec{\pmb a}/\pi_c}\,C^\epsilon_{\vec{\pmb a}}\Big[{\bf I}^{\epsilon,\W\epsilon}_{{\rm ssEYM};\circ}(\vec{{\pmb a}}^g;{\pmb a}^h_n\setminus{\pmb a}^h)+(-)^{|{\pmb a}|}{\bf I}^{\epsilon,\W\epsilon}_{{\rm ssEYM};\circ}(\overleftarrow{\pmb a}^g;{\pmb a}^h_n\setminus{\pmb a}^h)\Big]+R\,.~~~~\label{exp-GR-2}
\eea
Quite surprisingly, the above formula is the correct solution to the equations \eref{D-GR} and \eref{loop-mp-main} if ${\cal D}R=0$ and ${\cal C}^\epsilon_{\vec{\pmb a}_m}R=0$. The verification is based on the following three equalities,
\bea
{\cal D}\,\Big[C^\epsilon_{\vec{\pmb a}}\,[{\bf I}^{\epsilon,\W\epsilon}_{{\rm ssEYM};\circ}(\vec{{\pmb a}}^g;{\pmb a}^h_n\setminus{\pmb a}^h)+(-)^{|{\pmb a}|}{\bf I}^{\epsilon,\W\epsilon}_{{\rm ssEYM};\circ}(\overleftarrow{\pmb a}^g;{\pmb a}^h_n\setminus{\pmb a}^h)]\Big]=0\,.~~~~\label{equ-3}
\eea
\bea
{\cal C}^\epsilon_{\vec{\pmb a}_m}\,\Big[(d-2){\bf I}^{\epsilon,\W\epsilon}_{\rm ssEYM;\circ}({\pmb a}^h_n)\Big]=0\,,~~~~\label{equ-1}
\eea
and
\bea
& &{\cal C}^\epsilon_{\vec{\pmb s}_m}\,\Big[C^\epsilon_{\vec{\pmb a}}\,[{\bf I}^{\epsilon,\W\epsilon}_{{\rm ssEYM};\circ}(\vec{{\pmb a}}^g;{\pmb a}^h_n\setminus{\pmb a}^h)+(-)^{|{\pmb a}|}{\bf I}^{\epsilon,\W\epsilon}_{{\rm ssEYM};\circ}(\overleftarrow{\pmb a}^g;{\pmb a}^h_n\setminus{\pmb a}^h)]\Big]\nn
&=&\delta^\pi_{\vec{\pmb s}_m,\vec{\pmb a}}[{\bf I}^{\epsilon,\W\epsilon}_{{\rm ssEYM};\circ}(\vec{{\pmb a}}^g;{\pmb s}^h_n\setminus{\pmb a}^h)+(-)^{|{\pmb a}|}{\bf I}^{\epsilon,\W\epsilon}_{{\rm ssEYM};\circ}(\overleftarrow{\pmb a}^g;{\pmb s}^h_n\setminus{\pmb a}^h)]\,.~~~~\label{equ-2}
\eea
Here $\delta^\pi_{\vec{\pmb s}_m,\vec{\pmb a}}$ is understood as $1$ when $\vec{\pmb s}_m=\vec{\pmb a}$ up to a cyclic permutation, and $0$ otherwise.
The equality \eref{equ-3} holds obviously, and ensures that ${\bf I}^{\epsilon,\W\epsilon}_{\rm GR;\circ}({\pmb a}^h_n)$ in \eref{exp-GR-2} satisfies equation \eref{D-GR}
if ${\cal D}R=0$.
To see \eref{equ-1} and \eref{equ-2}, we first show that
\bea
{\cal C}^\epsilon_{\vec{\pmb s}_m}\,{\bf I}^{\epsilon,\W\epsilon}_{{\rm ssEYM};\circ}(\vec{{\pmb a}}^g;{\pmb a}^h_n\setminus{\pmb a}^h)=0\,,~~~~\label{c-anni-I}
\eea
for arbitrary $\vec{{\pmb a}}$ (including the special case ${\pmb a}=\emptyset$). If ${\pmb s}_m\cap{\pmb a}\neq\emptyset$, for each $i\in{\pmb s}_m\cap{\pmb a}$,
the corresponding polarization vector $\epsilon_i$ does not appear in ${\bf I}^{\epsilon,\W\epsilon}_{{\rm ssEYM};\circ}(\vec{{\pmb a}}^g;{\pmb a}^h_n\setminus{\pmb a}^h)$, thus ${\bf I}^{\epsilon,\W\epsilon}_{{\rm ssEYM};\circ}(\vec{{\pmb a}}^g;{\pmb a}^h_n\setminus{\pmb a}^h)$ is annihilated by
$\partial_{\epsilon_i\cdot k_j}$ in ${\cal C}^\epsilon_{\vec{\pmb s}_m}$. If ${\pmb s}_m\cap{\pmb a}=\emptyset$, it means ${\pmb s}_m\subseteq{\pmb a}_n\setminus{\pmb a}$.
For this case, one effective way to see \eref{c-anni-I} is to use
\bea
{\bf I}^{\epsilon,\W\epsilon}_{{\rm ssEYM};\circ}(\vec{{\pmb a}}^g;{\pmb a}^h_n\setminus{\pmb a}^h)&=&\sum_{\pi_c}\,\Big[{1\over\ell^2}\,{\cal F}\,{\cal A}^{\epsilon,\W\epsilon}_{{\rm sEYM};\circ}(+^g,\pi_c(\vec{{\pmb a}}^g),-^g;{\pmb a}^h_n\setminus{\pmb a}^h)\Big]\nn
&=&{1\over\ell^2}\,{\cal F}\,\sum_{\pi_c}\,{\cal T}^\epsilon_{+\vec{\pmb a}-}\,{\cal A}^{\epsilon,\W\epsilon}_{{\rm GR};\circ}({\pmb a}^h_n\cup\{+^h,-^h\})\,.~~~~\label{proof-c-anni}
\eea
Since the operator ${\cal T}^\epsilon_{+\vec{\pmb a}-}$ removes the polarization vectors $\epsilon_+$ and $\epsilon_-$ in ${\cal A}^{\epsilon,\W\epsilon}_{{\rm GR};\circ}({\pmb a}^h_n\cup\{+^h,-^h\})$, the manipulation ${\cal F}$ will not create any $\epsilon_i\triangleright k_j$
or $\epsilon_i\triangleleft k_j$, thus ${\cal C}^\epsilon_{\vec{\pmb s}_m}$ is commutable with ${\cal F}$.
Since ${\pmb s}_m\cap{\pmb a}=\emptyset$, ${\cal C}^\epsilon_{\vec{\pmb s}_m}$ is commutable with ${\cal T}^\epsilon_{+\vec{\pmb a}-}$. Then, the equality \eref{c-anni-I} is ensured by \eref{c-ani-A}. On the other hand, it is straightforward to see
\bea
{\cal C}^\epsilon_{\vec{\pmb s}_m}\,C^\epsilon_{\vec{\pmb a}}=\delta^\pi_{\vec{\pmb s}_m,\vec{\pmb a}}\,.~~~~\label{c-c}
\eea
Combining \eref{c-anni-I} and \eref{c-c} together gives \eref{equ-1} and \eref{equ-2}.
Using equalities  \eref{equ-3}, \eref{equ-1} and \eref{equ-2}, one see that
the formula \eref{exp-GR-2} is the solution to equations \eref{D-GR} and \eref{loop-mp-main}, if ${\cal D}R=0$ and ${\cal C}^\epsilon_{\vec{\pmb a}_m}R=0$ for each $\vec{\pmb a}_m$.

The remaining part $R$ serves as the un-fixed constant when solving the differential equations. It can be determined via the gauge invariance requirement. Here we employ the method similar as that in the previous subsection, with a slight modification. If an object $P$ is gauge invariant, i.e., ${\cal W}^\epsilon_iP=0$, then we obviously have
\bea
\partial_{\epsilon_q\cdot k_i}\,{\cal W}^\epsilon_i\,P=0,~~~~~~~~{\rm for}~\forall\,q\,,
\eea
which indicates
\bea
0&=&(\partial_{\epsilon_q\cdot k_i}\,{\cal W}^\epsilon_i)\,P+{\cal W}^\epsilon_i\,(\partial_{\epsilon_q\cdot k_i}\,P)\nn
&=&
\partial_{\epsilon_q\cdot\epsilon_i}\,P+{\cal W}^\epsilon_i\,(\partial_{\epsilon_q\cdot k_i}\,P)\,.
\eea
If we restrict our attentions on amplitudes and Feynman integrands, we can require $P$ to be linear in each polarization vector. With this condition,
one can immediately conclude that
\bea
-(\epsilon_q\cdot\epsilon_i)\,({\cal W}^\epsilon_i\,(\partial_{\epsilon_q\cdot k_i}\,P))\,\in\,P\,.~~~~\label{gauge-condition}
\eea

Now we apply the gauge invariance condition \eref{gauge-condition} to terms in \eref{exp-GR-2}.
Manifestly,
\bea
& &{\cal W}^\epsilon_i\,\Big[(d-2){\bf I}^{\epsilon,\W\epsilon}_{\rm ssEYM;\circ}({\pmb a}^h_n)\Big]=0\,,\nn
& &{\cal W}^\epsilon_i\,\Big[C^\epsilon_{\vec{\pmb a}}\,[{\bf I}^{\epsilon,\W\epsilon}_{{\rm ssEYM};\circ}(\vec{{\pmb a}}^g;{\pmb a}^h_n\setminus{\pmb a}^h)+(-)^{|{\pmb a}|}{\bf I}^{\epsilon,\W\epsilon}_{{\rm ssEYM};\circ}(\overleftarrow{\pmb a}^g;{\pmb a}^h_n\setminus{\pmb a}^h)]\Big]=0\,,~~~~{\rm if}~i\in{\pmb a}_n\setminus{\pmb a}\,,
\eea
due to the gauge invariance of the Feynman integrands ${\bf I}^{\epsilon,\W\epsilon}_{\rm ssEYM;\circ}({\pmb a}^h_n)$, ${\bf I}^{\epsilon,\W\epsilon}_{{\rm ssEYM};\circ}(\vec{{\pmb a}}^g;{\pmb a}^h_n\setminus{\pmb a}^h)$ and ${\bf I}^{\epsilon,\W\epsilon}_{{\rm ssEYM};\circ}(\overleftarrow{\pmb a}^g;{\pmb a}^h_n\setminus{\pmb a}^h)$. Thus, for the above parts, the condition \eref{gauge-condition} is satisfied by themselves.
Now we move to the case $i\in{\pmb a}$. Based on the definition of the coefficients $C^\epsilon_{\vec{\pmb a}}$ in \eref{defin-C2},
we observe that the part
\bea
\sum_{\substack{\vec{\pmb a}/\pi_c\\i\in{\pmb a}}}\,C^\epsilon_{\vec{\pmb a}}\,[{\bf I}^{\epsilon,\W\epsilon}_{{\rm EYM};\circ}(\vec{{\pmb a}};{\pmb H}_n\setminus{\pmb a})+(-)^{|{\pmb a}|}{\bf I}^{\epsilon,\W\epsilon}_{{\rm EYM};\circ}(\vec{{\pmb a}}^T;{\pmb H}_n\setminus{\pmb a})]
\eea
can be organized as
\bea
\sum_{\substack{\vec{\pmb a}/\pi_c\\i\in{\pmb a}}}\,C^\epsilon_{\vec{\pmb a}}\,[{\bf I}^{\epsilon,\W\epsilon}_{{\rm ssEYM};\circ}(\vec{{\pmb a}}^g;{\pmb a}^h_n\setminus{\pmb a}^h)+(-)^{|{\pmb a}|}{\bf I}^{\epsilon,\W\epsilon}_{{\rm ssEYM};\circ}(\overleftarrow{\pmb a}^g;{\pmb a}^h_n\setminus{\pmb a}^h)]=\sum_{j\neq i}\,(\epsilon_j\cdot k_i)\,(\epsilon_i\cdot B_{ij})\,.~~~~\label{organi-B}
\eea
We do not provide the explicit formulas of the vectors $B_{ij}^\mu$ here since they are irrelevant. The key point is,
\bea
{\cal W}^\epsilon_i\,\partial_{\epsilon_q\cdot k_i}\,\Big(\sum_{j\neq i}\,(\epsilon_j\cdot k_i)\,(\epsilon_i\cdot B_{ij})\Big)=k_i\cdot B_{iq}\,,
\eea
thus from \eref{gauge-condition} we know that
\bea
-(\epsilon_q\cdot \epsilon_i)\,(k_i\cdot B_{iq})\,\in\,{\bf I}^{\epsilon,\W\epsilon}_{\rm GR;\circ}({\pmb a}^h_n)\,.
\eea
In the formula \eref{exp-GR-2}, the monomial $-(\epsilon_q\cdot \epsilon_i)\,(k_i\cdot B_{iq})$ can only belong to the unknown $R$, since it is not contained in any already known part. Thus we have detect a piece of
the unknown part $R$.

To determine the full $R$, we first combine $(\epsilon_q\cdot k_i)\,(\epsilon_i\cdot B_{iq})$ and $-(\epsilon_q\cdot \epsilon_i)\,(k_i\cdot B_{iq})$
together to arrive at $(\epsilon_q\cdot f_i\cdot B_{iq})$. Since $q$ is chosen arbitrary, to preserve the gauge invariance of the leg $i$, we should replacing the tensor $k^\mu_i\epsilon^\nu_i$
by $f^{\mu\nu}_i$ at the r.h.s of \eref{organi-B}. The leg $i$ is also chosen arbitrary, and each $C^\epsilon_{\vec{\pmb a}}$ is invariant under the cyclic permutations for $\vec{{\pmb a}}$. Such symmetry requires us to replace all $k^\mu_{a_i}\epsilon^\nu_{a_i}$ in the cyclical factor $C^\epsilon_{\vec{\pmb a}}$
by $f^{\mu\nu}_{a_i}$. Thus we find the replacement
\bea
C^\epsilon_{\vec{\pmb a}}\,\to\,{\rm Tr}^\epsilon_{\overleftarrow{\pmb a}}={\rm Tr}(f_{a_{|{\pmb a}|}}\cdots f_{a_1})\,.~~~~\label{c-tr}
\eea
This replacement add various new terms to \eref{exp-GR-2}, and all these new terms vanish under the action of ${\cal D}$ and ${\cal C}^\epsilon_{\vec{\pmb s}_m}$. This observation supports our assumptions ${\cal D}R=0$ and ${\cal C}^\epsilon_{\vec{\pmb s}_m}R=0$.
After doing the replacement, the gauge invariance of arbitrary external graviton is manifest.

However, when doing the replacement  in \eref{c-tr}, an over counting arises. The term
\bea
{\rm Tr}^\epsilon_{\overleftarrow{\pmb a}}\,[{\bf I}^{\epsilon,\W\epsilon}_{{\rm ssEYM};\circ}(\vec{{\pmb a}}^g;{\pmb a}^h_n\setminus{\pmb a}^h)+(-)^{|{\pmb a}|}{\bf I}^{\epsilon,\W\epsilon}_{{\rm ssEYM};\circ}(\overleftarrow{\pmb a}^g;{\pmb a}^h_n\setminus{\pmb a}^h)]~~~~\label{tr1}
\eea
contains not only
\bea
C^\epsilon_{\vec{\pmb a}}\,[{\bf I}^{\epsilon,\W\epsilon}_{{\rm ssEYM};\circ}(\vec{{\pmb a}}^g;{\pmb a}^h_n\setminus{\pmb a}^h)+(-)^{|{\pmb a}|}{\bf I}^{\epsilon,\W\epsilon}_{{\rm ssEYM};\circ}(\overleftarrow{\pmb a}^g;{\pmb a}^h_n\setminus{\pmb a}^h)]\,,
\eea
but also
\bea
C^\epsilon_{\overleftarrow{\pmb a}}\,[{\bf I}^{\epsilon,\W\epsilon}_{{\rm ssEYM};\circ}(\overleftarrow{\pmb a}^g;{\pmb a}^h_n\setminus{\pmb a}^h)+(-)^{|{\pmb a}|}{\bf I}^{\epsilon,\W\epsilon}_{{\rm ssEYM};\circ}(\vec{{\pmb a}}^g;{\pmb a}^h_n\setminus{\pmb a}^h)]\,,
\eea
and summing over $\vec{{\pmb a}}$ counts both $\vec{{\pmb a}}$ and $\overleftarrow{\pmb a}$. To handle this, we
recognize the second term in \eref{tr1}
as the first term in
\bea
{\rm Tr}^\epsilon_{\vec{\pmb a}}\,[{\bf I}^{\epsilon,\W\epsilon}_{{\rm ssEYM};\circ}(\overleftarrow{\pmb a}^g;{\pmb a}^h_n\setminus{\pmb a}^h)+(-)^{|{\pmb a}|}{\bf I}^{\epsilon,\W\epsilon}_{{\rm ssEYM};\circ}(\vec{{\pmb a}}^g;{\pmb a}^h_n\setminus{\pmb a}^h)]\,,~~~~\label{tr2}
\eea
and the first term in \eref{tr1} as the second term in \eref{tr2}, since ${\rm Tr}^\epsilon_{\vec{\pmb a}}=(-)^{|{\pmb a}|}{\rm Tr}^\epsilon_{\overleftarrow{\pmb a}}$, due to the anti-symmetry of the tensors $f^{\mu\nu}_i$. Thus one can remove the over counting and get the expansion
\bea
{\bf I}^{\epsilon,\W\epsilon}_{\rm GR;\circ}({\pmb a}^h_n)=(d-2){\bf I}^{\epsilon,\W\epsilon}_{\rm ssEYM;\circ}({\pmb a}^h_n)+\sum_{\vec{\pmb a}/\pi_c}\,{\rm Tr}^\epsilon_{\overleftarrow{\pmb a}}\,{\bf I}^{\epsilon,\W\epsilon}_{{\rm ssEYM};\circ}(\vec{{\pmb a}}^g;{\pmb a}^h_n\setminus{\pmb a}^h)\,.~~~~\label{exp-GR-final}
\eea

In the expansion \eref{exp-GR-final}, all coefficients ${\bf I}^{\epsilon,\W\epsilon}_\emptyset$ and ${\bf I}^{\epsilon,\W\epsilon}_{\vec{\pmb a}}({\rm sign})$ in the first line of \eref{exp-GR-1} are fixed, thus \eref{exp-GR-final} is indeed the correct expanded formula for ${\bf I}^{\epsilon,\W\epsilon}_{\rm GR;\circ}({\pmb a}^h_n)$, which is coincide
with the result found in \cite{Geyer:2017ela}. The coefficients ${\rm Tr}^\epsilon_{\overleftarrow{\pmb a}}$ vanish when the length of ${\pmb a}$ is $1$, this feature supports the observation in subsection. \ref{construction of the operator} that the operator ${\cal C}^\epsilon_{\vec{\pmb a}_m}$ does not make sense when $m=1$. Notice that without the general formula \eref{exp-GR-1} obtained via the forward limit procedure, one can not conclude
the solution \eref{exp-GR-final} has detect all terms of the full GR Feynman integrand. For example, suppose we turn the factor $d-2$ in \eref{exp-GR-final} to $d$, or add the tree amplitude ${\cal A}^{\epsilon,\W\epsilon}_{\rm GR}({\pmb a}^h_n\cup\{+^h,-^h\})$ to the r.h.s of \eref{exp-GR-final}, the obtained results are still solutions to equations \eref{D-GR} and \eref{loop-mp-main}, with the Lorentz and gauge invariance. Such modifications are excluded
by the general formula \eref{exp-GR-1}.

The expansion \eref{exp-GR-final} is equivalent to
\bea
{\bf I}^{\epsilon,\W\epsilon}_{\rm GR;\circ}({\pmb a}^h_n)=(d-2){\bf I}^{\epsilon,\W\epsilon}_{\rm ssEYM}(+^g,-^g;{\pmb a}^h_n)+\sum_{\vec{\pmb a}}\,{\rm Tr}^\epsilon_{\overleftarrow{\pmb a}}\,{\bf I}^{\epsilon,\W\epsilon}_{{\rm ssEYM}}(+^g,\vec{{\pmb a}}^g,-^g;{\pmb a}^h_n\setminus{\pmb a}^h)\,,~~~~\label{exp-GR-final2}
\eea
which expands ${\bf I}^{\epsilon,\W\epsilon}_{\rm GR;\circ}({\pmb a}^h_n)$ to ssEYM partial Feynman integrands rather than full ones.
To see the equivalence, let us consider two parts at the r.h.s of \eref{exp-GR-final} in turn.
For the first part, we have used the observation ${\bf I}^{\epsilon,\W\epsilon}_{\rm ssEYM}(+^g,-^g;{\pmb a}^h_n)={\bf I}^{\epsilon,\W\epsilon}_{\rm ssEYM;\circ}({\pmb a}^h_n)$, as discussed below equation \eref{D-GR}. For the second part at the r.h.s, we use the relation among full ssEYM Feynman integrands and partial Feynman integrands to get
\bea
\sum_{\vec{\pmb a}/\pi_c}\,{\rm Tr}^\epsilon_{\overleftarrow{\pmb a}}\,{\bf I}^{\epsilon,\W\epsilon}_{{\rm ssEYM};\circ}(\vec{{\pmb a}}^g;{\pmb a}^h_n\setminus{\pmb a}^h)&=&\sum_{\vec{\pmb a}/\pi_c}\,{\rm Tr}^\epsilon_{\overleftarrow{\pmb a}}\,\Big[\sum_{\pi_c}{\bf I}^{\epsilon,\W\epsilon}_{{\rm ssEYM}}(+^g,\pi_c(\vec{{\pmb a}}^g),-^g;{\pmb a}^h_n\setminus{\pmb a}^h)\Big]\nn
&=&\sum_{\vec{\pmb a}}\,{\rm Tr}^\epsilon_{\overleftarrow{\pmb a}}\,{\bf I}^{\epsilon,\W\epsilon}_{{\rm ssEYM}}(+^g,\vec{{\pmb a}}^g,-^g;{\pmb a}^h_n\setminus{\pmb a}^h)\,,
\eea
where in the second equality we have used the observation ${\rm Tr}^\epsilon_{\overleftarrow{\pmb a}}$ is independent of the cyclic permutations $\pi_c$.
Combining two parts together gives the expansion \eref{exp-GR-final2}.
It is more convenient to use \eref{exp-GR-final2} to expand the GR Feynman integrands to YM partial Feynman integrands, as will be seen in the next subsection.

In the expansions \eref{exp-GR-final} and \eref{exp-GR-final2}, all coefficients of partial ssEYM integrands are independent of the loop momentum $\ell$, thus will not be altered by the integration over loop momentum. Consequently, these expansions also
hold at the $1$-loop amplitudes level.

\subsection{Coefficients of YM partial Feynman integrands}
\label{subsec-coe}

From expansions \eref{exp-EYM-final} and \eref{exp-GR-final2}, it is straightforward to observe that the $1$-loop ssEYM partial Feynman integrands and
GR Feynman integrands can be expanded to $1$-loop KK basis of YM partial integrands by applying \eref{exp-EYM-final} recursively, formally expressed
as
\bea
{\bf I}^{\epsilon,\W\epsilon}_{\rm GR;\circ}({\pmb a}^h_n)=\sum_\sigma\,{\bf C}^\epsilon_1(\sigma)\,{\bf I}^{\W\epsilon}_{\rm YM}(+^g,\sigma(\vec{\pmb a}^g_n),-^g)\,,~~~\label{exp-GR-KK}
\eea
and
\bea
{\bf I}^{\epsilon,\W\epsilon}_{\rm ssEYM}(+,\vec{\pmb a}^g_m,-;{\pmb a}^h_{n-m})=\sum_\sigma\,{\bf C}^\epsilon_2(\sigma,\vec{\pmb a}_m)\,{\bf I}^{\W\epsilon}_{\rm YM}(+^g,\sigma(\vec{\pmb s}^g_n),-^g)\,,~~~~\label{exp-EYM-kk}
\eea
where $\sigma$ stands for the permutations.
Coefficients ${\bf C}^\epsilon_1(\sigma)$ and ${\bf C}^\epsilon_2(\sigma,\vec{\pmb a}^g_m)$ depend on configurations of external legs under consideration, as well as particular color-orderings of the YM partial integrands. It is hard to find the general expressions for ${\bf C}^\epsilon_1(\sigma)$ and ${\bf C}^\epsilon_2(\sigma,\vec{\pmb a}^g_m)$, which are correct for all cases. Instead, the systematic algorithms for evaluating them, which can be applied to any configuration of external particles and any color-ordering, can be provided, as will be shown in this subsection.

We first consider the coefficients $C^{\W\epsilon}_1(\sigma)$, which serve as the $1$-loop level BCJ numerators.
To expand ${\bf I}^{\epsilon,\W\epsilon}_{\rm GR;\circ}({\pmb a}^h_n)$ to ${\bf I}^{\W\epsilon}_{\rm YM}(+^g,\sigma(\vec{\pmb a}^g_n),-^g)$, the expansion \eref{exp-GR-final2} require us to decompose the set of external legs ${\pmb a}_n$ into subsects ${\pmb a}$ and ${\pmb a}_n\setminus{\pmb a}$, while applying the expansion \eref{exp-EYM-final} recursively indicates further decompositions of ${\pmb a}_n\setminus{\pmb a}$.
The successive decompositions lead to the concept which is called ordered splitting, defined for each fixed color ordering $\sigma(\vec{\pmb a}_n)=\langle\sigma_1,\cdots,\sigma_n\rangle$ \cite{Fu:2017uzt}. To illustrate it, some notations need to be introduced. We denote the color-ordering
as $+\dot{<} \sigma_1\dot{<}\cdots\dot{<} \sigma_n\dot{<}-$. One also need to chose a reference ordering $-\prec j_1\prec\cdots\prec j_n$,
with $-$ fixed at the lowest position. This reference ordering is denoted by $\pmb{\cal R}$.
The correct ordered splittings consist with
the desired color-ordering, are constructed through the following procedure:
\begin{itemize}
\item At the first step, we construct all possible ordered subsets $\vec{\pmb{a}}^0=\langle a^0_1,\cdots,a^0_{|0|}\rangle$, which satisfy two conditions: (1) ${\pmb a}^0\subseteq\pmb{a}_n$; (2) $a^0_1\dot{<}a^0_2\dot{<}\cdots\dot{<}a^0_{|0|}$, respecting to the color-ordering of the YM amplitude.
    We call each ordered subset $\vec{\pmb{a}}^0$ a root \footnote{Here we borrow the language from the framework of increasing spanning trees.}.
    Here $|i|$ denotes the length of the set ${\pmb a}^i$.
\item For each root $\vec{\pmb{a}}^0$, we eliminate its elements in $\pmb{a}_n$ and $\pmb{\cal R}$, resulting in a reduced set $\pmb{a}_n\setminus{\pmb a}^0$, and a reduce reference ordering $\pmb{\cal R}\setminus{\pmb a}^0$. Suppose $R_1$ is the lowest element in the reduce reference ordering $\pmb{\cal R}\setminus{\pmb a}^0$, we construct all possible ordered subsets $\vec{\pmb a}^1$ as $\vec{\pmb{a}}^1=\langle a_1^1,a_2^1,\cdots,a_{|1|-1}^1,R_1\rangle$, with $a_1^1\dot{<}a_2^1\dot{<}\cdots\dot{<}a_{|1|-1}^1\dot{<}R_1$, regarding to the color-ordering.
\item By iterating the second step, one can construct $\vec{\pmb{a}}^2,\vec{\pmb{a}}^3,\cdots$, until ${\pmb a}^0\cup{\pmb a}^1\cup\cdots\cup{\pmb a}^r={\pmb a}_n$.
\end{itemize}
Each ordered splitting is given as an ordered set $\vec{{\bf S}}=\langle\vec{\pmb a}^0,\vec{\pmb a}^1,\cdots,\vec{\pmb a}^r\rangle$, where ordered sets
$\vec{\pmb a}^i$ serve as elements. For a given root $\vec{\pmb a}^0$, an ordered set $\vec{{\bf B}}=\langle\vec{\pmb a}^1,\vec{\pmb a}^2,\cdots,\vec{\pmb a}^r\rangle$ is called a branch. Notice that ${\pmb a}^0$ can be empty, while each ${\pmb a}^i$ with $i\neq 0$ contains at least one element $R_i$.

Now we give the corresponding kinematic factors
for each ordered set $\vec{\pmb a}^i$, by using \eref{exp-EYM-final} and \eref{exp-GR-final2}. For a given ordered splitting, the root $\vec{\pmb a}^0$ carries the factor
\bea
T^\epsilon_{\vec{\pmb a}^0}=\left\{ \begin{array}{ll} {\rm Tr}^\epsilon_{\overleftarrow{\pmb a^0}}={\rm Tr}(f_{a^0_{|0|}},\cdots,f_{a^0_1})\,, & ~~~~{\rm if}~{\pmb a}^0\neq\emptyset\,, \\ d-2\,,& ~~~~{\rm
if}~{\pmb a}^0=\emptyset\,. \end{array}\right.~~~~~\label{defin-T}
\eea
Other ordered sets $\vec{\pmb a}^i$ with $i\neq 0$ carry
\bea
K^\epsilon_{\vec{\pmb a}^i}=\epsilon_{R_i}\cdot f_{a^i_{|i|-1}}\cdots f_{a^i_2}\cdot f_{a^i_1}\cdot Z_{a^i_1}\,.~~~~\label{defin-K}
\eea
The combinatory momentum $Z_{a^i_1}$ is the sum of momenta of external legs satisfying two conditions: (1) legs
at the l.h.s of $a^i_1$ in the color-ordering, (2) legs belong to $\vec{\pmb a}^j$ at the l.h.s of $\vec{\pmb a}^i$
in the ordered splitting, i.e., $j<i$. The coefficient of the YM partial Feynman integrand ${\bf I}^{\W\epsilon}_{\rm YM}(+^g,\sigma(\vec{a}^g_n),-^g)$
is the sum of contributions from all proper ordered splittings.

For the ssEYM partial Feynman integrand ${\bf I}^{\epsilon,\W\epsilon}_{\rm ssEYM}(+^g,\vec{\pmb a}^g_m,-^g;{\pmb a}^h_{n-m})$,
the coefficient of ${\bf I}^{\W\epsilon}_{\rm YM}(+^g,\sigma(\vec{\pmb s}^g_n),-^g)$ is obviously the sum of contributions from all branches for the root
$\vec{\pmb a}^0=\vec{\pmb a}_m$.

Before ending this subsection, we point out the differential operators transmute the $1$-loop GR Feynman integrand to $1$-loop YM partial integrands are the same as the operators which transmute the YM partial integrand ${\bf I}^{\epsilon}_{\rm YM}(+^g,\vec{{\pmb a}}^g_n,-^g)$ to BAS double-partial integrands ${\bf I}_{\rm BAS}(+^s,\sigma(\vec{\pmb s}^s_n),-^s\parallel+^s,\vec{{\pmb a}}^s_n,-^s)$, as can be seen in Table. \ref{tab:unifying-loop}. Furthermore, all seven principles/assumptions listed at the beginning of this section hold for the YM and BAS case. The third assumption makes sense in the following way: this assumption together with the operator ${\cal T}^{\W\epsilon}_{+,\vec{\pmb a}_n,-}$ completely determine that each external gluon $i$ carries the polarization vector $\epsilon_i$, and the YM partial integrands carry the color ordering $+,\vec{\pmb a}_n,-$, as discussed at the beginning of this section. Such character of YM partial integrands plays the role of the original third assumption.  Thus we conclude
\bea
{\bf I}^{\epsilon}_{\rm YM}(+^g,\vec{{\pmb a}}^g_n,-^g)=\sum_\sigma\,{\bf C}^\epsilon_1(\sigma)\,{\bf I}_{\rm BAS}(+^s,\sigma(\vec{\pmb s}^s_n),-^s\parallel+^s,\vec{{\pmb a}}^s_n,-^s)\,.~~~\label{exp-YM-KK}
\eea
Similar argument for ssYMS partial Feynman integrands yields
\bea
{\bf I}^{\epsilon}_{\rm ssYMS}(+^s,\vec{\pmb a}^s_m,-^s;{\pmb a}^g_{n-m}\parallel+^A,\vec{{\pmb a}}^A_n,-^A)=\sum_\sigma\,{\bf C}^\epsilon_2(\sigma,\vec{\pmb a}_m)\,{\bf I}_{\rm BAS}(+^s,\sigma(\vec{\pmb s}^s_n),-^s\parallel+^s,\vec{{\pmb a}}^s_n,-^s)\,.~~~~\label{exp-YMS-kk}
\eea
%

\section{Unified web for expansions}
\label{sec-uni}

The expansions found in the previous section are based on the transmutational relations provided by differential operators. Since the differential
operators provide a unified web includes a wide range of theories, it is natural to expect the expansions can also be extended to other theories.
In this section, we discuss how to reach this goal, and establish the complete unified web for expansions. In particular, we give the systematic rules for constructing coefficients in the expansions.

From the expansion of $1$-loop GR Feynman integrand \eref{exp-GR-KK}, one can generate various new expansions by applying differential operators.
Suppose we act the operators defined via polarization vectors in $\{\W\epsilon_i\}$ at two sides of \eref{exp-GR-KK} simultaneously, such operators
transmute both ${\bf I}^{\epsilon,\W\epsilon}_{\rm GR;\circ}({\pmb a}^h_n)$ and ${\bf I}^{\W\epsilon}_{\rm YM}(+^g,\sigma(\vec{\pmb a}^g_n),-^g)$ to integrands of other theories, while keeping the coefficients ${\bf C}^\epsilon_1(\sigma)$ unmodified. For instance, using
\bea
{\bf I}^{\epsilon}_{\rm BI;\circ}({\pmb a}^p_n)&=&{\cal L}^{\W\epsilon}\,\W{\cal D}\,{\bf I}^{\epsilon,\W\epsilon}_{\rm GR;\circ}({\pmb a}^h_n)\,,\nn
{\bf I}_{\rm NLSM}(+^s,\sigma(\vec{\pmb a}^s_n),-^s)&=&{\cal L}^{\W\epsilon}\,\W{\cal D}\,{\bf I}^{\W\epsilon}_{\rm YM}(+^g,\sigma(\vec{\pmb a}^g_n),-^g)\,,
\eea
we obtain
\bea
{\bf I}^{\epsilon}_{\rm BI;\circ}({\pmb a}^p_n)=\sum_\sigma\,{\bf C}^\epsilon_1(\sigma)\,{\bf I}_{\rm NLSM}(+^s,\sigma(\vec{\pmb a}^s_n),-^s)\,.
\eea
Notice that the set of NLSM partial integrands ${\bf I}_{\rm NLSM}(+^s,\sigma(\vec{\pmb a}^s_n),-^s)$ can also be regarded as the $1$-loop
KK basis, since the tree level KK relation is among color ordered amplitudes without regarding to any other information
of external particles.
A more interesting case is applying the operators defined via $\{\epsilon_i\}$ rather than
$\{\W\epsilon_i\}$. These operators also transmute ${\bf I}^{\epsilon,\W\epsilon}_{\rm GR;\circ}({\pmb a}^h_n)$ at the l.h.s to the integrands of other theories. When acting on the r.h.s, they
modify the coefficients ${\bf C}^\epsilon_1(\sigma)$, while keeping the YM partial integrands unaltered. The above manipulation allows us to generate the following expansions
\bea
{\bf I}^{\epsilon,\W\epsilon}_{\rm ssEYM}(+^g,\vec{\pmb a}^g_m,-^g;{\pmb a}^h_{n-m})&=&\sum_\sigma\,{\bf C}^\epsilon_2(\sigma,\vec{\pmb a}_m)\,{\bf I}^{\W\epsilon}_{\rm YM}(+^g,\sigma(\vec{\pmb a}^g_n),-^g)\,,\nn
{\bf I}^{\W\epsilon}_{\rm BI;\circ}({\pmb a}^p_n)&=&\sum_\sigma\,{\bf C}_3(\sigma)\,{\bf I}^{\W\epsilon}_{\rm YM}(+^g,\sigma(\vec{\pmb a}^g_n),-^g)\,,\nn
{\bf I}^{\epsilon,\W\epsilon}_{\rm EM;\circ}({\pmb a}^p_{2m};{\pmb a}^h_{n-2m})&=&\sum_\sigma\,{\bf C}^\epsilon_4(\sigma,X_{2m})\,{\bf I}^{\W\epsilon}_{\rm YM}(+^g,\sigma(\vec{\pmb a}^g_n),-^g)\,,\nn
{\bf I}^{\epsilon,\W\epsilon}_{\rm EMf;\circ}({\pmb a}^p_{2m};{\pmb a}^h_{n-2m})&=&\sum_\sigma\,{\bf C}^\epsilon_5(\sigma,{\cal X}_{2m})\,{\bf I}^{\W\epsilon}_{\rm YM}(+^g,\sigma(\vec{\pmb a}^g_n),-^g)\,.
\eea
The rule for writing down ${\bf C}^\epsilon_2(\sigma,\vec{\pmb a}_m)$ is already provided in the previous section. In subsection \ref{exp-BI-EM} we will discuss the rules for constructing coefficients ${\bf C}^\epsilon_i(\sigma)$ with $i\in\{3,4,5\}$.

The full web for expansions can be established by applying differential operators further. We will not do this procedure. Instead, we will use a more compact way to describe the unified web for expansions. Here we briefly review the idea, the details are exhibited in subsection \ref{double-exp}. One can replace $\epsilon$ in the expansion \eref{exp-YM-KK} by $\W\epsilon$, and substitute it into \eref{exp-GR-KK}, then get the expansion to BAS KK basis as
\bea
{\bf I}^{\epsilon,\W\epsilon}_{\rm GR;\circ}({\pmb a}^h_n)=\sum_\sigma\,\sum_{\sigma'}\,{\bf C}^\epsilon_1(\sigma)\,{\bf I}_{\rm BAS}(+^s,\sigma(\vec{\pmb a}^s_n),-^s\parallel+^s,\sigma'(\vec{\pmb s}^s_n),-^s)\,{\bf C}^{\W\epsilon}_1(\sigma')\,,~~~\label{exp-GR-BAS}
\eea
with two coefficients ${\bf C}^{\epsilon}_1(\sigma)$ and ${\bf C}^{\W\epsilon}_1(\sigma')$. We call such expansion the double-expansion.
The differential operators transmute the l.h.s of \eref{exp-GR-BAS} to Feynman integrands of other theories, and transmute ${\bf C}^{\epsilon}_1(\sigma)$ or ${\bf C}^{\W\epsilon}_1(\sigma')$ to other ${\bf C}^{\epsilon}_i(\sigma)$ or ${\bf C}^{\W\epsilon}_j(\sigma')$
while acting on the r.h.s. Thus the double-expansions for all theories in Table. \ref{tab:unifying-loop} are obtained. Then, all expansions in the unified web can be
obtained by summing over $\sigma$ or $\sigma'$. The double-expanded formulas also manifest the duality between transmutational relations and expansions, as will be discussed in subsection \ref{double-exp}.

With the general ideas discussed above, now we begin to study the corresponding details.

\subsection{Expansions of BI, EM, and EMf to YM}
\label{exp-BI-EM}

As discussed above, the $1$-loop BI, EM and EMf Feynman integrands can also be expanded to the $1$-loop YM KK basis. The purpose of this subsection
is to give the rules for constructing corresponding coefficients ${\bf C}_3(\sigma)$, ${\bf C}^\epsilon_4(\sigma,X_{2m})$ and ${\bf C}^\epsilon_5(\sigma,{\cal X}_{2m})$.

We begin by considering the BI Feynman integrands, which can be generated from the GR integrands via the operator ${\cal L}^\epsilon{\cal D}$.
Applying this operator to two sides of \eref{exp-GR-KK}, the l.h.s gives ${\bf I}^{\W\epsilon}_{\rm BI;\circ}({\pmb a}^p_n)$. At the r.h.s,
the coefficients ${\bf C}^\epsilon_1(\sigma)$ are transmuted to ${\bf C}_3(\sigma)$, while the YM partial integrands are unmodified, since the operator ${\cal L}^\epsilon{\cal D}$ is defined via polarization vectors in $\{\epsilon_i\}$. Thus ${\bf C}_3(\sigma)$ is generated from ${\bf C}^\epsilon_1(\sigma)$,
namely,
\bea
{\bf C}_3(\sigma)={\cal L}^\epsilon\,{\cal D}\,{\bf C}^\epsilon_1(\sigma)\,.
\eea
Consequently, the rule for constructing ${\bf C}_3(\sigma)$ can be generated from the rule for ${\bf C}^\epsilon_1(\sigma)$.
We first consider the effect of the operator ${\cal D}$. This operator annihilates terms do not contain the factor $d-2$, thus we only need to find ordered splittings with ${\pmb a}^0=\emptyset$. It is direct to observe that the operator ${\cal D}$ transmutes ${\bf C}^\epsilon_1(\sigma)$ to ${\bf C}^\epsilon_2(\sigma,\emptyset)$ for the root ${\pmb a}^0=\emptyset$.

To continue, we perform the operator ${\cal L}^\epsilon$. There are two definitions
for the operator ${\cal L}^\epsilon$, which are un-equivalent at the algebraic level, but lead to the same physical result in the case under consideration.
We first consider the definition ${\cal L}^\epsilon\equiv\prod_i{\cal L}_i^\epsilon$. The operator
${\cal L}^\epsilon$ turns $\epsilon_i\cdot k_j$ to
$k_i\cdot k_j$, therefore only terms with the form
$\prod_i\,\epsilon_i\cdot K_i$ can survive under the action, where $K_i$ are combinations of external
and loop momenta. In ${\bf C}^\epsilon_2(\sigma,,\emptyset)$, such part is found to be
$\prod_i\,\epsilon_i\cdot X_i$, where $X_i$ is defined as the summation of $k_j$ with $j\dot{<}i$ in the color ordering. Thus the effect of ${\bf C}_3(\sigma)$ is given as
\bea {\bf C}_3(\sigma)={\cal
L}^\epsilon\,{\bf C}^\epsilon_2(\sigma,\emptyset)\,
=\prod_i\,k_i\cdot
X_i\,, ~~~~\label{c31}
\eea
which is a very  compact result.

Now we consider another
definition of the operator ${\cal L}^\epsilon$,
\bea {\cal L}^\epsilon=\sum_{\rho\in{\rm
pair}}\,\prod_{\{i,j\}\in\rho}\,{\cal L}_{ij}^\epsilon\,. \eea
Applying this ${\cal L}^\epsilon$ to ${\bf C}^\epsilon_2(\sigma,\emptyset)$,
the survived terms are those each polarization
vector $\epsilon_i$ is contracted with another one
$\epsilon_j$. Using the definition of ${\bf C}^\epsilon_2(\sigma,{\cal L}^\epsilon)$, such part is found to be
\bea & &\sum_{\vec{\pmb B}:|i|_{\rm
even}}\,\Big(\prod_{i=1}^t\,(-)^{{|i|\over2}}M_i(\sigma,\vec{\pmb B})\Big)\,,~~~~\label{ee-term} \eea
where the summation is over all possible branches $\vec{\pmb B}$ those the lengths of
all subsets ${\pmb a}^i$ are even, and the number of subsets included in each branch is denoted by $t$. The monomial $M_i(\sigma,\vec{\pmb B})$ for the
subset $\vec{{\pmb a}}^i$ is given
as
\bea M_i(\sigma,\vec{\pmb B})=(\epsilon_{a^i_{|i|}}\cdot
\epsilon_{a^i_{|i|-1}})(k_{a^i_{|i|-1}}\cdot k_{a^i_{|i|-2}})
(\epsilon_{a^i_{|i|-2}}\cdot\epsilon_{a^i_{|i|-3}})\cdots
(k_{a^i_3}\cdot k_{a^i_2})(\epsilon_{a^i_2}\cdot
\epsilon_{a^i_1}) (k_{a^i_1}\cdot Z_{a^i_1})\,. \eea
Under the action of ${\cal L}^\epsilon$, we find
\bea {\bf C}_3(\sigma)={\cal
L}^\epsilon\,{\bf C}^\epsilon_2(\sigma,\emptyset)\,
=\sum_{\vec{\pmb B}:|i|_{\rm
even}}\,\Big(\prod_{i=1}^t\,(-)^{{|i|\over2}}N_i(\sigma,\vec{\pmb B})\Big)\,, ~~~~\label{c32}
\eea
where
\bea N_i(\sigma,\vec{\pmb B})=\Big(\prod_{k=1}^{|i|-1}\,k_{a^i_k}\cdot k_{a^i_{k+1}}\Big) (k_{a^i_1}\cdot
Z_{a^i_1})\,. \eea
The equivalence between \eref{c31} and \eref{c32} can be verified for simple cases, and we have checked it for the $3$-point integrand. The general proof is an interesting challenge, which we leave as the feature work.

Then we turn to EM and EMf Feynman integrands. The EM Feynman integrands can be generated from the GR integrands via the operator ${\cal T}^\epsilon_{X_{2m}}({\cal D}+1)$, thus the argument similar as for the BI case gives
\bea
{\bf C}^\epsilon_4(\sigma,X_{2m})={\cal T}^\epsilon_{X_{2m}}\,({\cal D}+1)\,{\bf C}^\epsilon_1(\sigma)\,.
\eea
The operator $({\cal D}+1)$ transmutes ${\bf C}^\epsilon_1(\sigma)$ as
\bea
({\cal D}+1)\,{\bf C}^\epsilon_1(\sigma)={\bf C}^\epsilon_2(\sigma,\emptyset)+{\bf C}^\epsilon_1(\sigma)\,.
\eea
Then we need to perform ${\cal T}^\epsilon_{X_{2m}}$ on ${\bf C}^\epsilon_2(\sigma,\emptyset)$ and ${\bf C}^\epsilon_1(\sigma)$. Let us consider
${\cal T}^\epsilon_{X_{2m}}{\bf C}^\epsilon_1(\sigma)$ first. Recall that${\cal T}^\epsilon_{X_{2m}}$ is defined as summing over $\prod_{i_k,j_k\in\rho}{\cal T}^\epsilon_{i_kj_k}$ for different partitions,
where each partition groups the $2m$ external photons into $m$ pairs as $\{(i_1,j_1),(i_2,j_2),\cdots(i_m,j_m)\}$, with $i_1<i_2<\cdots<i_m$ and $i_k<j_k$ for $\forall k$. Thus we can consider the effect of operator $\prod_{i_k,j_k\in\rho}{\cal T}^\epsilon_{i_kj_k}$ for a given partition. This operator annihilates all terms do not contain $\prod_{i_k,j_k\in\rho}(\epsilon_{i_k}\cdot\epsilon_{j_k})$. Henceforth, one can start with ordered splittings for ${\bf C}^\epsilon(\sigma)$, and select ordered splittings by the condition each pair in the partition appears in one subset as a single element, i.e., two photons are adjacent. Then, for a selected ordered splitting, we now consider the effect of applying $\prod_{i_k,j_k\in\rho}{\cal T}^\epsilon_{i_kj_k}$ to the corresponding kinematic factor
${\bf T}^\epsilon_{\vec{\pmb a}^0}\Big(\prod_{i=1}^t\,K^\epsilon_{\vec{\pmb a}^i}\Big)$.
For ${\bf T}^\epsilon_{\vec{\pmb a}^0}$, we turn all $(f_{i_k}\cdot f_{j_k})^{\mu\nu}$ to $-k_{i_k}^\mu k_{j_k}^\nu$. For $K^\epsilon_{\vec{\pmb a}^i}$, we turn all $(f_{i_k}\cdot f_{j_k})^{\mu\nu}$ to $-k_{i_k}^\mu k_{j_k}^\nu$ when $i_k\neq R_i$, $j_k\neq R_i$, and turn $(\epsilon_{i_k}\cdot f_{j_k})^\mu$ to $-k_{j_k}^\mu$
when $i_k=R_i$, or turn $(\epsilon_{j_k}\cdot f_{i_k})^\mu$ to $-k_{i_k}^\mu$
when $j_k=R_i$. The resulting object of ${\cal T}^\epsilon_{X_{2m}}{\bf C}^\epsilon_1(\sigma)$ is obtained by summing over contributions from selected ordered splittings, then summing over all proper partitions. Another part ${\cal T}^\epsilon_{X_{2m}}{\bf C}^\epsilon_2(\sigma,\emptyset)$
can be obtained by performing the above manipulation to branches for the root ${\vec{\pmb a}^0}=\emptyset$.

The EMf Feynman integrands are generated form the GR ones via the operator ${\cal T}^\epsilon_{{\cal X}_{2m}}(N{\cal D}+1)$, where $N$ stands for
the number of different flavors. Thus we have
\bea
{\bf C}^\epsilon_5(\sigma,{\cal X}_{2m})&=&{\cal T}^\epsilon_{{\cal X}_{2m}}\,(N{\cal D}+1)\,{\bf C}^\epsilon_1(\sigma)\nn
&=&N\,{\cal T}^\epsilon_{{\cal X}_{2m}}\,{\bf C}^\epsilon_2(\sigma,\emptyset)+{\cal T}^\epsilon_{{\cal X}_{2m}}\,{\bf C}^\epsilon_1(\sigma)\,.
\eea
The consideration for ${\cal T}^\epsilon_{{\cal X}_{2m}}\,{\bf C}^\epsilon_2(\sigma,\emptyset)$ and ${\cal T}^\epsilon_{{\cal X}_{2m}}\,{\bf C}^\epsilon_1(\sigma)$ is analogous as for ${\cal T}^\epsilon_{{X}_{2m}}\,{\bf C}^\epsilon_2(\sigma,\emptyset)$ and ${\cal T}^\epsilon_{{ X}_{2m}}\,{\bf C}^\epsilon_1(\sigma)$. The only difference is that the appropriate partitions are reduced: a partition is allowed if and only if $\delta_{I_{i_k}I_{j_k}}\neq0$ for all pairs $(i_k,j_k)$.

\subsection{Double-expansion and unified web}
\label{double-exp}

As discussed previously, the $1$-loop GR Feynman integrands can be double-expanded as in \eref{exp-GR-BAS}. The BAS KK basis contributes the propagators, while coefficients ${\bf C}^\epsilon_1(\sigma)$ and ${\bf C}^{\W\epsilon}_1(\sigma')$ serve as BCJ numerators. On the other hand,
we have
\bea
{\cal T}^\epsilon_{+\vec{\pmb a}_m-}\,{\bf C}^\epsilon_1(\sigma)&=&{\bf C}^\epsilon_2(\sigma,\vec{\pmb a}_m)\,,\nn
{\cal L}^\epsilon\,{\cal D}\,{\bf C}^\epsilon_1(\sigma)&=&{\bf C}_3(\sigma)\,,\nn
{\cal T}^\epsilon_{X_{2m}}\,({\cal D}+1)\,{\bf C}^\epsilon_1(\sigma)&=&{\bf C}^\epsilon_4(\sigma,X_{2m})\,,\nn
{\cal T}^\epsilon_{{\cal X}_{2m}}\,(N{\cal D}+1)\,{\bf C}^\epsilon_1(\sigma)&=&{\bf C}^\epsilon_5(\sigma,{\cal X}_{2m})\,,\nn
{\cal T}^\epsilon_{+\vec{\pmb s}_n-}\,{\bf C}^\epsilon_1(\sigma)&=&{\bf C}_6(\sigma,\vec{\pmb s}_n)\,.
\eea
In the first line, the length $m$ of the set ${\pmb a}_m$ is required to be $0\leq m<n$. The rules for constructing ${\bf C}^\epsilon_i(\sigma)$
with $i\in\{2,3,4,5\}$ are provided before. Here we added ${\bf C}_6(\sigma,\vec{\pmb s}_n)=\delta_{\vec{\pmb s}_n\sigma(\vec{\pmb a}_n)}$ to the list, since
\bea
{\cal T}^\epsilon_{+\vec{\pmb s}_n-}\,{\bf I}^{\epsilon,\W\epsilon}_{\rm GR}({\pmb a}^h_n)
&=&{\bf I}^{\W\epsilon}_{\rm YM}(+^g,\vec{\pmb s}^g_n,-^g)\nn
&=&\sum_\sigma\,\delta_{\vec{\pmb s}_n\sigma(\vec{\pmb a}_n)}\,{\bf I}^{\W\epsilon}_{\rm YM}(+^g,\sigma(\vec{\pmb a}^g_n),-^g)\,.
\eea
Here $\delta_{\vec{\pmb s}_n\sigma(\vec{\pmb a}_n)}$ is understood as $1$ when $\vec{\pmb s}_n=\sigma(\vec{\pmb a}_n)$ and $0$ otherwise, which is different from $\delta^\pi$ defined below \eref{equ-2}. The relations among ${\bf C}^{\W\epsilon}_i(\sigma')$ with $i\in\{1,2,3,4,5,6\}$ are completely analogous. Let us simplify the notations as
\bea
{\cal O}^\epsilon_i\,{\bf C}^\epsilon_1(\sigma)={\bf C}^\epsilon_i(\sigma)\,,~~~~\label{dual-map}
\eea
where
\bea
{\cal O}^\epsilon_1=\mathbb{I}\,,&~~~~&{\cal O}^\epsilon_2={\cal T}^\epsilon_{+\vec{\pmb a}_m-}\,,\nn
{\cal O}^\epsilon_3={\cal L}^\epsilon\,{\cal D}\,,&~~~~&{\cal O}^\epsilon_4={\cal T}^\epsilon_{X_{2m}}\,({\cal D}+1)\,,\nn
{\cal O}^\epsilon_5={\cal T}^\epsilon_{{\cal X}_{2m}}\,,&~~~~&{\cal O}^\epsilon_6={\cal T}^\epsilon_{+\vec{\pmb s}_n-}\,,
\eea
and introduce the analogous notations ${\cal O}^{\W\epsilon}_i$ for ${\bf C}^{\W\epsilon}_i(\sigma')$. Applying the above operators
to the double-expanded GR Feynman integrand in \eref{exp-GR-BAS}, we get
\bea
{\bf I}_{ij}=\sum_{\sigma}\,\sum_{\sigma'}\,{\bf C}^\epsilon_i(\sigma)\,{\bf I}_{\rm BAS}(+^s,\sigma(\vec{\pmb a}^s_n),-^s\parallel+^s\sigma'(\vec{\pmb s}^s_n),-^s)\,{\bf C}^{\W\epsilon}_j(\sigma')\,,~~~~\label{double-exp-theo}
\eea
where
\bea
{\bf I}_{ij}={\cal O}^\epsilon_i\,{\cal O}^{\W\epsilon}_j\,{\bf I}^{\epsilon,\W\epsilon}_{\rm GR}({\pmb a}^h_n)\,.
\eea
The physical interpretation for each ${\bf I}_{ij}$ can be seen in Table. \ref{tab:unifying-loop}. Thus \eref{double-exp-theo}
is indeed the double-expanded formula for Feynman integrands for a variety of theories, as listed in Table. \ref{tab:unifying-exp2}.
\begin{table}[!h]
\begin{center}
\begin{tabular}{c|c|c}
${\bf I}_{ij}$& $i$  & $j$ \\
\hline
${\bf I}_{{\rm GR}}^{\epsilon,\W\epsilon}(\pmb{a}^h_n)$ & $1$ & $1$  \\
${\bf I}_{{\rm ssEYM}}^{\epsilon,\W\epsilon}(+^g,\vec{\pmb a}^g_m,-^g;\pmb{a}^h_{n-m})$ & $2$ & $1$ \\
${\bf I}_{{\rm BI}}^{\W\epsilon}(\pmb{a}^p_n)$ & $3$ & $1$  \\
${\bf I}_{{\rm EM}}^{\epsilon,\W\epsilon}(\pmb{a}^p_{2m};\pmb{a}^h_{n-2m})$& $4$ & $1$  \\
${\bf I}_{{\rm EMf}}^{\epsilon,\W\epsilon}(\pmb{a}^p_{2m};\pmb{a}^h_{n-2m})$& $5$ & $1$  \\
${\bf I}_{{\rm YM}}^{\W\epsilon}(+^g,\vec{\pmb a}^g_n,-^g)$& $6$ & $1$  \\
${\bf I}_{{\rm ssYMS}}^{\W\epsilon}(+^s,\vec{\pmb a}^s_m,-^s;\pmb{a}^g_{n-m}\parallel +^A,\vec{\pmb a}^A_n,-^A)$ & $6$ & $2$\\
${\bf I}_{{\rm NLSM}}(+^s,\vec{\pmb a}^s_n,-^s)$ & $6$ & $3$ \\
${\bf I}_{{\rm SYMS}}^{\W\epsilon}(\pmb{a}^s_{2m};\pmb{a}^g_{n-2m}\parallel +^A,\vec{\pmb a}^A_n,-^A)$ & $6$ & $5$ \\
${\bf I}_{{\rm ssEDBI}}^{\W\epsilon}(\vec{\pmb a}^s_m;\pmb{a}^p_{n-m})$ & $3$ & $2$ \\
${\bf I}_{{\rm SG}}({\pmb a}^s_n)$ & $3$ & $3$ \\
${\bf I}_{{\rm DBI}}^{\W\epsilon}({\pmb a}^s_{2m};{\pmb a}^p_{n-2m})$ & $3$ & $5$ \\
${\bf I}_{{\rm BAS}}(+^s,\vec{\pmb a}^s_n,-^s\parallel +^s,\vec{\pmb s}^s_n,-^s)$ & $6$ & $6$ \\
\end{tabular}
\end{center}
\caption{\label{tab:unifying-exp2}${\bf I}_{ij}$ for different $i$ and $j$}
\end{table}

\begin{figure}
  \centering
  \includegraphics[width=12cm]{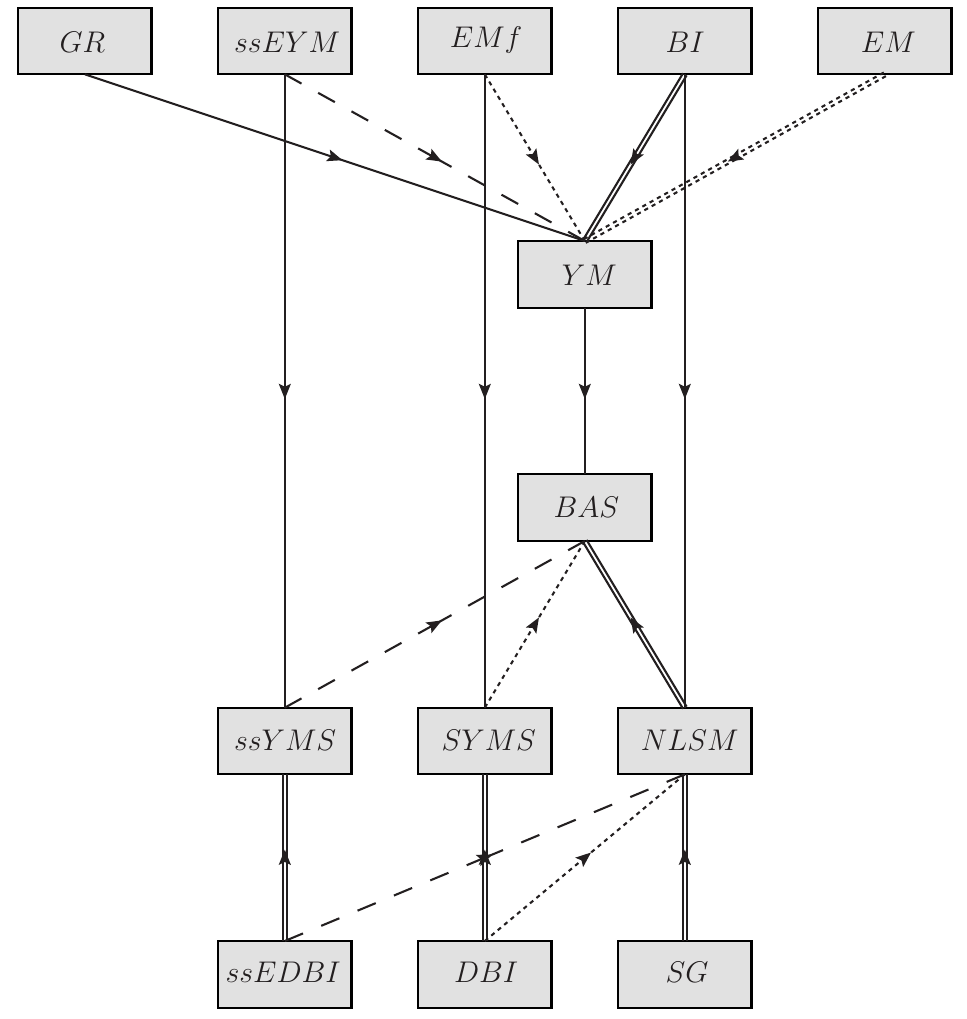} \\
  \caption{Unified web for expansions of $1$-loop Feynman integrands. The straight lines denote the coefficients ${\bf C}_1^\epsilon(\sigma)$, the
  dashed lines denote ${\bf C}_2^\epsilon(\sigma,\vec{\pmb a}_m)$, the double straight lines denote ${\bf C}_3^\epsilon(\sigma)$,
  the thin dashed lines denote ${\bf C}_5^\epsilon(\sigma,{\cal X}_{2m})$, the double thin dashed lines denote ${\bf C}_4^\epsilon(\sigma,{ X}_{2m})$.}\label{web}
\end{figure}

The full unified web for expansions can be constructed from the double-expansions in \eref{double-exp-theo} and Table. \ref{tab:unifying-exp2},
by summing over $\sigma$ or $\sigma'$. To do this, we first summing over $\sigma$ for ${\bf C}_6(\sigma,\vec{\pmb a}_n)$ to get
\bea
{\bf I}_{{\rm YM}}^{\W\epsilon}(+^g,\vec{\pmb a}^g_n,-^g)&=&\sum_{\sigma'}\,{\bf C}^{\W\epsilon}_1(\sigma')\,{\bf I}_{{\rm BAS}}(+^s,\vec{\pmb a}^s_n,-^s\parallel+^s,\sigma'(\vec{\pmb s}^s_n),-^s)\,,\nn
{\bf I}_{{\rm ssYMS}}^{\W\epsilon}(+^s,\vec{\pmb a}^s_m,-^s;\pmb{a}^g_{n-m}\parallel +^A,\vec{\pmb a}^A_n,-^A)&=&\sum_{\sigma'}\,{\bf C}^{\W\epsilon}_2(\sigma',\vec{\pmb a}^s_m)\,{\bf I}_{{\rm BAS}}(+^s,\vec{\pmb a}^s_n,-^s\parallel+^s,\sigma'(\vec{\pmb s}^s_n),-^s)\,,\nn
{\bf I}_{{\rm NLSM}}(+^s,\vec{\pmb a}^s_n,-^s)&=&\sum_{\sigma'}\,{\bf C}_3(\sigma')\,{\bf I}_{{\rm BAS}}(+^s,\vec{\pmb a}^s_n,-^s\parallel+^s,\sigma'(\vec{\pmb s}^s_n),-^s)\,,\nn
{\bf I}_{{\rm SYMS}}^{\W\epsilon}(\pmb{a}^s_{2m};\pmb{a}^g_{n-2m}\parallel +^A,\vec{\pmb a}^A_n,-^A)&=&\sum_{\sigma'}\,C_5^{\W\epsilon}\,(\sigma',{\cal X}_{2m})\,{\bf I}_{{\rm BAS}}(+,\vec{\pmb a}^s_n,-\parallel+^s,\sigma'(\vec{\pmb s}^s_n),-^s)\,.~~~~\label{sum-BAS}
\eea
The definition of ${\bf C}_6(\sigma,\vec{\pmb a}_n)$ makes the summation to be straightforward. Using the expansions in \eref{sum-BAS},
one can obtain the expansions for other theories. For example, from Table. \ref{tab:unifying-exp2} we see that
\bea
{\bf I}_{{\rm DBI}}^{\W\epsilon}({\pmb a}^s_{2m};{\pmb a}^p_{n-2m})=\sum_{\sigma}\,\sum_{\sigma'}\,{\bf C}_3(\sigma)\,{\bf I}_{\rm BAS}(+^s,\sigma(\vec{\pmb a}^s_n),-^s\parallel+^s\sigma'(\vec{\pmb s}^s_n),-^s)\,{\bf C}^{\W\epsilon}_5(\sigma',{\cal X}_{2m})\,.~~~~\label{double-exp-DBI}
\eea
Substituting the equality in the third line of \eref{sum-BAS} into \eref{double-exp-DBI} gives
\bea
{\bf I}_{{\rm DBI}}^{\W\epsilon}({\pmb a}^s_{2m};{\pmb a}^p_{n-2m})=\sum_{\sigma}\,{\bf C}^{\W\epsilon}_5(\sigma,{\cal X}_{2m})\,{\bf I}_{{\rm NLSM}}(+^s,\sigma(\vec{\pmb a}^s_n),-^s)\,,
\eea
while substituting the equality in the fourth line of \eref{sum-BAS} provides
\bea
{\bf I}_{{\rm DBI}}^{\W\epsilon}({\pmb a}^s_{2m};{\pmb a}^p_{n-2m})=\sum_{\sigma}\,{\bf C}_3(\sigma)\,{\bf I}_{{\rm SYMS}}^{\W\epsilon}(\pmb{a}^s_{2m};\pmb{a}^g_{n-2m}\parallel +^A,\sigma(\vec{\pmb a}^A_n),-^A)\,.
\eea
The unified web can be established via the above method, and is represented diagrammatically in Fig. \ref{web}.

The double-expanded formula is dual to the transmutational formula \eref{uni-diff-loop}, due to the following three reasons. First, the formula \eref{double-exp-theo} is constructed from \eref{uni-diff-loop}, together with some very general principles/assumptions mentioned at the beginning of section \ref{solve-coefficient}. Secondly, they include the same list of theories. Thirdly, operators ${\cal O}^\epsilon_i$ and coefficients ${\bf C}^\epsilon_i(\sigma)$ are linked by acting operators on BCJ numerators ${\bf C}_1^\epsilon(\sigma)$ as in \eref{dual-map}, and so do operators ${\cal O}^{\W\epsilon}_i$ and coefficients ${\bf C}^{\W\epsilon}_i(\sigma')$. Thus we have one-to-one dual mappings between operators and coefficients,
as shown in Table. \ref{tab:dual mapping}. In this table, the exact form of ${\bf C}_6(\sigma,\vec{\pmb s}_n)$ is provided since it is special.
Based on the duality discussed above, one can claim that the expansions of $1$-loop Feynman integrands is the dual version
of transmutations of Feynman integrands via the differential operators.
\begin{table}[!h]
\begin{center}
\begin{tabular}{c|c}
${\cal O}^\epsilon_i$  & ${\bf C}^\epsilon_i(\sigma)$ \\
\hline
$\mathbb{I}$ & ${\bf C}_1^\epsilon(\sigma)$ \\
${\cal T}^\epsilon_{+\vec{\pmb a}_m-}$ & ${\bf C}_2^\epsilon(\sigma,\vec{\pmb a}_m)$ \\
${\cal L}^\epsilon\,{\cal D}$ & ${\bf C}_3(\sigma)$ \\
${\cal T}^\epsilon_{X_{2m}}\,({\cal D}+1)$ & ${\bf C}_4^\epsilon(\sigma,X_{2m})$ \\
${\cal T}^\epsilon_{{\cal X}_{2m}}\,(N{\cal D}+1)$ & ${\bf C}_5^\epsilon(\sigma,{\cal X}_{2m})$ \\
${\cal T}^\epsilon_{+\vec{\pmb s}_n-}$ & ${\bf C}_6(\sigma,\vec{\pmb s}_n)=\delta_{\vec{\pmb s}_n\sigma(\vec{\pmb a}_n)}$ \\
\end{tabular}
\end{center}
\caption{\label{tab:dual mapping}Dual mappings between operators and coefficients}
\end{table}
%

\section{Summary and discussions}
\label{secconclu}

In this paper, we investigated the connections among $1$-loop Feynman integrands of a large variety of theories with massless external states. The work includes two parts. First, we constructed a new class of $1$-loop level differential operators ${\cal C}^\epsilon_{\vec{\pmb a}_m}$, which transmute the $1$-loop GR Feynman integrands to $1$-loop ssEYM Feynman integrands. The advantage of these new operators is, they are commutable with the integration of loop momentum, thus the transmutational relation holds at not only the integrand level, but also the amplitude level.

Secondly, via the $1$-loop level transmutational relations, as well as some general principles/assumptions such as gauge invariance, we constructed the unified web for expansions of $1$-loop Feynman integrands for a wide range of theories including GR, ssEYM, EM, EMf, BI, YM, ssYMS, SYMS, NLSM, DBI, EDBI, SG. We showed that the $1$-loop Feynman integrands of all above theories can be double-expanded to the BAS $1$-loop KK basis, and provided the systematic rules for constructing the coefficients in the expansions. Through out the whole process, we only used the knowledge of transmutational relations among $1$-loop Feynman integrands of different theories, as well as some very general requirements listed at the beginning of section. \ref{solve-coefficient}, without knowing any detail about the Feynman integrands under consideration. Based on this character, together with the one-to-one mappings between transmutation operators and coefficients in expansions, we claimed that the transmutational relations and expansions dual to each other.

In this paper and the previous work in \cite{Zhou:2021kzv}, the consideration for the EYM partial Feynman integrands is not complete. We restricted ourselves to the special single-trace case that the virtual particle propagating in the loop is only a gluon. This special case corresponds to only a part of the color ordered $1$-loop EYM partial Feynman integrands, since the virtual particles in the loop can be either gluons or gravitons, and the multiple-trace configurations are allowed. We have not considered the general case due to some technical difficulty, and leave the complete solution to this problem as the future work.

The expansions of $1$-loop Feynman integrands also indicate a new method for calculating the $1$-loop Feynman integrands of various theories. One can evaluate the BAS Feynman integrands at the first step, then use the rules for constructing coefficients to get the expressions for Feynman integrands of other theories in the double-expanded formulas. In principle, one can also calculate the GR Feynman integrands at the first step, then
use the differential operators to generate others. However, in practice the GR integrands are most complicated ones in the unified webs. On the other hand, the BAS integrands are easiest ones, since they only contain propagators without carrying any kinematic numerator.

We have shown that the $1$-loop level expansions are determined by $1$-loop level transmutational relations as well as some general requirements, where the $1$-loop level transmutation operators are constructed from tree level operators via the forward limit operation. The reader may ask, which principles/assumptions determine the tree level operators. In \cite{Cheung:2017ems}, these tree level operators are constructed by general principles/assumptions which are first six requirements listed at the beginning of section. \ref{solve-coefficient}, as well as the requirement for preserving the momentum conservation\footnote{The assumption of double-copy structure has not been emphasized in \cite{Cheung:2017ems}, but this assumption is necessary for explaining why they do not consider operators $\partial_{\epsilon_i\cdot\W\epsilon_j}$.}. When saying preserving the momentum conservation, it means the effects of performing operators will not be modified if one use the momentum conservation to replace $\epsilon_i\cdot k_j$ by
$-\epsilon_i\cdot(\sum_{l\neq j}k_l)$, as discussed in subsection. \ref{exp-EYM} in section. \ref{solve-coefficient}. Thus the connections among tree amplitudes and $1$-loop Feynman integrands, either transmutational relations or expansions, are fully determined by the seven principles/assumptions listed at the beginning of section. \ref{solve-coefficient}. In \cite{Zhou:2018wvn,Bollmann:2018edb}, CHY formulas are used for proving the transmutational relations, it is because of we want to relate the generated amplitudes to the classical Lagrangians. Indeed, to relate the CHY integrands to the classical theories, one still need to use the Feynman rules or on-shell recursions, since the three manipulations
compactifying, squeezing and generalized dimensional reduction do not indicate the corresponding classical Lagrangians for resulting objects. On the other hand, without knowing any classical counterpart, the amplitudes/Feynman integrands generated from the GR ones via differential operators, or from the BAS ones via the expansions, are also physically acceptable, since they satisfy the general physical requirements (along the first line, the locality and unitaraty are inherit from the GR ones, while along the second line, they are inherit from the BAS ones).

The above discussion motivates two interesting future directions. First, as discussed in subsection. \ref{exp-EYM} in section. \ref{solve-coefficient}, the tree level transmutation operators preserve the momentum conservation, while the $1$-loop level operators do not, i.e., for operators ${\cal I}^\epsilon_{+a_ia_{i+1}}$ and ${\cal C}^\epsilon_{\vec{\pmb a}_m}$ the transmutational relations do not hold when replacing $\epsilon_i\cdot k_j$ by
$-\epsilon_i\cdot(\sum_{l\neq j}k_l)$ in the Feynman integrands.
It means as long as the $1$-loop level operators make sense, the Feynman integrands should have special formulas with external momenta are represented in special ways. Such special formulas are found in \eref{exp-EYM-final} and \eref{exp-GR-final}. However, the transmutation operators which are consistent with momentum conservation are more attractive for us. Seeking such new $1$-loop level operators is a crucial future direction.

Secondly, our unified webs have not included all possible combinations of differential operators or coefficients in the expansions. For instance, ${\bf I}_{55}$ is not included in Table. \ref{tab:unifying-exp2}. On the other hand, all combinations lead to objects which satisfy the general physical requirements, as discussed above. Furthermore, these omitted objects have well defined CHY integrands, which can be obtained by applying the corresponding differential operators to the GR CHY integrands (recall that each coefficient in the expansion dual to a differential operator).
It is interesting to study if an omitted physically acceptable object corresponds to any classical Lagrangian.

\section*{Acknowledgments}

The author would thank Xiaodi Li for valuable discussions. This
work is supported by Chinese NSF funding under
contracts No.11805163, as well as NSF of Jiangsu Province under Grant No.BK20180897.

\appendix

\section{CHY formulas at tree and $1$-loop levels}
\label{sec-CHY}

In the CHY framework, tree amplitudes for $n$ massless particles in arbitrary dimensions arise from a multi-dimensional contour integral over
the moduli space of genus zero Riemann surfaces with $n$ punctures, ${\cal M}_{0,n}$ \cite{Cachazo:2013gna,Cachazo:2013hca,Cachazo:2013iea,Cachazo:2014nsa,Cachazo:2014xea}, formulated as
\bea
{\cal A}_n=\int d\mu_n\,{\cal I}^L(\{k_i,\epsilon_i,z_i\}){\cal I}^R(\{k_i,\W\epsilon_i,z_i\})\,,~~~~\label{CHY}
\eea
which possesses the M\"obius ${\rm SL}(2,\mathbb{C})$ invariance. Here $k_i$, $\epsilon_i$ and $z_i$ are the momentum, polarization vector, and puncture location for $i^{\rm th}$ external
particle, respectively. The measure part is defined as
\bea
d\mu_n\equiv{d^n z\over{\rm vol}\,{\rm SL}(2,\mathbb{C})}\prod_i{'}\delta(\xi_i)\,.
\eea
The $\delta$-functions impose the scattering equations
\bea
E_i\equiv\sum_{j\in\{1,2,\ldots,n\}\setminus\{i\}}{k_i\cdot k_j\over z_{ij}}=0\,,
\eea
where $z_{ij}\equiv z_i-z_j$. The scattering equations define the map from the punctures on the moduli space ${\cal M}_{0,n}$ to vectors on the light cone, and fully localize the integral on
their solutions.
The measure part is universal ,while the integrand in \eref{CHY} depends on the theory under consideration. For any theory known to have a CHY representation, the corresponding integrand can be split into two
parts ${\cal I}^L$ and ${\cal I}^R$, as can be seen in \eref{CHY}. Either of them are weight-$2$ for each variable $z_i$
under the M\"obius transformation. In Table.\ref{tab:theories}, we list the tree level CHY integrands which will be used in this paper \cite{Cachazo:2014xea}\footnote{For theories contain gauge or flavor groups, we only show
the integrands for color-ordered partial amplitudes instead of full ones.}.
\begin{table}[!h]
    \begin{center}
        \begin{tabular}{c|c|c}
            Theory& ${\cal I}^L(k_i,\epsilon_i,z_i)$ & ${\cal I}^R(k_i,\W\epsilon_i,z_i)$ \\
            \hline
            GR & ${\bf Pf}'\Psi$ & ${\bf Pf}'{\Psi}$ \\
            YM & $PT(\sigma_1,\cdots,\sigma_n)$ & ${\bf Pf}' \Psi$ \\
            BAS & $PT(\sigma_1,\cdots,\sigma_n)$ & $PT(\sigma'_1,\cdots,\sigma'_n)$ \\
        \end{tabular}
    \end{center}
    \caption{\label{tab:theories}Form of the integrands}
\end{table}

We now explain building blocks appearing in Table \ref{tab:theories} in turn. The $2n\times2n$ antisymmetric matrix $\Psi$ is given by
\bea\label{Psi}
\Psi = \left(
         \begin{array}{c|c}
           ~~A~~ &  ~~C~~ \\
           \hline
           -C^{\rm T} & B \\
         \end{array}
       \right)\,,
\eea
where
\bea
& &A_{ij} = \begin{cases} \displaystyle {k_{i}\cdot k_j\over z_{ij}} & i\neq j\,,\\
\displaystyle  ~~~ 0 & i=j\,,\end{cases} \qquad\qquad\qquad\qquad B_{ij} = \begin{cases} \displaystyle {\epsilon_i\cdot\epsilon_j\over z_{ij}} & i\neq j\,,\\
\displaystyle ~~~ 0 & i=j\,,\end{cases} \nn
& &C_{ij} = \begin{cases} \displaystyle {k_i \cdot \epsilon_j\over z_{ij}} &\quad i\neq j\,,\\
\displaystyle -\sum_{l=1,\,l\neq j}^n\hspace{-.5em}{k_l \cdot \epsilon_j\over z_{lj}} &\quad i=j\,,\end{cases}
\label{ABCmatrix}
\eea
The notation ${\bf Pf}$ stands for the polynomial called Pfaffian. For a $2n\times 2n$ skew symmetric matrix $S$, Pfaffian is defined as
\bea
{\bf Pf}S={1\over 2^n n!}\sum_{\sigma\in S_{2n}} {\bf sgn}(\sigma)\prod_{i=1}^n\,a_{\sigma(2i-1),\sigma(2i)}\,,~~~\label{pfa-1}
\eea
where $S_{2n}$ is the permutation group of $2n$ elements and ${\bf sgn}(\sigma)$ is the signature of $\sigma$.
More explicitly, let $\Pi$ be the set of all partitions of $\{1,2,\cdots, 2n\}$ into pairs without regard to the order.
An element $\a$ in $\Pi$ can be written as
\bea
\a=\{(i_1,j_1),(i_2,j_2),\cdots,(i_n,j_n) \}\,,
\eea
with $i_k<j_k$ and $i_1<i_2<\cdots<i_n$. Now let
\bea
\sigma_{\a} = \left(
         \begin{array}{c}
           ~~~1~~~ 2~~~3~~~4~~\cdots~2n-1~~2n~~ \\
           \,\,i_1~~j_1~~i_2~~j_2~~\cdots~~~i_n~~~~~~j_n \\
         \end{array}
       \right)
\eea
be the associated permutation of the partition $\a$. If we define
\bea
S_{\a}={\bf sgn}(\sigma_{\a})\,a_{i_1j_1}a_{i_2j_2}\cdots a_{i_nj_n}\,,
\eea
then the Pfaffian of the matrix $S$ is given as
\bea
{\bf Pf}S=\sum_{\a\in\Pi}S_{\a}\,.~~~~~\label{pfa}
\eea
With the definition of Pfaffian provided above, the reduced Pfaffian of the matrix $\Psi$ is defined as
\bea
{\bf Pf}'\Psi={(-)^{i+j}\over z_{ij}}\,{\bf Pf}\Psi^{ij}_{ij},
\eea
where the notation $\Psi^{ij}_{ij}$ means the $i^{\rm th}$ and $j^{\rm th}$ rows and columns of the matrix $\Psi$
have been removed (with $1\leq i,j\leq n$). It can be proved that this definition is independent of the choice of $i$ and $j$.

The Parke-Taylor factor for $PT(\sigma_1,\cdots,\sigma_n)$ is given as
\bea
PT(\sigma_1,\cdots,\sigma_n)={1\over z_{\sigma_1\sigma_2}z_{\sigma_2\sigma_3}\cdots z_{\sigma_{n-1}\sigma_n}z_{\sigma_n\sigma_1}}\,,
\eea
it implies the color ordering $\sigma_1,\cdots,\sigma_n$ for the color ordered amplitude.

The $1$-loop CHY formulas can be obtained via either the underlying ambitwistor string theory \cite{Adamo:2013tca,Mason:2013sva,Adamo:2013tsa,Casali:2014hfa,Geyer:2015bja,Geyer:2015jch,Geyer:2017ela,Adamo:2015hoa}, or the forward limit procedure \cite{He:2015yua,Cachazo:2015aol,Feng:2016nrf,Feng:2019xiq}. Here we only introduce the latter one.
The $1$-loop level scattering equations are found to be
\bea
& &E_i\equiv\sum_{j\in\{1,2,\ldots,n\}\setminus\{i\}}{k_i\cdot k_j\over z_{ij}}+{k_i\cdot \ell\over z_{i+}}-{k_i\cdot \ell\over z_{i-}}=0\,,~~~~~~~~i\in\{1,\cdots,n\}\nn
& &E_+\equiv\sum_{j=1}^n{\ell\cdot k_j\over z_{+j}}=0\,,~~~~~~~~
E_-\equiv\sum_{j=1}^n{-\ell\cdot k_j\over z_{-j}}=0\,.~~~~\label{SE-loop}
\eea
These equations yield the massive propagators $1/((\ell+K)^2-\ell^2)$ in the loop, rather than
the desired massless ones $1/(\ell+K)^2$. However, these massive propagators relate to the massless ones through the
well known partial fraction identity
\bea
{1\over D_1\cdots D_m}=\sum_{i=1}^m\,{1\over D_i}\Big[\prod_{j\neq i}\,{1\over D_j-D_i}\Big]\,,
\eea
which implies
\bea
{1\over \ell^2(\ell+K_1)^2(\ell+K_1+K_2)^2\cdots (\ell+K_1+\cdots+K_{m-1})^2}\simeq{1\over \ell^2}\sum_{i=1}^m\Big[\prod_{j= i}^{i+m-2}\,{1\over (\ell+K_i+\cdots+K_j)^2-\ell^2}\Big]\,.~~~~\label{partial-fraction}
\eea
For each individual term at the r.h.s of the above relation, we have shifted the loop momentum without alternating the result of Feynman integral.
Here $\simeq$ means the l.h.s and r.h.s are not equivalent to each other at the integrand level, but are equivalent at the integration level.
The l.h.s of \eref{partial-fraction} is the standard propagators in the loop for an individual diagram, while each term at the r.h.s can be obtained via the forward limit method.

Thus, to obtain the correct $1$-loop Feynman integrand from the $1$-loop scattering equations in \eref{SE-loop}, one need to cut each propagator in the loop once, and sum over all resulting objects, as required by the partial fraction
relation \eref{partial-fraction}. For the amplitude without any color ordering, this requirement is satisfied automatically when summing over
all possible Feynman diagrams.
For the color ordered amplitude, this requirement is satisfied by summing over color orderings cyclically.

As an equivalent interpretation, the forward limit method can also be understood from the dimensional reduction point of view, as studied in \cite{Cachazo:2015aol}.

Let us take a brief glance at the CHY integrand at the $1$-loop level.
In the CHY framework, the forward limit operator ${\cal F}$ acts on the $(n+2)$-point tree amplitude as follows
\bea
{\cal F}\,{\cal A}_{n+2}&=&{\cal F}\,\int d\mu_{n+2}\,{\cal I}^L(\{k,\epsilon,z\}){\cal I}^R(\{k,\W\epsilon,z\})\nn
&=&\int d\mu'_{n+2}\,\Big({\cal F}\,{\cal I}^L(\{k,\epsilon,z\})\Big)\Big({\cal F}\,{\cal I}^R(\{k,\W\epsilon,z\})\Big)\,,~~~~\label{1loop-CHY}
\eea
where the measure $d\mu'_{n+2}$ is generated from $d\mu_{n+2}$ by turning the scattering equations to those in \eref{SE-loop}.
Thus the $1$-loop CHY integrand is determined by
\bea
{\cal I}^{L}_\circ(\{k,\epsilon,z\})={\cal F}\,{\cal I}^L(\{k,\epsilon,z\})\,,~~~~~~~~
{\cal I}^{R}_\circ(\{k,\epsilon,z\})={\cal F}\,{\cal I}^R(\{k,\epsilon,z\})\,.
\eea
Using this statement, the $1$-loop CHY integrands for GR, YM and BAS are given in Table \ref{tab:1-loop-theories}.
\begin{table}[!h]
    \begin{center}
        \begin{tabular}{c|c|c}
            Theory& ${\cal I}^L_\circ(k_i,\epsilon_i,z_i)$ & ${\cal I}^R_\circ(k_i,\W\epsilon_i,z_i)$ \\
            \hline
            GR & ${\cal F}\,{\bf Pf}'\Psi$ & ${\cal F}\,{\bf Pf}'{\Psi}$ \\
            YM & $PT_\circ(\sigma_1,\cdots,\sigma_n)$ & ${\cal F}\,{\bf Pf}' \Psi$ \\
            BAS & $PT_\circ(\sigma_1,\cdots,\sigma_n)$ & $PT_\circ(\sigma'_1,\cdots,\sigma'_n)$ \\
        \end{tabular}
    \end{center}
    \caption{\label{tab:1-loop-theories}$1$-loop CHY integrands }
\end{table}
Here $\Psi$ is a $2(n+2)\times2(n+2)$ matrix constituted by $\{k_1,\cdots,k_n,k_+,k_-\}$ and $\{\epsilon_1,\cdots,\epsilon_n,\epsilon_+,\epsilon_-\}$.
For simplicity, we assume the nodes $+$ and $-$ are located at $(n+1)^{\rm th}$ and $(n+2)^{\rm th}$ rows and columns, respectively, and the reduced Pfaffian is evaluated by removing them, i.e., ${\bf Pf}'\Psi={(-)\over z_{+-}}{\bf Pf}\Psi'$, with $\Psi'=\Psi^{+-}_{+-}$.
The $1$-loop Parke-Taylor factor $PT_\circ(\sigma_1,\cdots,\sigma_n)$ is obtained by summing over tree Parke-Taylor
factors cyclically,
\bea
PT_\circ(\sigma_1,\cdots,\sigma_n)=\sum_{i\in\{1,\cdots,n\}}\,PT(+,\sigma_i,\cdots,\sigma_{i-1},-)\,.~~~~\label{PT-C}
\eea
Notice that since the Parke-Taylor factor only depends on the coordinates of punctures, we have
\bea
{\cal F}\,PT(+,\sigma_i,\cdots,\sigma_{i-1},-)=PT(+,\sigma_i,\cdots,\sigma_{i-1},-)\,.
\eea
The tree Parke Taylor factor $PT(\cdots)$ at the r.h.s of \eref{PT-C} should be understood as
${\cal F}\,PT(\cdots)$.
The integrands in Table \ref{tab:1-loop-theories} can be found in \cite{Geyer:2015jch,Geyer:2017ela,He:2015yua,Cachazo:2015aol}.

The $1$-loop CHY formulas in \eref{1loop-CHY} suffer from the divergence in the forward limit. It was observed in \cite{He:2015yua}
that the solutions of $1$-loop scattering equations separate into three sectors which are called regular, singular ${\bf I}$ and singular ${\bf II}$,
according to the behavior of punctures $z_{\pm}$ in the limit $k_++k_-\to 0$. In this paper, we will bypass this subtle and crucial point by employing the conclusion in \cite{Cachazo:2015aol}, which can be summarized as follows: as long as the CHY integrand is homogeneous in $\ell^\mu$,
the singular solutions contribute to the scaleless integrals which vanish under the dimensional regularization. The homogeneity is manifest for the Parke-Taylor factor. For ${\cal F}\,{\bf Pf}'{\Psi}$, the only place that can violate the homogeneity in $\ell^\mu$ is the diagonal elements in the matrix $C$, since the deleted rows and columns are chosen to be $k_+$ and $k_-$. Singular solutions correspond to $z_+=z_-$, then it is direct to observe that the dependence on $\ell^\mu$ exactly cancel away, left with a homogeneous CHY integrand.
This observation allows us to ignore the problem of singular solutions.


\end{document}